\documentclass[aps,prc,twocolumn,nofootinbib,amsmath,amssymb]{revtex4-2}

\usepackage{graphicx}
\usepackage{dcolumn}
\usepackage{bm}
\usepackage{physics}
\usepackage{color}
\usepackage{todonotes}
\usepackage{placeins}
\usepackage{slashed}
\usepackage{amssymb}
\usepackage{hyperref}
\usepackage[capitalise]{cleveref}
\usepackage[caption=false]{subfig}

\newcommand*{\pr}{\mathrm{pr}}

\newcommand*{\yth}{y_\mathrm{th}}

\newcommand*{\lecs}{\bm{\alpha}}
\newcommand*{\thetacm}{\theta_\mathrm{c.m.}}
\newcommand*{\tS}{^3S_1 \mathrm{-}^3D_1}
\newcommand*{\tP}{^3P_2 \mathrm{-}^3F_2}

\begin{document}

\title{Bayesian Analysis of $\chi$EFT at Leading Order in a Modified Weinberg Power 
Counting Approach}
\author{Oliver Thim} \email{oliver.thim@chalmers.se}
\author{Eleanor May}
\author{Andreas Ekström}
\author{Christian Forssén}
\affiliation{Department of Physics, Chalmers University of Technology, SE-412 96, Göteborg, Sweden}

\date{\today}

\begin{abstract}
We present a Bayesian analysis of renormalization-group invariant nucleon-nucleon interactions at leading order in chiral effective field theory ($\chi$EFT) with momentum cutoffs in the range 400--4000~MeV. We use history matching to identify relevant regions in the parameter space of low-energy constants (LECs) and subsequently infer the posterior probability density of their values using Markov chain Monte Carlo. All posteriors are conditioned on experimental data for neutron-proton scattering observables and we estimate the $\chi$EFT truncation error in an uncorrelated limit. We do not detect any significant cutoff dependence in the posterior predictive distributions for two-nucleon observables. For all cutoff values we find a multimodal LEC posterior with an insignificant mode harboring a bound $^1{S}_0$ state. The $^3P_0$ and $^3P_2$ phase shifts emerging from the Bayesian analysis are less constrained and typically more repulsive compared to the results of a phase shift optimization. We expect that our inference will impact predictions for nuclei. This work demonstrates how to perform inference in the presence of limit-cycle-like behavior and spurious bound states, and lays the foundation for a Bayesian analysis of renormalization-group invariant $\chi$EFT interactions beyond leading order.
\end{abstract}

\maketitle
\section{\label{sec:introduction}Introduction}
\noindent
The foundational principles of chiral effective field theory ($\chi$EFT)~\cite{Weinberg:1978kz,Weinberg:1990rz,Weinberg:1991um} promise a systematically improvable description of the strong force between nucleons that is consistent with quantum chromodynamics (QCD). However, establishing a power counting of chiral nuclear interactions that fulfills renormalization requirements presents several challenges---see, e.g.,~\citet{vanKolck:2020llt} for an overview. Recently,~\citet{Yang:2020pgi} analyzed nuclear ground-state energies using renormalization-group (RG) invariant formulations of $\chi$EFT up to (perturbative) next-to-leading order (NLO) corrections. Apparently, the essential mechanism for nuclear-binding tends to fail already at leading order (LO) for selected $A>4$ nuclei when using available RG-invariant power counting schemes. In fact, similar results had already been found using $\chi$EFT~\cite{Carlsson:2015vda} based on the canonical Weinberg power counting (WPC)~\cite{Bedaque:2002mn,Epelbaum:2008ga,Machleidt:2011zz,Hammer:2019poc}, and pionless EFT~\cite{Stetcu:2006ey,Contessi:2017rww,Bansal:2017pwn}.
\citet{Yang:2020pgi} put forward three possible reasons for these shortcomings at LO in $\chi$EFT: i) One (or more) scales critical to the physical description of finite nuclei might not be correctly captured by the contact terms at LO~\cite{Schafer:2020ivj}; ii) The LO nucleon-nucleon ($NN$) interaction should possibly be complemented with other interaction terms such as sub-leading pion exchange~\cite{Mishra:2021luw} and many-nucleon interactions~\cite{Kievsky:2016kzb,Bazak:2018qnu,Yang:2019hkn,Yang:2021vxa}; iii) The description of the nuclear interaction might be finely tuned and therefore require careful calibration of the low energy constants (LECs). Indeed, \citet{Yang:2020pgi} renormalized the relevant LECs by demanding exact reproduction of selected phase shifts at a single scattering energy. This procedure resulted in point estimates of the LECs without the possibility to analyze possible fine-tuning effects. 

In this work we tackle overfitting and expose possible fine tuning using Bayesian methods. Specifically, we infer a posterior probability density function (pdf) for the values of the LECs conditioned on neutron-proton ($np$) scattering data. The advantages of a Bayesian approach are several. First, we obtain a probabilistic measure of our uncertainty about the values of the LECs and subsequent predictions, something that is not obtained when doing maximum likelihood estimation~\cite{Carlsson:2015vda}. Second, with the Bayesian approach we can utilize the expected systematicity of $\chi$EFT as prior knowledge about the truncation error~\cite{Furnstahl:2015rha}. There exist several Bayesian studies of $\chi$EFT interactions~\cite{Melendez:2017phj,Melendez:2019izc,Wesolowski:2021cni,Svensson:2022,Svensson:2022kkj} and quantified posterior predictive distributions (ppds) of nuclear properties~\cite{Drischler:2020hwi,Djarv:2021hcj,Hu:2021trw}. However, so far all such studies are grounded in $\chi$EFT based on WPC.

In EFTs that are perturbative at all orders, power counting usually follows the momentum scaling of the different Feynman diagrams. This is known as naive dimensional analysis (NDA)~\cite{Manohar:1983md}. However, in $\chi$EFT we must account for bound multi-nucleon states and therefore face non-perturbative physics with the consequence of infrared enhancement of purely nucleonic intermediate states. To deal with this, Weinberg~\cite{Weinberg:1990rz,Weinberg:1991um} suggested to apply NDA to the potential which is then iterated to all orders by solving, e.g., the Lippmann-Schwinger equation. This prescription assumes that the iteration to infinite order does not introduce additional divergences with the need for higher-order counterterms---an assumption that did not hold upon closer inspection. Indeed, taking the momentum cutoff $\Lambda$ of the regulator very large, i.e., far beyond the anticipated breakdown scale of $\chi$EFT, \citet{Nogga:2005hy} found that WPC does not yield RG-invariant amplitudes in attractive spin-triplet partial-waves. By now there exist several proposals on how to modify WPC, see, e.g., Refs.~\cite{Long:2012ve,Long:2013cya,PhysRevC.85.034002,Yang:2020pgi, Yang:2019hkn,Valderrama:2008kj,Birse:2005um,Kaplan:1998tg,Long:2007vp}. We refer to all such proposals as modified Weinberg power counting (MWPC). One can argue that WPC provides a consistent EFT framework as long as $\Lambda$ is kept in the vicinity of the breakdown scale $\Lambda_{\chi}$~\cite{Epelbaum:2006pt,Epelbaum:2009sd,Epelbaum:2018zli,Gasparyan:2022isg}. In that scheme, all orders are typically summed up non-perturbatively and, starting at third order, it can provide realistic predictions for selected nuclei~\cite{Hergert:2020bxy} and nuclear matter~\cite{Hebeler:2020ocj}. However, this achievement does not imply that WPC provides the foundation for an EFT of QCD.

This paper is organized as follows: In Sec.~\ref{sec:theory} the LO potential and non-relativistic $np$ scattering are addressed along with relevant background regarding power counting and the renormalization of singular potentials. In Sec.~\ref{sec:bayes} we discuss our Bayesian approach and in Sec.~\ref{sec:sampling} we review the numerical sampling of the posterior pdf for the LECs. Finally, in Sec.~\ref{sec:ppd} ppds are presented and analyzed for some $np$-scattering observables and the deuteron ground-state energy. A conclusion and outlook is presented in Sec.~\ref{sec:conclusion}.

\section{\label{sec:theory}Theory}
To keep this work self-contained we first discuss how counterterms (and associated LECs) are introduced to renormalize the singular nature of the one-pion exchange (OPE) potential at LO in $\chi$EFT. We also study limit-cycle-like behavior in some detail since they play an important role during the Bayesian inference.

\subsection{Effective field theory expansion}
The EFT approach promises an order-by-order improvable description of a nuclear observable $y$ residing in the low-energy domain below the relevant breakdown scale. In general, up to some finite order $n$ we can expect an expansion of the form~\cite{Hammer:2019poc,Griesshammer:2020fwr}
\begin{align}
    y^{(n)}_\mathrm{th} &= y_\text{ref} \sum_{k=0}^n  b_k\left(\frac{Q}{\Lambda_\chi}\right)^k\ +   y_\text{ref} \left(\frac{Q}{\Lambda_\chi}\right)^{n+1} \mathcal{D}_{n+1}\left(\Lambda\right),
    \label{eq:EFT_expansion}
\end{align}
and we specialize to $\chi$EFT assuming a breakdown scale $\Lambda_\chi=600$ MeV as per previous analyses~\cite{Epelbaum:2014efa,Melendez:2017phj} and denote the momentum cutoff by $\Lambda$. We also assume a soft scale $Q = \max(p,m_\pi)$ with $p$ denoting the external momentum and $m_\pi$ denoting the pion mass. The unspecified function $\mathcal{D}_{n+1}$ depends on ratios of the relevant low energy scales and, importantly, absorbs the residual cutoff dependence. In this work we set the reference scale $y_\text{ref}$ to the corresponding experimental value $y_\text{exp}$ of $y$. Pulling out $y_\text{ref}$ leads to dimensionless expansion coefficients $b_k$, and if the theory is renormalized order-by-order, these should not exhibit any cutoff dependence although they do depend on ratios of relevant scales. Also, along the lines of naturalness~\cite{vanKolck_2020}, we expect the $b_k$-values to be of order unity.

For a perturbative EFT---where NDA applied to the Feynman diagrams carries over to the amplitudes and hence to observables---the power counting in \cref{eq:EFT_expansion} follows straightforwardly from the Lagrangian. The relation between the Lagrangian and observables in a non-perturbative theory like $\chi$EFT is less direct. The non-perturbative effects in combination with the need to treat the problem numerically pose challenges to finding a consistent power counting. When the amplitudes cannot be obtained perturbatively, the LO $T$-matrix must scale as $T \sim Q^{-1}$~\cite{Barford:2001sx,Barford:2002je,Griesshammer:2020fwr}. The summation in \cref{eq:EFT_expansion} can still start at $k=0$, since the $Q^{-1}$ dependence can be absorbed into $y_\mathrm{ref}$. Terms with $k=1,2,\ldots$ then correspond to higher-order corrections with respect to LO. We employ MWPC where amplitudes of the LO potential are computed non-perturbatively and sub-leading corrections should be accounted for using perturbation theory \cite{vanKolck:2020llt,Valderrama:2009ei,PavonValderrama:2011fcz,PhysRevC.85.034002,PhysRevC.84.057001}. 

\subsection{\label{sec:potential}Two-nucleon potential and scattering amplitudes}
The momentum-space and isospin-symmetric LO potential considered in this work is adopted from Ref.~\cite{Yang:2020pgi}, and using our conventions it reads 
\begin{align}
    V(\bm{p}',\bm{p}) &= \frac{1}{(2\pi)^3} \Bigg[-\frac{g^2_A}{4f^2_\pi} \frac{\left(\bm{\sigma}_1\cdot \bm{q}\right)\left(\bm{\sigma}_2\cdot \bm{q}\right)}{\bm{q}^2 + m^2_\pi} \left(\bm{\tau}_1\cdot\bm{\tau}_2\right) + \nonumber \\ & + \tilde{C}_{^1S_0} + \tilde{C}_{^3S_1} + \left(C_{^3P_0} + C_{^3P_2}\right)p'p \Bigg].
    \label{eq:LO_potential}
\end{align}
The first term is the OPE potential where $\bm{\tau}_i$ and $\bm{\sigma}_i$, for $i=1,2$, are the isospin and spin operators for the respective nucleon, $\bm{p}$ ($\bm{p}'$) are the ingoing (outgoing) relative momenta with normalization $\braket{\bm{p}'}{\bm{p}} = \delta^3\left(\bm{p}'-\bm{p}\right)$, $\bm{q} = \bm{p}'-\bm{p}$ is the momentum transfer, and we have $p \equiv |\bm{p}|$ and $p' \equiv |\bm{p}'|$. The contact LECs: $\tilde{C}_{^1S_0},\tilde{C}_{^3S_1},C_{^3P_0}, C_{^3P_2}$, carry implicit projection operators such that they act in the indicated partial-waves, $^{(2s+1)}l_j$, where $s,l,j$ denote the quantum numbers of the $NN$ spin, orbital angular momentum, and total angular momentum, respectively. We use the PDG~\cite{PDG_2022} values for the axial coupling constant $g_A = 1.275$, the pion decay constant $f_\pi = 92.1$~MeV, and averaged pion mass $m_\pi = 138.039$~MeV. Finally, for the partial-wave decomposed~\cite{Erkelenz:1971caz} potential, and operators, we use the notation $V_{l'l}^{sj}(p',p) \equiv \bra{l'sj;p'} V \ket{lsj;p}$. As discussed in Sec.~\ref{sec:PC} and Ref.~\cite{Yang:2020pgi}, this LO potential is restricted to low partial waves: $^1S_0$, $\tS$, $^1P_1$, $^3P_0$, $^3P_1$, $\tP$. Consequently, it is set to zero for all partial waves with $l>1$ that have no coupling to $l \leq 1$. 

To render the integrals of the Lippmann-Schwinger equation finite, the relative momenta of in- and out-going nucleons are regulated using the function, 
\begin{eqnarray}
    f_{\Lambda}(p) = \exp\left[ -\frac{p^4}{\Lambda^4}\right]\times \theta(\Lambda+ (300 \ \mathrm{MeV})-p),
\end{eqnarray}
which was also used by \citet{Yang:2020pgi}\footnote{The exact form of the regulator was not given in Ref~\cite{Yang:2020pgi}. Thus, we define it here.}, where
\begin{equation}
    \theta(x) = \begin{cases} 1, \quad x > 0 \\ 0, \quad x \leq 0 \end{cases}.
\end{equation}
We straightforwardly account for the small effects of relativistic kinematics using the minimal relativity prescription\cite{Brown:1969tfp, Machleidt:2011zz}. Combined with the momentum regulators $f_{\Lambda}$, the LO potential in \cref{eq:LO_potential} is thus modified according to
\begin{eqnarray}
    V(\bm{p}',\bm{p}) \to f_{\Lambda}(p') \sqrt{\frac{m_N}{ E(p')}} V(\bm{p}',\bm{p}) \sqrt{\frac{m_N}{ E(p)}} f_{\Lambda}(p),
\label{eq:Full_LO_potential}
\end{eqnarray}
where $E(p) = \sqrt{m_N^2 + p^2}$ and $m_N= \frac{2m_pm_n}{m_p+m_n}$ is the nucleon mass with proton and neutron masses $m_p = 938.272$~MeV and $m_n = 939.918$~MeV, respectively~\cite{PDG_2022}.

We condition all inferences on $np$ scattering data and must therefore solve the corresponding Lippmann-Schwinger equation for the $T$-matrix
\begin{align}
    &T^{sj}_{l'l}(p', p) = V^{sj}_{l'l}(p',p) \ + \nonumber \\ &+ \sum_{l''}\int_0^\infty dk \ k^2 \ V^{sj}_{l'l''}(p',k) \frac{m_N}{p^2-k^2 + i\epsilon} T^{sj}_{l''l}(k,p).
    \label{eq:LS_pw}
\end{align}
We solve this numerically in momentum space using a standard matrix-inversion method by first converting to a real equation for the reaction matrix \cite{Haftel:1970zz}. Furthermore, we sum the partial-wave amplitudes to construct the spin scattering matrix, $M^s_{m'_sm_s}(p,\thetacm)$, see Appendix \ref{app:np_scattering}. Here, $m_s$, ($m'_s$) is the in-(out-)going total spin projection and $\thetacm$ is the center-of-mass scattering angle. All scattering observables can be computed from this spin scattering matrix~\cite{Bystricky:1976jr}. The types of observables that we condition our inferences on are listed in \cref{tab:exp_data} and discussed in Sec.~\ref{sec:likelihood}.

\subsection{\label{sec:PC}Singular potentials and limit-cycle-like behavior}
\label{sec:limit_cycles}
An attractive potential is \emph{singular} near the origin if it behaves as $-\lambda/r^n$ with $n>2$ (or $n=2$ with sufficiently large $\lambda>0$)~\cite{Case:1950an,Frank:1971xx}. Two particles interacting only via a singular potential will collapse towards the origin with increasing velocity. Akin to the infinities in quantum field theory, the singularity at the origin is cured via renormalization. The OPE potential is singular in attractive spin-triplet partial waves, e.g., $^3P_0$, $\tS$, $^3D_2$ and $\tP$~\cite{Nogga:2005hy}. From an EFT perspective, the singular character of the OPE becomes physically meaningful only with the addition of counterterms that parameterize, and cure, these short-range pathologies~\cite{PhysRevA.64.042103,HAMMER2006306}.

To avoid the introduction of infinitely many counterterms, the OPE potential in Eq.~\eqref{eq:LO_potential} is truncated to act only in partial waves with orbital angular momentum $l<l_c$. It is not obvious, however, where the limit $l_c$ should be set. Higher-order terms can be included using the distorted-wave Born approximation and there is evidence that this does not necessitate  the introduction of additional counterterms, see Refs.~\cite{Long:2007vp,Gasparyan:2022isg} for contrasting views. According to previous studies, OPE contributions to the $NN$ scattering amplitude in partial waves with orbital angular momentum $l >1$ can be treated perturbatively up to at least $p\sim300$~MeV~\cite{Birse:2005um,PhysRevC.99.024003,vanKolck:2020llt}. We therefore truncate the OPE potential such that it is nonzero only for channels that contain a partial wave component with $l \leq l_c = 1$.

\begin{figure}
\includegraphics[width=\columnwidth]{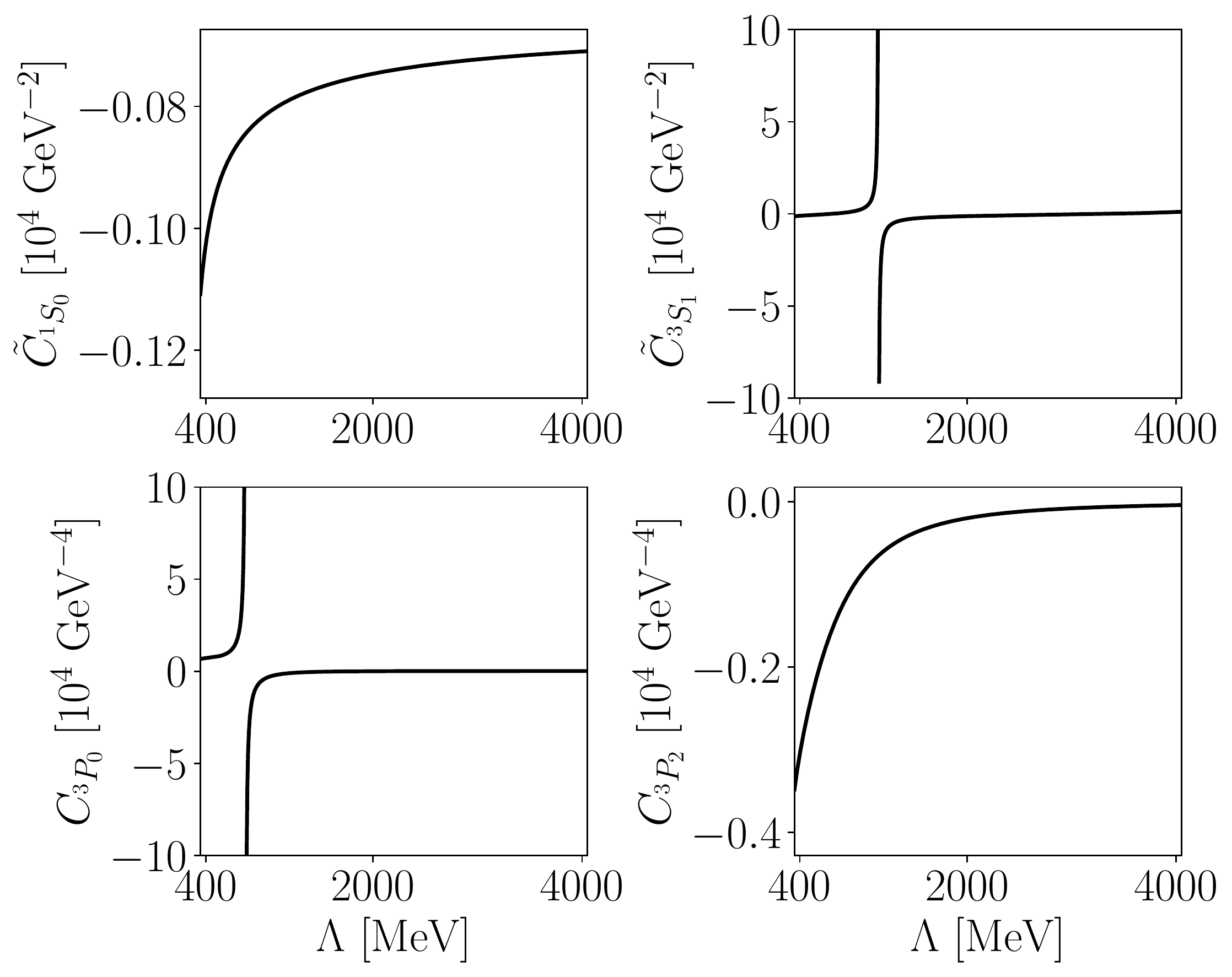}
\caption{LECs renormalized to reproduce the phase shift from the Nijmegen partial-wave analysis~\cite{Stoks:1993tb} at $T_\text{lab}=50$~MeV as a function of the cutoff $\Lambda$. The limit-cycle-like behavior in $\tilde{C}_{^3S_1}$ is seen at $\Lambda = 1150$ MeV and in $C_{^3P_0}$ at $\Lambda = 780$ MeV.}
\label{fig:running_all}
\end{figure}

LECs associated with counterterms that renormalize a singular potential with $n>2$ usually exhibit a limit-cycle-like behavior~\cite{PhysRevA.64.042103,HAMMER2006306,Nogga:2005hy}, i.e., provided that there is only one counterterm per partial wave, the corresponding LEC will exhibit periodic discontinuities as a function of the regulator value. This was extensively investigated for one-pion exchange in Ref.~\cite{Nogga:2005hy}. We can reproduce those results exactly, but we also observe a slight shift in the location of the limit-cycles-like behavior when including the minimal relativity correction in Eq.~\eqref{eq:Full_LO_potential}. In \cref{fig:running_all} we show how the LECs in the $^1S_0$, $^3S_1$, $^3P_0$ and $^3P_2$ partial waves run with $\Lambda$ when renormalized to reproduce the Nijmegen partial-wave phase shifts~\cite{Stoks:1993tb} for the laboratory kinetic energy $T_\mathrm{lab} = 50$~MeV of the projectile nucleon. There is no limit-cycle-like behavior observed in the $^3P_2$ wave for the cutoff region studied here. 

\begin{figure}
\includegraphics[width=\columnwidth]{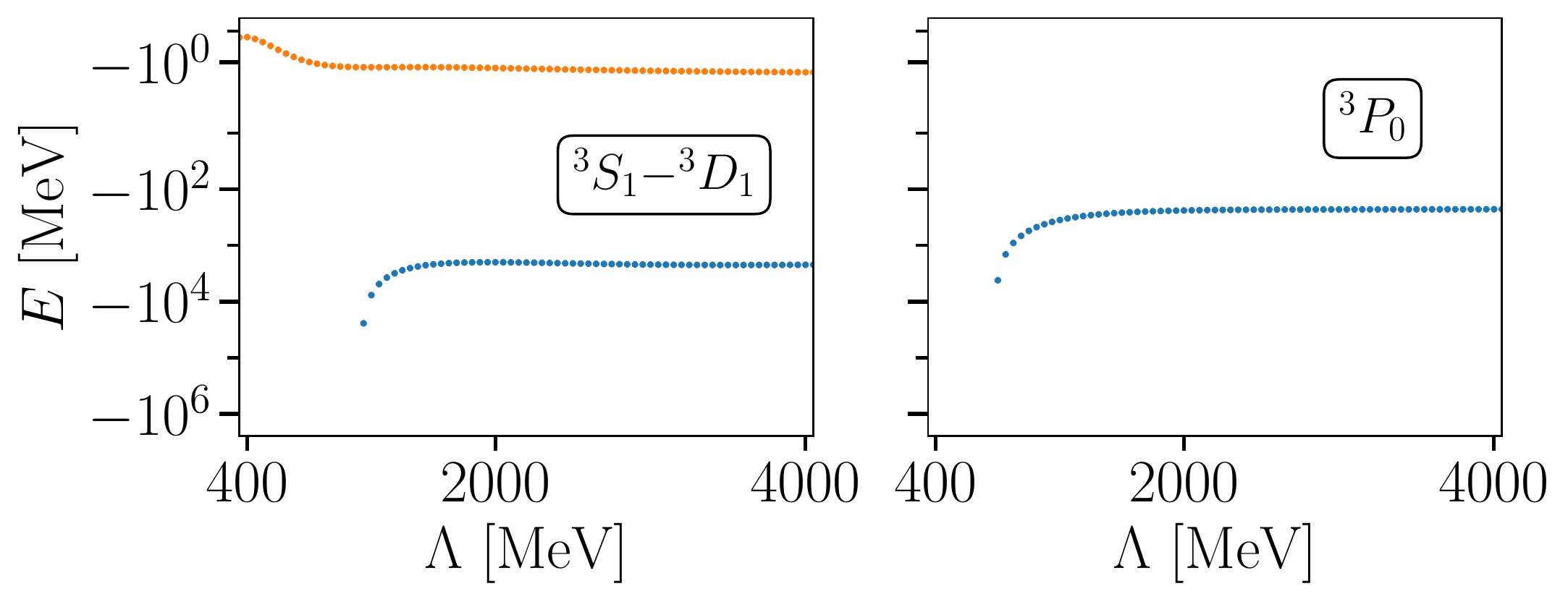}
\caption{Bound-state energies in selected channels with the LECs renormalized to reproduce the phase shift from the Nijmegen partial-wave analysis \cite{Stoks:1993tb} at $T_\text{lab}=50$ MeV. There are two bound states in the $\tS$ channel, where the upper one (orange dots) is the physical deuteron state while the lower one (blue dots) is spurious. The latter, deeply-bound state appears at $\Lambda \geq 1140$ MeV. In the $^3P_0$ channel there is a spurious, deeply-bound state for $\Lambda \geq 770$ MeV.}
\label{fig:bound_states}
\end{figure}

When this renormalization procedure is used, spurious and deeply bound states appear in the singular and attractive partial-waves as the cutoff $\Lambda$ is increased. This is not a problem \emph{if} the states are deeply bound and thus outside the applicable domain of the EFT. In practice, one can project these states out of the spectrum in a phase-equivalent fashion~\cite{Nogga:2005hy}. In \cref{fig:bound_states} we show how the spurious states in  $^3P_0$ and $\tS$ appear at the threshold value of $\Lambda$ corresponding to where the limit-cycle-like behavior is observed in \cref{fig:running_all}. 

In this work we infer the LEC posteriors at fixed values of the cutoff $\Lambda$. There exist threshold LEC values for which the total potential becomes sufficiently attractive such that a spurious bound state appears. The ($\Lambda$-dependent) LEC values for which this happens are tabulated in Appendix \ref{app:tables_bound_states} for the respective channels. Note that a spurious state is not deeply bound in the immediate vicinity of these threshold values and we stress that the behavior depicted in \cref{fig:bound_states} is obtained when requiring the exact reproduction of a specific phase shift.

\begin{figure}
\includegraphics[width=\columnwidth]{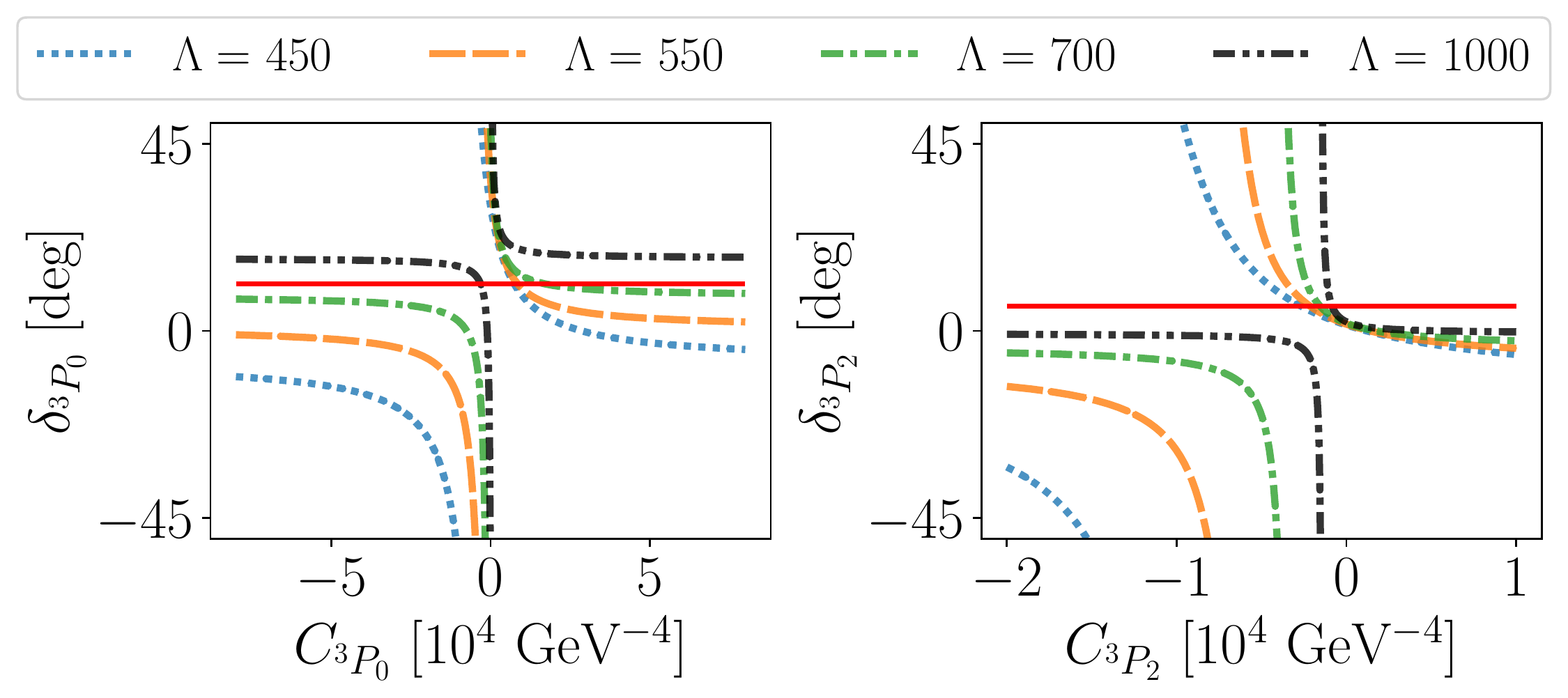}
\caption{The phase shifts $\delta_{^3P_0}$ and $\delta_{^3P_2}$ as a function of the respective LEC, for four different cutoff values $\Lambda=450,550,700,1000$~MeV at $T_\mathrm{lab} = 50$ MeV. The horizontal (red solid) line is the empirical value of the phase shift~\cite{Stoks:1993tb}.}
\label{fig:limit_cycles}
\end{figure}

As the bound state pole moves near the threshold when the corresponding LEC is varied, the phase shift will change dramatically by Levinson's theorem \cite{Taylor72}. Rapidly varying phase shifts lead to particular challenges in the inference process since scattering observables that are included in the likelihood, and that depend sensitively on partial-wave amplitudes, might then constrain the LEC values to very narrow regions of parameter space. In~\cref{fig:limit_cycles} we exemplify this by showing phase shifts at $T_\mathrm{lab} = 50$ MeV as a function of the LEC in the partial waves $^3P_0$ and $^3P_2$ for different cutoff values. The red, horizontal line indicates the empirical value of the phase shift according to the Nijmegen analysis~\cite{Stoks:1993tb}. \cref{fig:limit_cycles} also shows that for certain cutoffs we can obtain a wide range of LEC values that reproduce the empirical phase shift with reasonable precision. 

Figures \ref{fig:running_all} and \ref{fig:limit_cycles} provide complementary views on limit-cycles-like behavior. For example, in \cref{fig:running_all} we see that $C_{^3P_0}$ changes from a large positive value to a large negative one as we approach $\Lambda \approx 780$~MeV from below. The same information is contained in \cref{fig:limit_cycles} where the LO $^3P_0$ phase shift intersects with the Nijmegen result for a positive (negative) value of $C_{^3P_0}$ for $\Lambda=700$ $(1000)$~MeV. Enforcing the exact reproduction of a phase shift therefore implies that the spurious bound state will be deeply bound exactly when the limit-cycle-like behavior appears as shown in \cref{fig:running_all}. For the $^3P_2$ wave, the  phase shift only intersects with the Nijmegen result on one side of the discontinuity, at least for the low values of $\Lambda$ shown here. Consequently, there is no spurious bound state, or limit-cycle-like behavior appearing.

\section{\label{sec:bayes}Bayesian parameter estimation}
Bayes' theorem expresses the posterior pdf, $\pr\left(\bm{\theta}|D,I\right)$, for the relevant model parameters, $\bm{\theta}$, conditioned on experimental data, $D$, and other pertinent information, $I$, in terms of quantities that we can evaluate
\begin{equation}
    \pr\left(\bm{\theta}|D,I\right) = \frac{\pr\left(D|\bm{\theta},I\right) \times \pr\left(\bm{\theta}|I\right)}{\pr\left(D|I\right)}.
    \label{eq:Bayes}
\end{equation}
Here, $\pr\left(\bm{\theta}|I\right)$ is the \textit{prior}, $\pr\left(D|\bm{\theta},I\right)$ is the \textit{likelihood}, and $\pr\left(D|I\right)$ is the \textit{model evidence}. The latter is independent of $\bm{\theta}$ and can be omitted for parameter estimation purposes leaving us with the simpler expression
\begin{equation}
    \pr\left(\bm{\theta}|D,I\right) \propto  \pr\left(D|\bm{\theta},I\right) \times \pr\left(\bm{\theta}|I\right).
    \label{eq:posterior_theta}
\end{equation}
The set of model parameters in this study is collected in a vector
\begin{eqnarray}
    \bm{\theta} = \left(\bm{\alpha},\bar{c}\right),
\end{eqnarray}
where $\bm{\alpha}$ denotes the LECs of $\chi$EFT at LO and $\bar{c}$ governs the magnitude of the EFT truncation error, as described below in~\cref{eq:model_error}.
The elements of the vector $\bm{\alpha}$ at LO are 
\begin{equation}
    \bm{\alpha} = \left(\tilde{C}_{^1S_0},\tilde{C}_{^3S_1},C_{^3P_0}, C_{^3P_2}\right)
    \label{eq:lecs_def}
\end{equation}
in units of $10^4$ GeV$^{-2}$ and $10^4$ GeV$^{-4}$ for the $S$- and $P$-wave LECs, respectively.

\subsection{Likelihood}
\label{sec:likelihood}
When relating theory $y^{(n)}_\text{th}$ and experiment $y_\text{exp}$ for some observable $y$ we must account for both the theoretical and experimental uncertainties. This can be achieved with a statistical model
\begin{equation}
    y_\text{exp} = y^{(n)}_\mathrm{th} + \delta y_\mathrm{th} + \delta y_\text{exp},
    \label{eq:stat_model}
  \end{equation}
  in which the uncertainties are modeled by random variables $\delta y_\text{th}$ and $\delta y_\text{exp}$, respectively, and where we suppressed all parameter dependencies for simplicity.
One of the great benefits of working with EFTs is the expected systematicity of the truncation error. Assuming that QCD accounts for all relevant physics, and that our EFT is sound, i.e., it systematically approximates QCD in the low-energy regime, then we can express the value of some nuclear observable $y$ as
\begin{equation}
    y = y_\mathrm{ref} \sum_{k=0}^\infty b_k \left(\frac{Q}{\Lambda_\chi}\right)^k,
    \label{eq:y_EFT}
\end{equation}
in the notation of \cref{eq:EFT_expansion}. Thus, the LO theoretical prediction can be expressed as
\begin{equation}
    \yth^{(0)} = y_\mathrm{ref} \Bigg[b_0
    +\left(\frac{Q}{\Lambda_\chi}\right) \mathcal{D}_{1}\Bigg],
  \end{equation}
where the second term contains the residual cutoff dependence through $\mathcal{D}_1$. For the present case we assume that $b_1=0$ by time-reversal and parity invariance~\cite{Epelbaum:2008ga,Machleidt:2011zz,Hammer:2019poc}. With this, \cref{eq:y_EFT} can be written as
\begin{equation}
    y = y_\text{th}^{(0)} + y_\text{ref} \sum_{k=2}^\infty c_{k}  \left(\frac{Q}{\Lambda_\chi}\right)^k.
    \label{eq:EFT_div}
\end{equation}
The coefficients $c_k$ are related to $b_k$ and the power series expansion of $\mathcal{D}_1$. Since $b_1 = 0$, it implies that $c_1 = 0$ if the first term in the power-series expansion of $\mathcal{D}_1$ vanishes, which we assume it does by the same symmetry argument as for $b_1$. The coefficients $c_k$, like $b_k$, are also assumed to be of natural size, i.e., of order one. Finally, we note that $c_k$ might have a residual cutoff dependence inherited from $\mathcal{D}_1$.

We can now express the LO truncation error as
\begin{equation}
    \delta y_\mathrm{th} \equiv y_\text{ref} \sum_{k=2}^\infty c_{k} \left(\frac{Q}{\Lambda_\chi}\right)^k.
\end{equation}
Quantitatively, this is unknown to us since we have not computed anything beyond LO. However, we can evaluate the probability distribution of the random variable $\delta y_\mathrm{th}$ if we assume a distribution for the expansion coefficients $c_k$ guided by domain knowledge included in $I$. 
In this work we follow Ref. \cite{Furnstahl:2015rha} and assume that 
\begin{equation}
    \pr\left(c_k|I\right) = \mathcal{N}\left(0,\bar{c}^2\right),
    \label{eq:c_k_c_bar}
\end{equation} 
where $\mathcal{N}(\mu,\sigma^2)$ denotes a normal distribution with mean $\mu$ and variance $\sigma^2$. Assuming independent and identically distributed $c_k$ coefficients the distribution for $\delta y_\mathrm{th}$ at each kinematical point is given by \cite{Wesolowski_2019}
\begin{align}
  \pr\left(\delta y_\mathrm{th}|I\right) &= \mathcal{N}\left(0,\sigma^2_\mathrm{th}\right), \nonumber \\
  \text{with }\sigma^2_\mathrm{th} &= \frac{y_\mathrm{ref}^2 \bar{c}^2 \left(\frac{Q}{\Lambda_\chi}\right)^4}{1-\left(\frac{Q}{\Lambda_\chi}\right)^2}.
    \label{eq:model_error}
\end{align}
Hence, the variance of the truncation error is governed by $\bar{c}^2$, which is unknown \textit{a priori}. However, operating with an EFT we expect $\bar{c}$ to be of order one. We will quantify this in Sec.~\ref{sec:prior}. 

The experimental error for each datum is assumed to follow a normal distribution with variance $\sigma^2_\mathrm{exp}$, i.e.,
\begin{equation}    
\pr\left(\delta y_\mathrm{exp}|I\right) = \mathcal{N}\left(0,\sigma^2_\mathrm{exp}\right).
\end{equation}

Assuming that the EFT truncation error and the experimental errors are independent the likelihood for a single observation reads
\begin{align}
    &\pr\left(y_\mathrm{exp} | \bm{\theta}, \sigma_\mathrm{exp},I\right)=\mathcal{N}\left(\yth^{(0)}(\lecs), \sigma^2_\mathrm{th}(\bar{c}) + \sigma^2_\mathrm{exp}\right),
    \label{eq:pr_one_obs}
\end{align}
by \cref{eq:stat_model}, where we now explicitly indicate the relevant parameter dependencies.
Following Refs.~\cite{Svensson:2022,Epelbaum:2014efa} we further assume that EFT errors of scattering observables at different energies and angles are independent. 

The experimental data set, $D = \{ y_{\mathrm{exp},i}\}_{i=1}^N$, on which we condition the MCMC inference is listed in \cref{tab:exp_data} and contains $N=1043$ scattering observables. This is not all the data in the Granada database~\cite{Granada_1,Granada_2}. We attempted an inference using all $np$ observables with $T_{\mathrm{lab}}<100$ MeV, but a model check revealed that most calibration data were poorly reproduced. For example, low-energy total cross sections deviated significantly from experimental values, which in turn yielded significantly overbound deuteron states for most cutoffs. This is probably due to a misspecified EFT truncation error, leading to unnaturally large values for $\bar{c}$ and thus overestimated EFT errors for low-energy cross sections (that are expected to be reproduced relatively well at LO). There are several suggestions~\cite{Melendez:2019izc} for how to construct more sophisticated EFT error models that would  accommodate the incorporation of more data in the likelihood. We postpone such developments until we have incorporated sub-leading orders in MWPC.

\begin{table}[b]
\caption{Total and differential $np$ scattering observables used to condition the Bayesian inference in this work. The total number of data points is 1043. This is a subset of the data in the Granada database \cite{Granada_1,Granada_2}, where the normalizations determined from the Granada analysis are included. The uncertainties in the normalizations are not taken into account. The definitions of the various scattering observables can be found in Ref. \cite{Bystricky:1976jr}.}
\begin{ruledtabular}
\begin{tabular}{ccc}
Observable &$T_\mathrm{lab}$ [MeV]& \# of data points\\
\colrule
$\sigma$ & $10^{-6}$ -- 99.0 & 324  \\
${d\sigma}/{d\Omega}$ & 2.72 -- 99.0 & 698  \\
$\sigma_T$ & 3.65 -- 17.1 & 13  \\
$\sigma_L$ & 4.98 -- 66.0 & 8 
\end{tabular}
\end{ruledtabular}
\label{tab:exp_data}
\end{table}

\subsection{Priors}
\label{sec:prior}
\begin{figure}
\includegraphics[width=0.7\columnwidth]{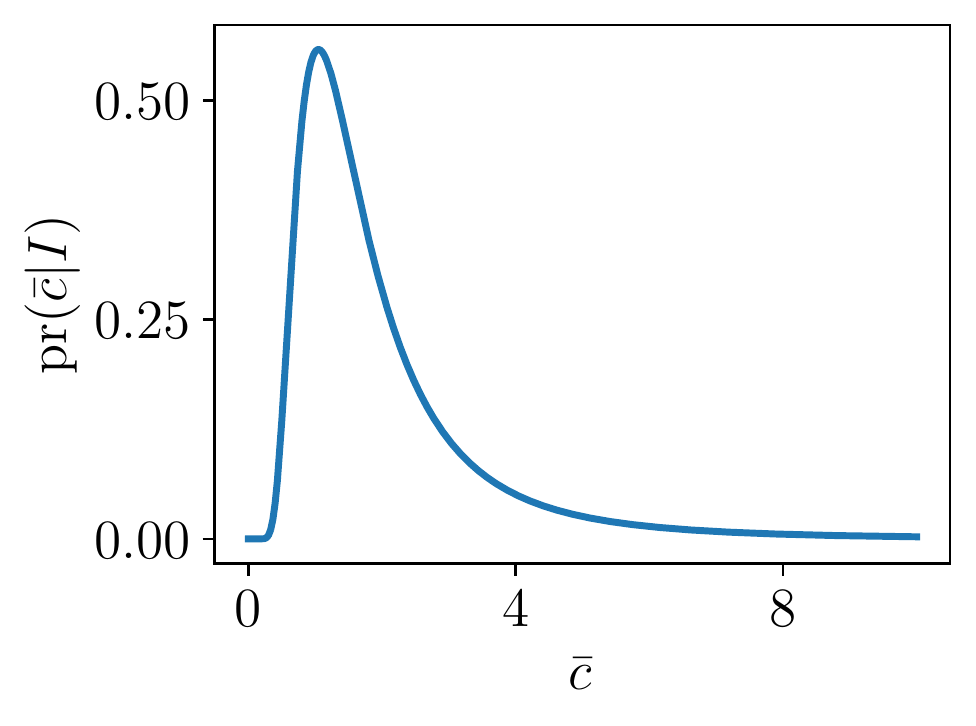}
\caption{The prior pdf for $\bar{c}$, $\pr\left(\bar{c}|I\right)$.}
\label{fig:prior}
\end{figure}

We treat $\bm{\alpha}$ and $\bar{c}$ as random variables and must therefore place priors on them.   For the LECs we employ independent and normally distributed priors with a standard deviation of four times their naturalness estimates: $  \frac{4\pi}{f^2_\pi}  \approx 0.1 \times 10^4 \ \mathrm{GeV}^{-2}$ and $\frac{4\pi}{f^2_\pi \Lambda^2_\chi} \approx 0.4 \times 10^4 \ \mathrm{GeV}^{-4}$ for the $S$- and $P$-wave LECs, respectively \cite{Epelbaum:2009sd,Machleidt:2011zz,Hammer:2019poc}. Similarly, we expect $\bar{c}$ to be of natural size and always positive. As shown in \cref{fig:prior}, we employ an inverse gamma distribution 
\begin{equation}
    \pr\left(\bar{c}|I\right) = \frac{\beta^\alpha}{\Gamma(\alpha)}(1/\bar{c})^{\alpha+1}\exp\left(-\beta/\bar{c}\right),
    \label{eq:cbar_prior}
\end{equation}
with parameters $\alpha=3$ and $\beta=4.2$. This distribution has a mode around one, but also a heavy tail that allows for a significant variation.

\section{\label{sec:sampling}Sampling the posterior at different cutoff values}
The parameter inference proceeds in two steps. First, we employ a Bayes linear approach known as history matching~\cite{Vernon:2010, Vernon:2014, Vernon:2018, Hu:2021trw} to identify relevant domains for the LECs. Second, we employ an affine invariant MCMC algorithm called \texttt{emcee}~\cite{emcee} to numerically draw samples from the posterior pdf. The history-matching domains identified in the first step are used to initialize the MCMC sampling.  Furthermore, to avoid detrimental influence of the limit-cycle-like behavior during sampling we select momentum cutoffs based on the analysis in Sec.~\ref{sec:PC}.

\subsection{History Matching}
\label{sec:hm}
History matching is an iterative scheme that can be employed to identify and exclude regions of parameter space that produce model outputs inconsistent with observational data, taking relevant uncertainties into account. See, e.g., Ref.~\cite{Hu:2021trw} for details of the algorithm and an application in nuclear physics. At the end, history matching returns the region(s) of parameter space that could not be ruled out, the so-called \emph{non-implausible} domain.

In this work, however, we employ history matching as a precursor to a full Bayesian analysis. A similar use of history matching can be found in Ref.~\cite{Kondo:2023lty}. Our aim is then to identify regions of parameter space where we expect to find large contributions to the posterior probability mass. We will be less concerned with the observational property of the data that we utilize in this step (we will employ scattering phase shifts), or with the rigour of our uncertainty estimates, as the sole purpose is to pick promising starting points for the MCMC algorithm. For this reason we refer to excluded samples as \emph{inefficient} starting points, whereas the final domain is at least \emph{non-inefficient}. These designations replace the standard implausible and non-implausible labels that are found in the history-matching literature. 

We employ an ``inefficiency'' measure $I_M(\bm{\alpha})$ that gauges the performance of the theoretical predictions for a selected subset of data given values for the LECs
\begin{align}
    I_M(\lecs) \equiv \max_i \sqrt{ \frac{ \abs{y^{(0)}_{\mathrm{th},i}\left(\lecs\right) - y_{\mathrm{exp},i}}^2}{ \sigma^2_{\mathrm{exp},i} + \sigma^2_{\mathrm{th},i}}}.
    \label{eqn:implaus}
\end{align}
The index, $i$, enumerates the set of experimental observations that is included in the history matching. The measure $I_M(\lecs)$ is defined as the maximum over the observations $\{y_{\mathrm{exp},i}\}$ and is therefore governed by the datum that is reproduced the worst. The experimental and theoretical errors are incorporated via the variances $\sigma^2_{\mathrm{exp},i}$ and $\sigma^2_{\mathrm{th},i}$, respectively. A second measure, $I_{2M}(\bm{\alpha})$, is sometimes used to construct an additional constraint. It is analogously defined as the second maximum over the observations and obviously fulfills $I_{2M}(\lecs) \leq I_M(\lecs)$.

\begin{table}[b]
\caption{$np$ scattering phase shifts in degrees, used as observational data within the first wave of history matching. Values for $T_{\mathrm{lab}}$ are given in MeV. The phase shift at $T_{\mathrm{lab}}=40$~MeV was obtained through spline interpolation. All other phase shifts were taken from Refs.~\cite{Granada_1, Granada_2}. The standard deviation for the theoretical error in Eq.~\eqref{eqn:implaus} is presented in the last column as a relative error.}
\begin{ruledtabular}
\begin{tabular}{cccccc}
$T_{\mathrm{lab}}$ & $\delta_{^1S_0}$ & $\delta_{^3S_1}$ & $\delta_{^3P_0}$ & $\delta_{^3P_2}$ & $\sigma_\mathrm{th}/\delta$\\
\colrule
     1 & 62.1 & 147.6 & 0.181 & 0.022 & 0.3\\
     5 & 63.7 & 118.0 & 1.66 & 0.258 & 0.3\\
     10 & 60.0 & 102.3 & 3.75 & 0.727 & 0.3\\
     25 & 51.1 & 80.2 & 8.48 & 2.63 & 0.3\\
     40 & 44.8 & 67.6 & 10.7 & 4.70 & 0.3\\
     100 & 27.2 & 42.7 & 9.14 & 11.06 & 0.8\\
\end{tabular}\\
\end{ruledtabular}
\label{tab:HM_scattering_phaseshifts}
\end{table}

\begin{figure}
\includegraphics[width=\columnwidth]{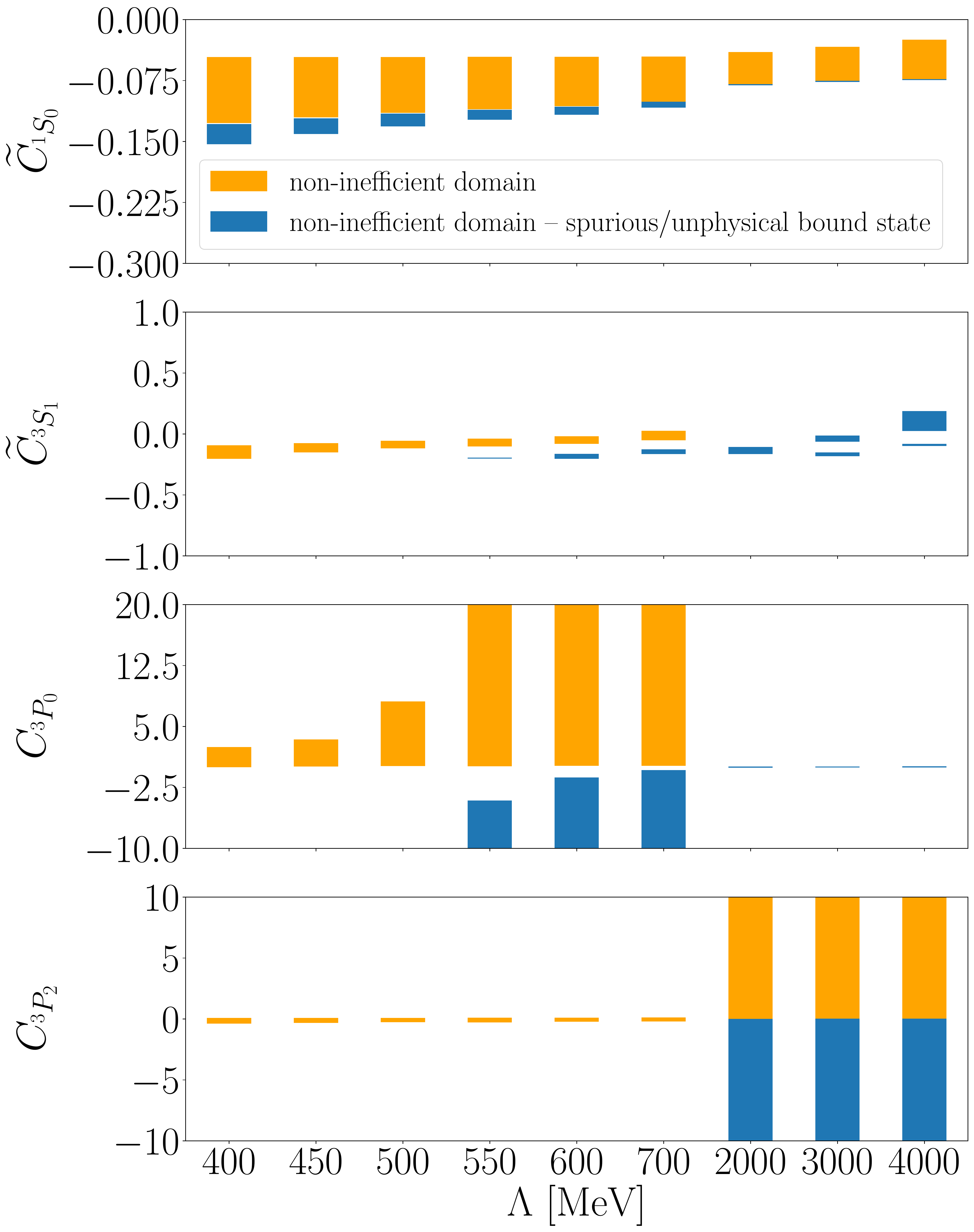}
\caption{The ranges of relevant starting points for each LEC, resulting from two waves of history matching for several cutoff values. The units used for the LECs are defined below \cref{eq:lecs_def}. Non-inefficient ranges are separated and color-coded according to if the channel contains spurious bound states. The range of each LEC axis indicates the region that was searched in the first wave of history matching.}
\label{fig:historymatch_results}
\end{figure}

We perform two waves of history matching. In the first wave, the scattering phase shifts listed in \cref{tab:HM_scattering_phaseshifts} are considered as observational data. Since all LECs act in distinct partial waves, each LEC could be constrained independently within one-dimensional sub-waves. We estimate conservative phase-shift errors $\sigma_\mathrm{th}$ from \cref{eq:model_error} using fixed $\bar{c} \approx 6$. This choice implies a $30\%$ error for $T_\mathrm{lab} < 40.6 $ MeV (i.e., $p<m_{\pi}$) and $80 \%$ for phase shifts with $T_\mathrm{lab} \approx 100$ MeV. For each sub-wave we use $10^4$ different LEC values in a space-filling Latin hypercube design~\cite{JosephRoshan} across a rather wide interval informed by the phase-shift analysis in Sec.~\ref{sec:limit_cycles} and corresponding to the ranges shown on the $y$-axes in \cref{fig:historymatch_results}. Samples for which the inefficiency measures are larger than some cutoff values are deemed as poor candidates for initializing the MCMC algorithm. In this analysis we use a sequence of cutoffs, $I_M>3.0$ combined with $I_{2M}>2.5$, which is similar to Ref.~\cite{Vernon:2018} (therein based on Pukelsheim's three sigma rule \cite{pukelsheim}). In the end, the relevant parameter volume is reduced by a large fraction. The ratio of the non-inefficient over initial volume was found to be as small as $10^{-5}$, although this ratio depends rather strongly on the value of the cutoff.

In the second wave, a set of 13 $np$ scattering observables is used to construct the inefficiency measures. In this wave all four LECs are active simultaneously.
Here we employ $10^{5}$ samples, using the same space-filling design, in the non-inefficient domain resulting from the first wave. The larger number of samples, and the much reduced volume helps to provide sufficient resolution for detecting narrow domains. The set of observables and corresponding model uncertainties, estimated using the same prescription as in the first wave, can be found in \cref{tab:HM_scattering_observables}.

\begin{table}[b]
\caption{$np$ scattering observables used within the second wave of history matching. Note that $\sigma$, without any subscripts, is denoting the total $np$ cross section, while $\sigma_\mathrm{th}/y_\mathrm{exp}$ is the (relative) model error for each observable.}
\begin{ruledtabular}
\begin{tabular}{ccccc}
Observable & $T_{\mathrm{lab}}$ [MeV] & $\thetacm$ [deg] & $y_\mathrm{exp}$    & $\sigma_\mathrm{th}/y_\mathrm{exp}$ \\
\colrule
    $\sigma$ & 0.12   & - & 12050.0 mb & 0.3\\
    $\sigma$ & 3.186  & - & 2206.0 mb   & 0.3\\
    $\sigma$ & 12.995 & - & 749.0 mb    & 0.3\\
    $\sigma$ & 28.0  & - & 321.5 mb    & 0.3\\
    $\sigma$ & 40.0   & - & 217.8 mb    & 0.3\\
    $\sigma$ & 97.2  & - & 76.0 mb     & 0.8\\
    $d \sigma/d\Omega$& 99.0 & 21.0  & 8.01 mb/sr   & 0.8\\
    $d \sigma/d\Omega$& 99.0 & 78.0  & 2.14 mb/sr   & 0.8\\
    $d \sigma/d\Omega$& 99.0 & 149.0 & 9.50 mb/sr   & 0.8\\
    $P_b$ & 25.0 & 33.1 & 0.047   & 0.3\\
    $P_b$ & 25.0 & 90.3 & 0.053   & 0.3\\
    $P_b$ & 95.0 & 29.8 & 0.170   & 0.8\\
    $P_b$ & 95.0 & 88.5 & 0.291   & 0.8\\
\end{tabular}
\end{ruledtabular}
\label{tab:HM_scattering_observables}
\end{table}

The resulting, relevant ranges for each of the LECs are shown in \cref{fig:historymatch_results}. The domains are classified, and color-coded, according to if the LEC gives rise to a spurious (unphysical) bound state in the respective channel. The rapid variation of the phase shift by $180^\circ$ where a bound state appears---discussed and illustrated in connection to \cref{fig:limit_cycles}---gives rise to disjoint domains. Therefore, we can anticipate multimodal structures of the LEC posterior pdf.

Depending on the cutoff, we find spurious states in all channels. In the $^{1}S_0$ channel there is always a region with an unphysical bound state, but it becomes more narrow for increasing values of the cutoff. The physical deuteron state is always present in $\tS$, and we observe additional (spurious) states in disjoint domains for several cutoffs. For $\Lambda \geq 3000$~MeV there are two disjoint domains that contain at least one spurious bound state each.

While there are no spurious bound states in $^3P_0$ for $\Lambda \leq 500$~MeV, a second domain with lower values of the LEC contains a spurious bound state for $\Lambda = 550,600,700$~MeV. For larger cutoffs, $\Lambda \geq 2000$~MeV, there is only one narrow domain and it gives rise to a spurious bound state. In $^3P_2$ we observe larger domains and spurious bound states for $\Lambda \geq 2000$ MeV.

We can understand the $\Lambda$-dependence of the $P$-wave domains by studying \cref{fig:limit_cycles} in light of the errors going into the inefficiency measure \eqref{eqn:implaus}. In doing so we see that for $^3P_0$ and $\Lambda = 550 - 700$~MeV the theoretical and empirical phase shifts match across wide regions of LEC values that qualitatively agree with the non-inefficient domains observed in \cref{fig:historymatch_results}. A similar argument holds for $^3P_2$ and greater values of $\Lambda$.

Unfortunately, history-matching fails to reduce the initial domain of $C_{^3P_0}$ and $C_{^3P_2}$ for some momentum cutoffs and thus provides limited information in these cases. In the following we perform a Bayesian analysis. The LEC posteriors presented in \cref{sec:mcmc} are conditioned on observational data (see Table~\ref{tab:exp_data}) and a more reliable error model. They provide more informative and credible inference compared to the domains shown in \cref{fig:historymatch_results}.

\subsection{Markov chain Monte Carlo sampling}
\label{sec:mcmc}
We sample the posterior pdf in Eq.~\eqref{eq:posterior_theta} using the affine-invariant ensamble sampler from the python package \texttt{emcee}~\cite{emcee} conditioned on the experimental data in Table~\ref{tab:exp_data}. As noted in Sec.~\ref{sec:hm}, there are disconnected non-inefficient regions for some LECs at several values of the cutoff. These regions might correspond to multimodal structures of the posteriors. In infinite time, the MCMC walkers will explore all parameter space, but in finite time they are likely trapped in local modes. To handle such convergence challenges we initiate sampling in several of the regions identified using history matching. Ideally, all combinations of such regions should be explored in order to locate the dominant posterior mode(s) with some certainty.

We initialize 50 MCMC walkers using a random subset of points from the non-inefficient domains for each cutoff and let them take 5000 steps each after a burn-in period of 1000 steps. This is typically sufficient for obtaining a good representation of each posterior mode of our five-dimensional posterior pdf. The initialization of walkers in the $P$-waves are chosen in the region of the domain without bound states, shown in orange in \cref{fig:historymatch_results}, except for the higher cutoffs in $^3P_0$. The same principle is applied in the $^3S_1$ partial wave, where we initialize in the region containing the least amount of spurious states. The posterior turns out to be relatively flat in the direction of $P$-wave LECs. Thus, the MCMC sampling is less sensitive to where the walkers are initialized in the rather large start domains. Moving forward, we focus on the possible multimodalities originating from the $^1S_0$ channel where we always found a parameter region giving rise to a shallow and unphysical bound state. The initialization of MCMC walkers in the $^1S_0$ channel is chosen either in the unbound (orange) or bound (blue) region. Proceeding like this for selected cutoff values between 400~MeV and 4000~MeV we indeed find multimodal structures in the posterior pdf, as summarized in \cref{tab:modes_all} in \cref{app:post}. 

To analyze the probability mass gathered in each mode we calculate the model evidence $\mathcal{Z}_K$ for all domains $K$ enclosing a mode. We do this using the Laplace approximation, which is accurate if the probability distribution can be locally approximated with a multivariate Gaussian, i.e., 
\begin{align}
    \mathcal{Z}_K = \pr\left(D|I\right)_K &\equiv \int_K  d \bm{\theta} \ \pr\left(\bm{\theta}|D,I\right) \approx \nonumber\\
    & \approx \pr\left(\bm{\theta}^*|D,I\right) \frac{\left(2\pi\right)^{N/2}}{\sqrt{\mathrm{det} \ \Sigma^{-1}}},
    \label{eq:Laplace_approx}
\end{align}
where $\Sigma^{-1}$ is the Hessian matrix of $\pr\left(\bm{\theta}|D,I\right)$ evaluated at the mode $\bm{\theta}^*$. Each mode listed in \cref{tab:modes_all} in \cref{app:post} is indeed found to be well approximated with a Gaussian. The numerical evaluation of \cref{eq:Laplace_approx} might be sensitive to the convergence of the MCMC chains. However, this sensitivity mainly impacts the evaluation of the determinant of the inverse Hessian which turns out to be a less significant contribution to the evidence compared to the density at the MAP point~\cite{gregory_2005}. The $\ln(\mathcal{Z}_K)$ values for all investigated modes are summarized in \cref{tab:modes_all} of \cref{app:post}. Fortunately, a well-defined ``best'' mode with highest evidence can be identified at all cutoffs except for $\Lambda=700$ and $\Lambda=4000$ MeV for which there are two $P$-wave modes with very similar evidences. All the significant modes identified in the evidence analysis are retained, and none of them contain an unphysical bound state in the $^1S_0$ channel.

To ensure that we obtain a high-quality representation of the relevant posteriors we perform the final sampling using 50 walkers taking 5000 steps each initialized in the vicinity of the largest $\ln(\mathcal{Z}_K)$ mode for each cutoff. This lays the foundation for accurate inferences of the ppds presented in Sec.~\ref{sec:ppd}. We use an autocorrelation analysis~\cite{Svensson:2022} and visual inspection of the traces to confirm MCMC convergence. At the cutoffs $\Lambda = 700$ MeV and $\Lambda = 4000$ MeV the two $P$-wave modes with similar evidences remain. The resulting values for the model evidences and MAP locations at several cutoffs are shown in \cref{tab:modes_high_ev}. 
\begin{table}[b]
\caption{The model evidences and maximum a posteriori (MAP) estimates ($\bm{\theta}^*$) of the LECs for all cutoffs $\Lambda$ (in MeV). A posterior bimodality with comparable probability mass per mode exists for both $\Lambda=700$ and $\Lambda=4000$ MeV.}
\begin{ruledtabular}
\begin{tabular}{ccl}
$\Lambda$& $\ln(\mathcal{Z}_K)$ & MAP ($\bm{\theta}^*)$\\
\colrule
400 & -3280 & (-0.1186, -0.1167,  5.386,  0.5962,  2.716)   \\ 
450 & -3312 & (-0.1128, -0.09640,  5.115,  0.5118,  2.813)  \\
500 & -3325 & (-0.1084, -0.07815,  4.713,  0.4560,  2.859)  \\
550 & -3305 & (-0.1048, -0.06069, -1.534,   0.7440,  2.808) \\
600 & -3296 & (-0.1018, -0.04352, -0.8692,  0.9547,  2.790) \\
700 & -3282 & (-0.09706, -0.006363, -0.3682, -2.848,  2.776) \\
700 & -3280 & (-0.09707,  -0.006305, -0.3803, 1.910,  2.761) \\
2000 & -3278 & (-0.07831, -0.1267,  0.01847, -0.02938,  2.745) \\
3000 & -3286 & (-0.07482, -0.03519,  0.02497, -0.01020,  2.763) \\
4000 & -3283 & (-0.07210,  0.09648, -1.140, -0.005023,  2.773) \\ 
4000 & -3282 & (-0.07210,  0.09641,  0.2060, -0.005050,  2.773) \\
\end{tabular}
\end{ruledtabular}
\label{tab:modes_high_ev}
\end{table}
The model evidences are of similar magnitude for all cutoffs. In fact, the biggest difference, on a logarithmic scale, is about $40$, i.e., the maximal ratio of evidences is $e^{40}$. The same difference can, for example, be obtained by shifting all predictions by about $5 \%$, which is quite small compared to the EFT truncation error. 

\begin{figure*}
\subfloat[$\Lambda=450$~MeV.]{%
      \includegraphics[width=\columnwidth]{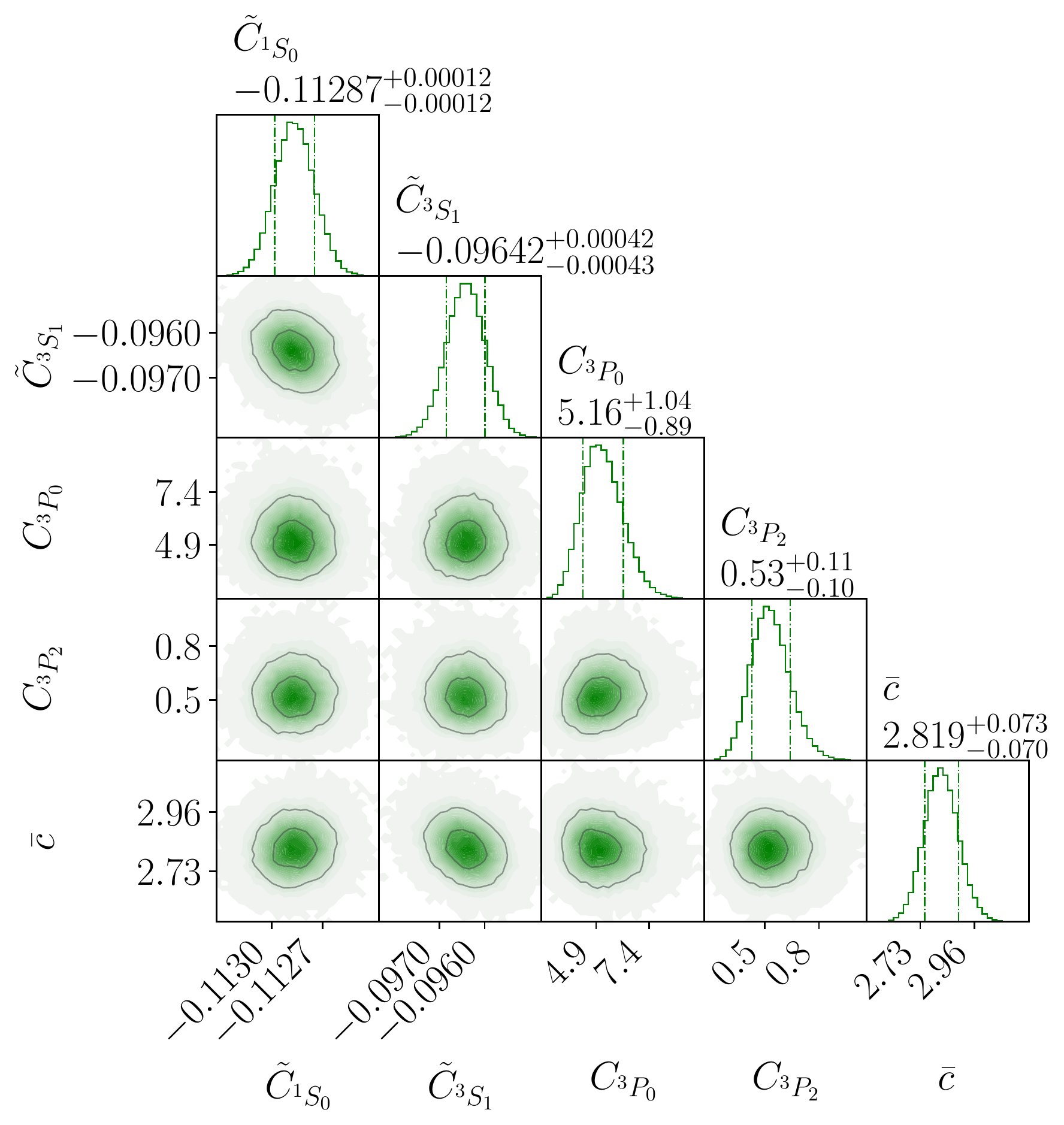}%
      \label{fig:pdf_450}
    }
\hfill
\subfloat[$\Lambda=700$~MeV.]{%
      \includegraphics[width=\columnwidth]{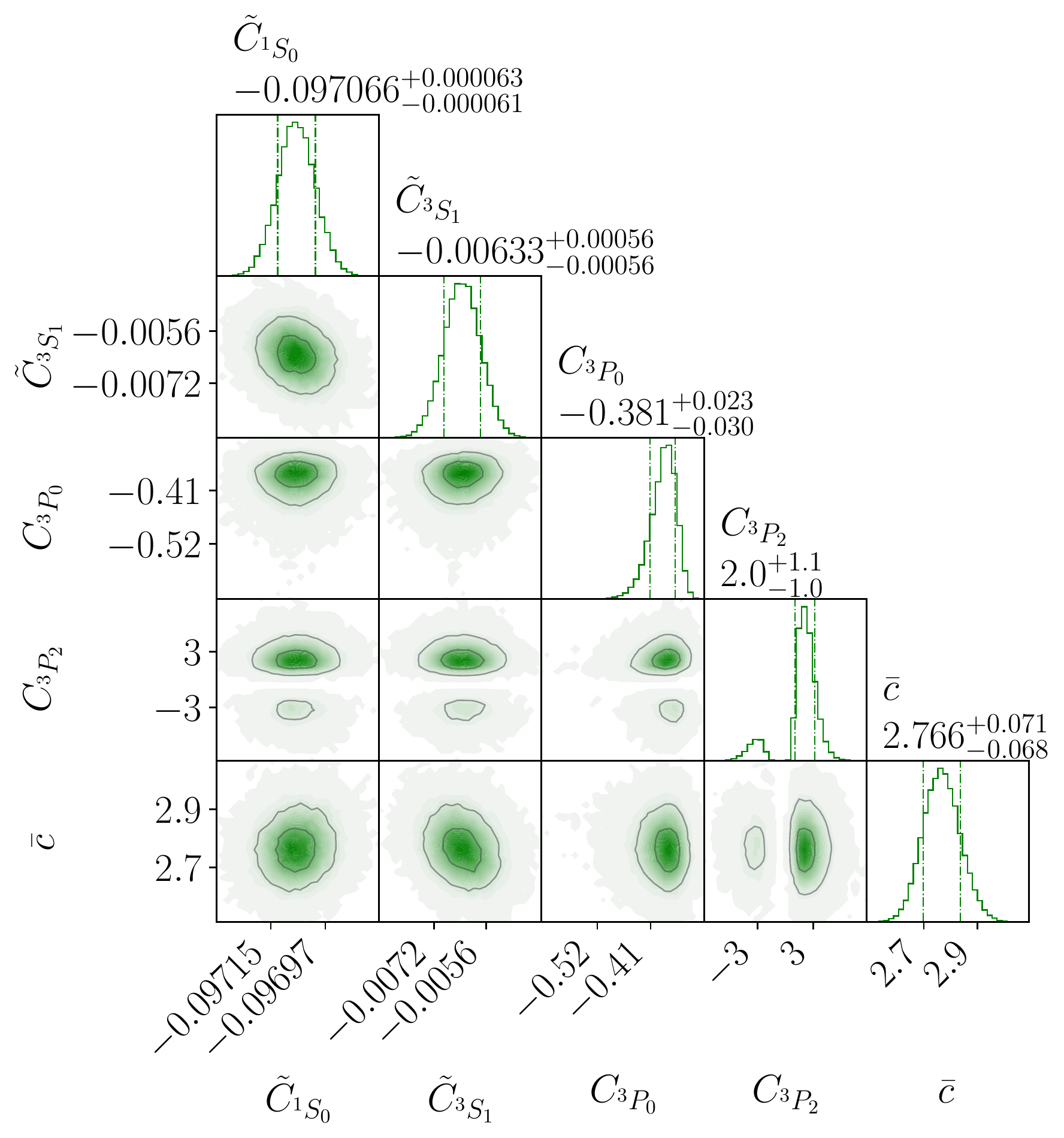}%
      \label{fig:pdf_700}
    }
\caption{Posterior pdfs for the parameters $\bm{\theta} = \left(\bm{\alpha},\bar{c}\right)$ using cutoffs $\Lambda=450$ MeV and $\Lambda=700$ MeV. The units of the LECs are $10^4$ GeV$^{-2}$ and $10^4$ GeV$^{-4}$ for the $S$- and $P$-waves respectively. The median and the 68\% equal-tailed credible interval are indicated for the univariate marginal pdfs.}
\end{figure*}

In Figs.~\ref{fig:pdf_450} and \ref{fig:pdf_700}, we show the marginal posterior pdfs for the cutoff values $\Lambda=450$ MeV and $\Lambda=700$ MeV. The posterior pdfs for the remaining cutoff values are shown in Appendix \ref{app:post}. The posteriors are consistent with our naturalness assumptions for both the LECs and the scale of the EFT truncation error as no tensions with our priors are observed. 

\begin{figure}
\includegraphics[width=\columnwidth]{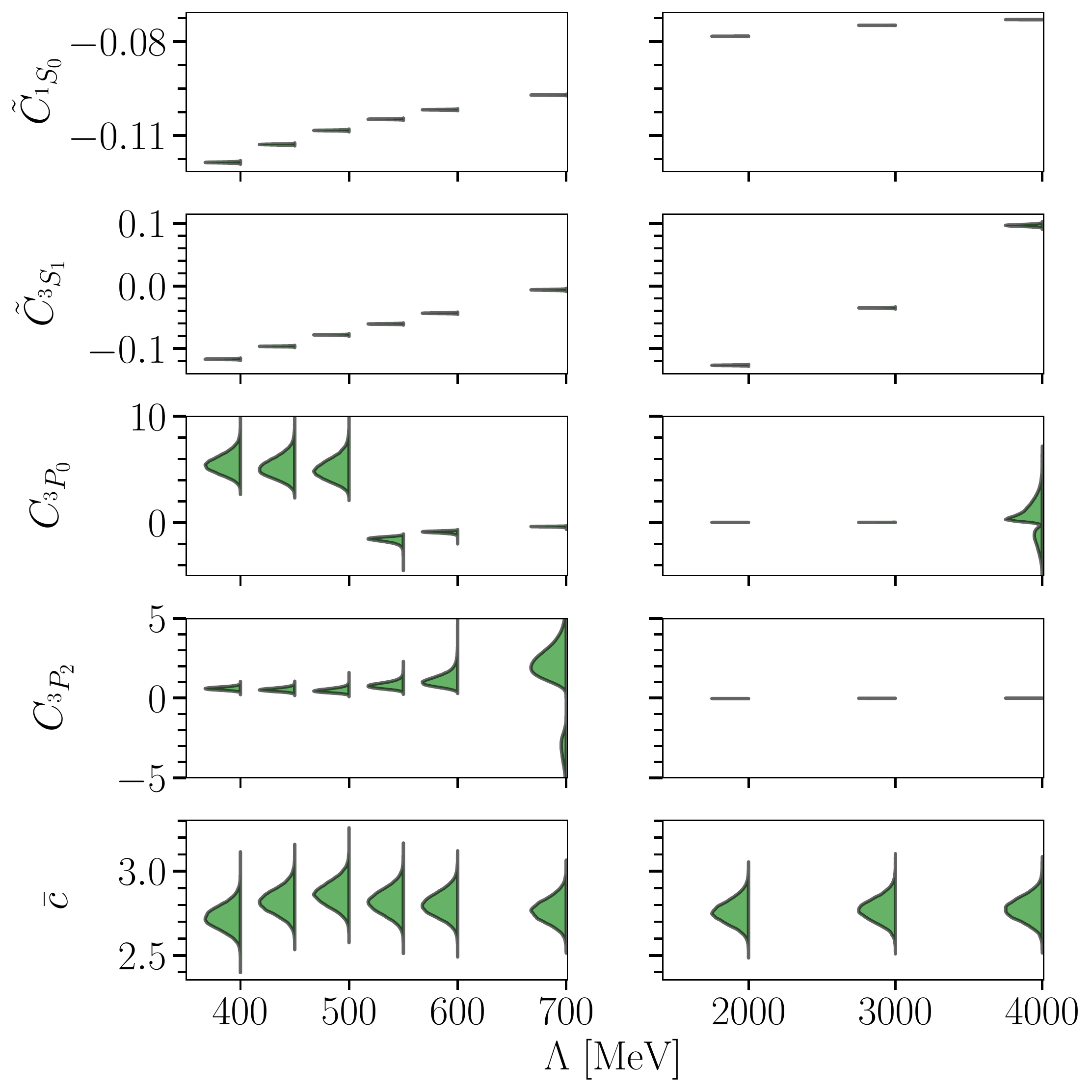}
\caption{The marginal posterior pdfs for the parameters $\bm{\theta}$ for cutoffs, $\Lambda$, in the range 400 to 4000 MeV. The units of the LECs are $10^4$ GeV$^{-2}$ and $10^4$ GeV$^{-4}$ for the $S$- and $P$-waves, respectively.}
\label{fig:pdf_marginal_lecs}
\end{figure}

The running of the inferred LECs can now be monitored. The cutoff dependence of the marginal posteriors for the LECs is shown in \cref{fig:pdf_marginal_lecs}. We conclude that $\tilde{C}_{^1S_0}(\Lambda)$ and $\tilde{C}_{^3S_1}(\Lambda)$ have a similar behavior as the running of those couplings determined from the phase shift fits in Sec.~\ref{sec:PC} and \cref{fig:running_all}. This finding can likely be attributed to the fact that these LECs are quite well constrained by the employed low-energy $np$ scattering cross sections. However, the $P$-wave LECs $C_{^3P_0}$ and $C_{^3P_2}$ are not that well constrained by the $np$ data used in the inference and run differently with $\Lambda$ in the Bayesian analysis. 

The inference of $\bar{c}$ is quite interesting in its own right. In~\cref{fig:pdf_marginal_lecs} we show its marginal posterior and see that it is rather insensitive to cutoff variation, in particular for $\Lambda > 700$. The breakdown scale $\Lambda_{\chi}$, $\bar{c}$ and the low-energy scale, $Q$, are connected through \cref{eq:model_error} and their posteriors have previously been investigated in WPC in Refs.~\cite{Melendez:2017phj,Wesolowski:2021cni}. 

\section{\label{sec:ppd}Posterior predictive distributions}
In this section we present ppds for scattering phase shifts as well as selected $np$ observables. The ppd for an observable $y$ is a marginalization of the model predictions over the posterior of the relevant parameters
\begin{equation}
    \pr\left(y|D,I\right) = \int d \bm{\theta} \ \pr\left(y|\bm{\theta},D,I\right) \pr\left(\bm{\theta}|D,I\right).
    \label{eq:ppd}
\end{equation}
When sampling the model prediction, $\pr\left(y|\bm{\theta},D,I\right)$, we can choose to include the EFT truncation error, in which case it takes the form
\begin{equation}
    \pr\left(y|\bm{\theta},D,I\right) = \mathcal{N}\left(y^{(0)}_\mathrm{th}(\bm{\alpha}),\sigma_\mathrm{th}^2(\bar{c})\right),
    \label{eq:pdf_model_err}
\end{equation}
or we can exclude it, which corresponds to just propagating the parametric LEC uncertainty, by using a delta distribution
\begin{equation}
    \pr\left(y|\bm{\theta},D,I\right) = \delta\left(y-y^{(0)}_\mathrm{th}\left(\bm{\alpha}\right)\right).
    \label{eq:prob_delta}
\end{equation}
The integral in \cref{eq:ppd} can be straightforwardly evaluated using the samples from the MCMC chains obtained in Sec.~\ref{sec:sampling}. In the following, ppds including (excluding) the EFT error are colored blue (purple).

\subsection{Phase shifts}
\begin{figure}
\includegraphics[width=\columnwidth]{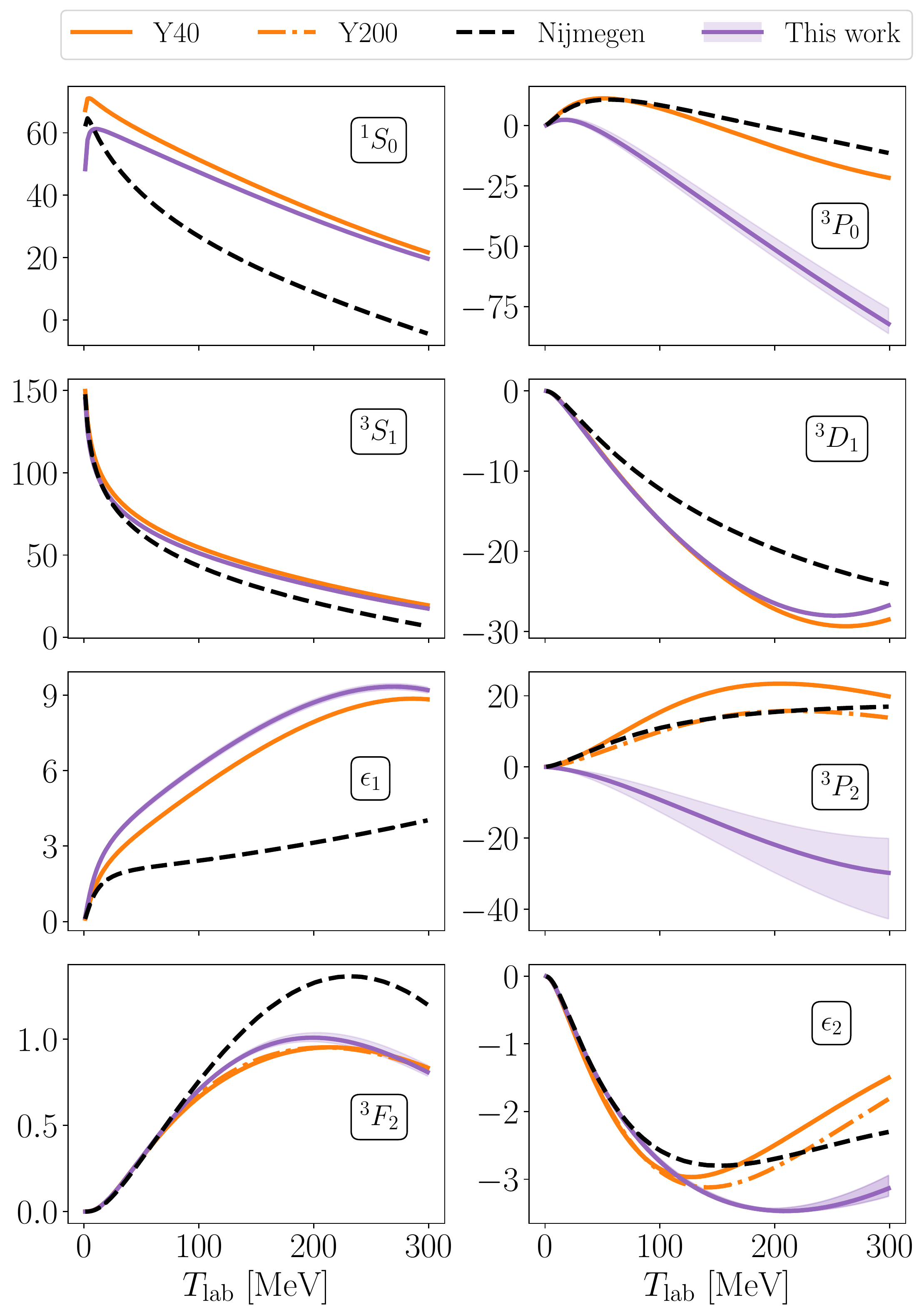}
\caption{Predicted phase shifts (in degrees) with $\Lambda=450$ MeV represented by the median of the ppd (solid purple line) and the 95\% equal-tailed credible interval (purple band). Note that the EFT error is not included. Our results are compared to the phase shifts of \citet{Yang:2020pgi} and the Nijmegen partial-wave analysis \cite{Stoks:1993tb}. The labels Y40 and Y200 indicate the two types of fits that were done by \citet{Yang:2020pgi} in the $\tP$ channel; renormalizing the corresponding LECs to reproduce the $^3P_2$ phase shifts at $T_\mathrm{lab}=40$ MeV and $T_\mathrm{lab}=200$ MeV, respectively.
\label{fig:ppd_phase}}
\end{figure}

\begin{figure}
\includegraphics[width=\columnwidth]{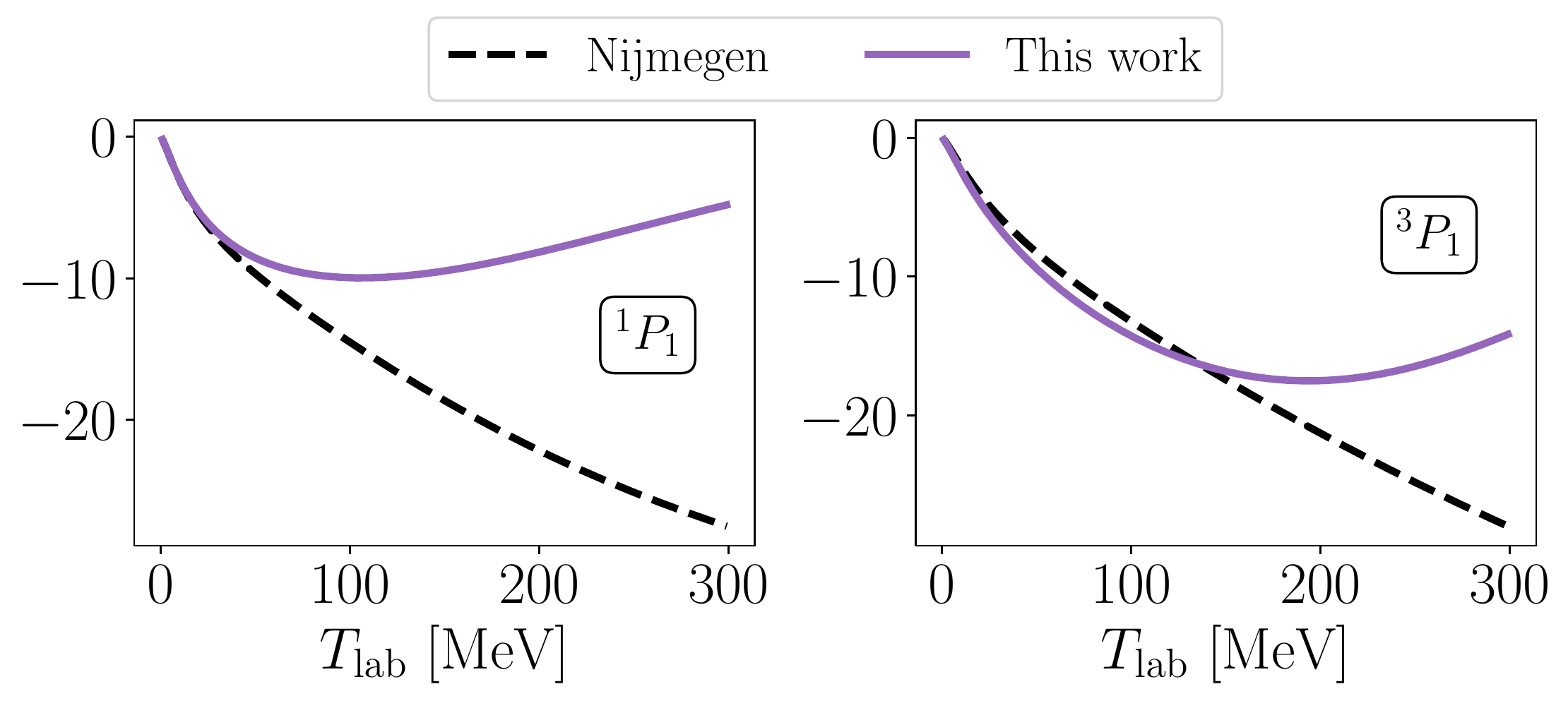}
\caption{Predicted phase shifts (in degrees) in the $P$-waves not part of the contact potential using the cutoff $\Lambda=450$~MeV. Our results are compared to the results from the Nijmegen partial-wave analysis \cite{Stoks:1993tb}.
\label{fig:1p1_3p1_phase}}
\end{figure}

The Bayesian inference in this work differs from the point estimates obtained from a standard phase shift optimization. In \cref{fig:ppd_phase} we compare the ppds from the present analysis with the optimized phase shifts from \citet{Yang:2020pgi} for $\Lambda=450$ MeV and the Nijmegen partial-wave analysis~\cite{Stoks:1993tb}. The EFT error is not included in the ppd. One finds that the parametric uncertainty of the phase shifts, stemming from the LEC posterior, is rather small and the $95\%$ credible intervals shown in the figure are mostly visible in the $^3P_2$ partial wave. We find good agreement with the results by \citet{Yang:2020pgi} except for the $P$-wave phase shifts which are more repulsive in the Bayesian analysis. Apparently, the $P$-wave LECs receive significant contributions when conditioning on scattering data. At $\Lambda=450$ MeV, The $P$-wave phase shifts $^1P_1$ and $^3P_1$, which are not part of the contact potential, are more attractive than the ones from the Nijmegen analysis, see \cref{fig:1p1_3p1_phase}. Note that the LO potential only includes low partial waves, for which the OPE is not perturbative. To prevent overfitting to higher order effects, which are better described by high partial waves, either the LO potential or the model for the LO EFT error likely needs to be revisited. We also foresee that the LO LECs will receive corrections at higher orders and this could significantly change the scattering amplitudes and the pole structure in a given channel. 
The $np$ phase shifts for the MAP estimate of the LECs for $\Lambda = 450$~MeV are tabulated in \cref{tab:phase_Lambda_450_map} in Appendix \ref{app:phase_shifts}. 

\subsection{\label{sec:ppd_obs}Observables in the $np$ sector}
\begin{figure}
\includegraphics[width=\columnwidth]{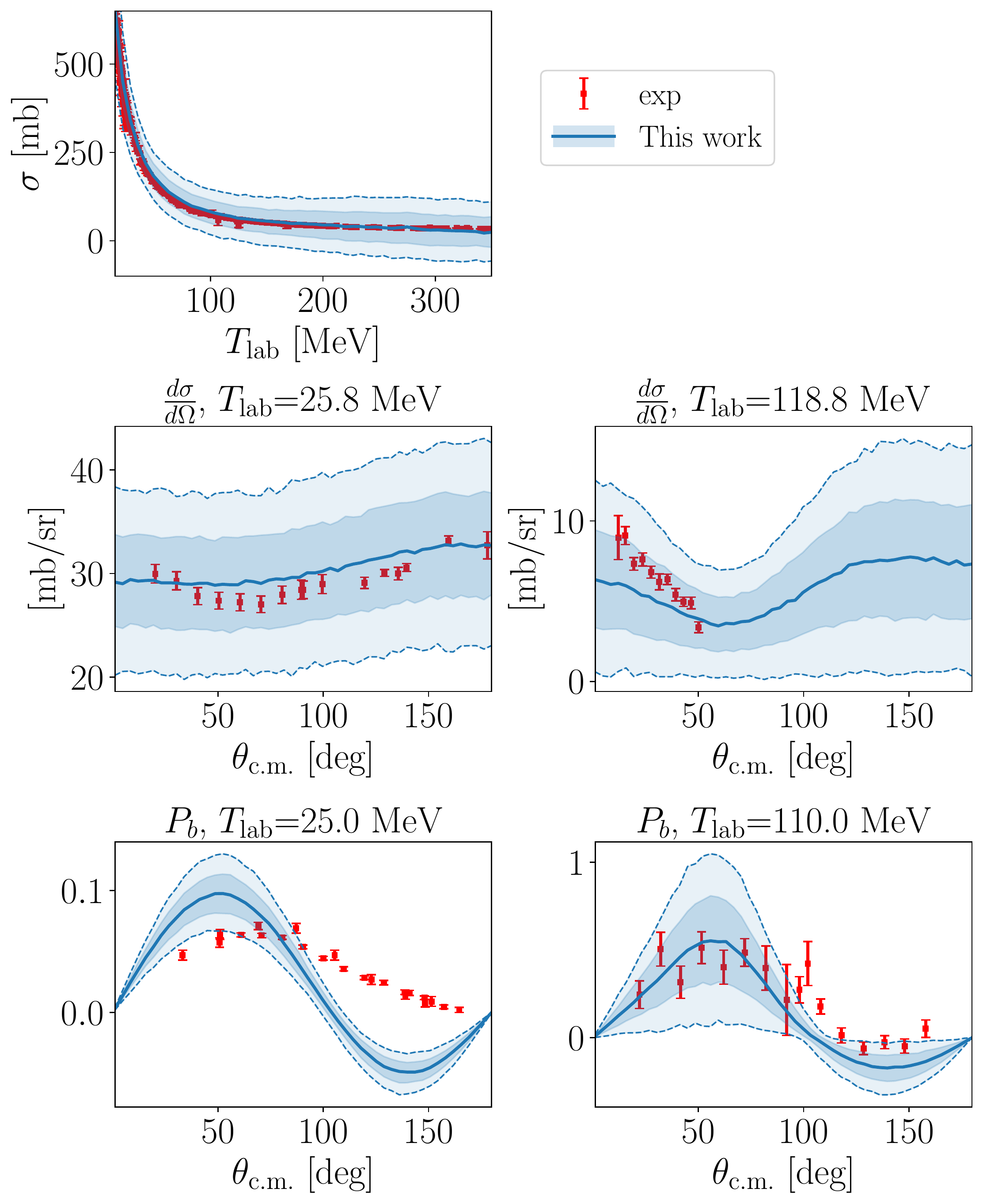}
\caption{Median (solid blue line) of the ppds for selected $np$ scattering observables using MWPC at LO with a cutoff $\Lambda = 450$ MeV. The EFT error is included. Shaded dark and light blue bands represent 68\% and 95\% equal-tailed credible intervals, respectively. Experimental data (exp) from Refs. \cite{Granada_1,Granada_2}.}
\label{fig:ppd_450}
\end{figure}

In \cref{fig:ppd_450} we compare experimental data for several $np$ scattering observables and the corresponding ppds, including the EFT truncation error, at LO and using the cutoff $\Lambda = 450$ MeV. Total cross sections, $\sigma$, and differential cross sections, ${d \sigma }/{ d \Omega}$, are reproduced quite well, although the ppd error bands are rather wide. Using \cref{eq:model_error} and $\bar{c} \approx 2.8$ from \cref{fig:pdf_450} we estimate that the EFT truncation error for $p \lesssim m_\pi$ is about $15\%$. The ppd for the spin-polarization observable $P_b$, which was not included in the MCMC sampling, reproduces experimental data rather poorly. This result is particularly striking in the ppd at $T_\mathrm{lab}=25.0$ MeV. The situation improves at higher energies, which is somewhat surprising for a low-energy EFT. As such, conditioning on low-energy $P_b$ data could improve the model. However, expanding the dataset in \cref{tab:exp_data} requires an improved model for the EFT truncation error ~\cite{Melendez:2019izc} and likely access to order-by-order calculations beyond LO to guide the inference of $\bar{c}$. 

\begin{figure}
\includegraphics[width=\columnwidth]{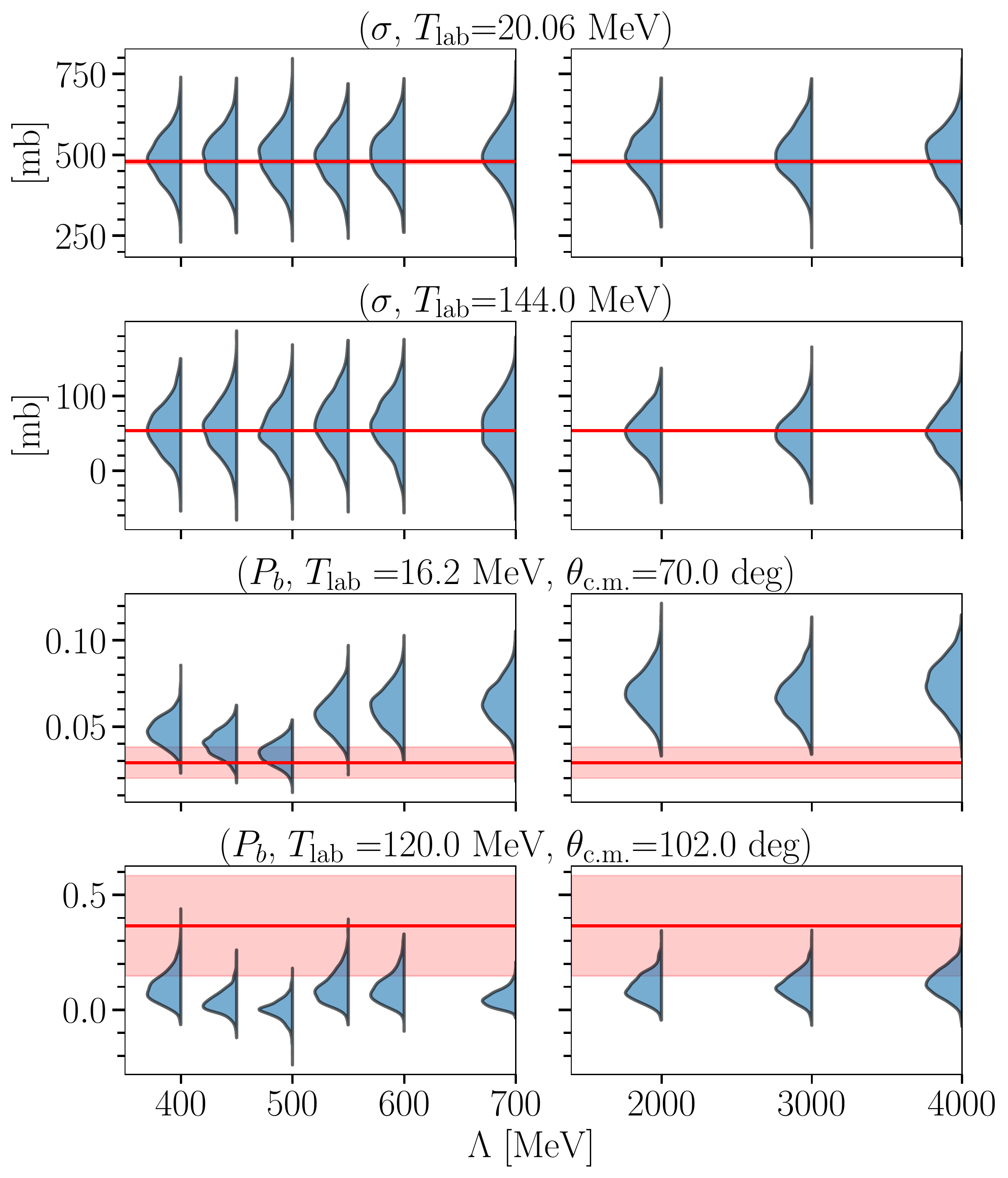}
\caption{The ppds for the scattering observables $\sigma$ and $P_b$ at two different lab energies as a function of the cutoff, $\Lambda$, and including the EFT error. The red horizontal line and band represent the experimental value and error, respectively \cite{Granada_1,Granada_2}. 
\label{fig:ppd_SGTandPB}}
\end{figure}

In \cref{fig:ppd_SGTandPB} we show the ppds for $\sigma$ and $P_b$ as a function of $\Lambda$, including the EFT error, for a handful of values of $T_\mathrm{lab}$ and $\thetacm$. Except for a slight variation of $P_b$ for $\Lambda < 700$, no significant cutoff variation is observed. It can also be seen that the experimental value is well reproduced for $\sigma$ and more poorly for $P_b$. The LO EFT error for $\sigma$ appears to be on the conservative side and might shrink as we learn it better.

\begin{figure}
\includegraphics[width=\columnwidth]{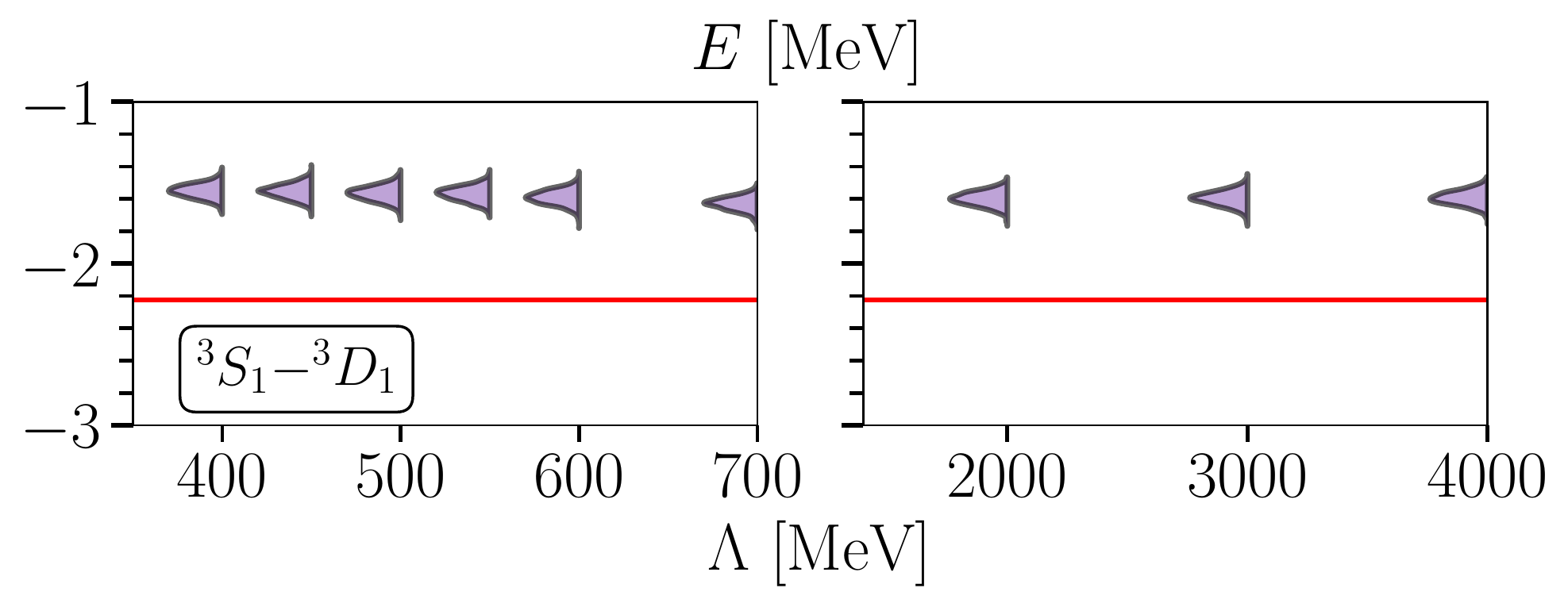}
\caption{The ppds for the energy of the deuteron ($\tS$) channel as a function of the cutoff, $\Lambda$, and excluding the EFT error. The red horizontal line indicates the experimental binding energy \cite{PDG_2022}.}
\label{fig:ppd_bindingD}
\end{figure}

The ppd, without the EFT error, for the deuteron ground-state energy is shown in \cref{fig:ppd_bindingD} and reveals an approximate 0.5 MeV underbinding, but no significant cutoff dependence. For some of the higher cutoffs additional spurious and deeply bound states appear at $E \approx -1$~GeV. Employing the same error model as for the scattering observables one can attempt to make an uncertainty estimate for the deuteron ground-state energy. Using the binding momentum $p_{gs} = \sqrt{m_N(-E_{gs})} \approx 50 \ \mathrm{MeV} <m_\pi$ as a proxy for the relevant soft scale we arrive at an estimated EFT error of around $15\%$ of the predicted value. Including this error makes the lower end of the ppds touch the experimental value.

\begin{figure}
\includegraphics[width=\columnwidth]{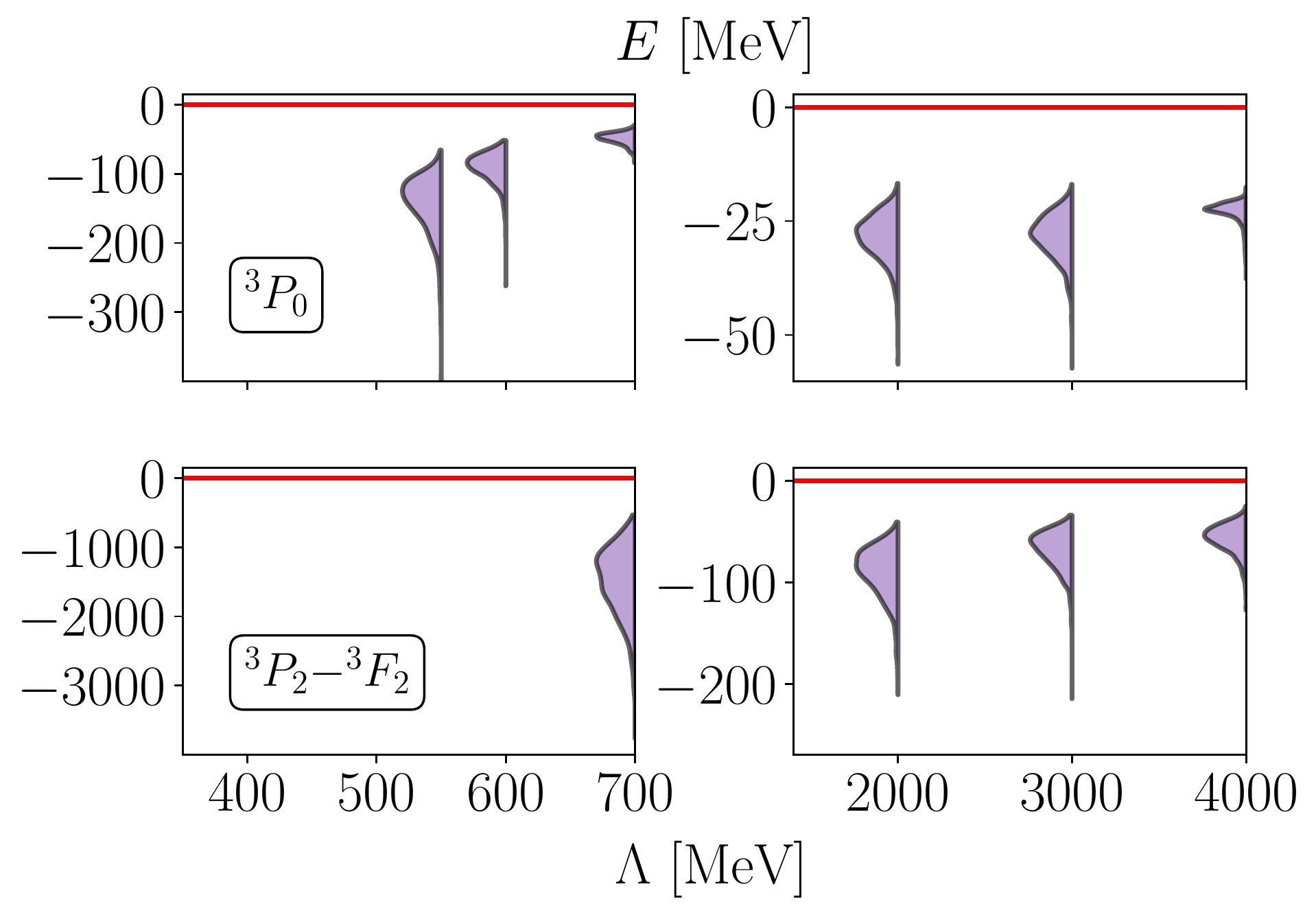}
\caption{The ppds for the energy of selected bound states in the $^3P_0$ (upper row) and $\tP$ (lower row) channels as a function of the cutoff, $\Lambda$, and excluding the EFT error. Note the different scales on the $y$-axes. The red horizontal line indicates the binding threshold. 
\label{fig:ppd_binding02}}
\end{figure}

The ppds for the energies of the unphysical bound states in the channels $^3P_0$ and $\tP$ are shown in \cref{fig:ppd_binding02}, except for a very deep ($E\approx -1$ TeV) state in $^3P_0$ for $\Lambda=4000$~MeV. Some of the energies are around $25$~MeV, which is well within the validity of the EFT. Note also that the ppds in \cref{fig:ppd_binding02} show a strong dependence on the cutoff, which is generally the case for spurious states, which was also seen in Sec.~\ref{sec:PC}. In the $\tP$ channel for the cutoff $\Lambda=700$ the deeply bound state at $E\approx -1$~GeV is produced by the very small tail of negative $C_{^3P_2}$-values shown in the marginal pdf in \cref{fig:pdf_marginal_lecs}. We see spurious states in all singular channels for $\Lambda \geq 2000$ MeV.

\section{Conclusions and outlook%
\label{sec:conclusion}}
In this work we have presented a detailed Bayesian study of a LO $\chi$EFT potential in MWPC. We have performed robust inferences of the LECs and the scale $\bar{c}$ of the EFT truncation error using history matching and MCMC sampling for several values of the momentum cutoff in the range from 400 to 4000~MeV. Below, we summarize the main takeaways.

\begin{itemize}
    \item Multimodal LEC posteriors are induced via the existence of bound states, virtual states and resonances that move around near threshold for varying values of the LECs and cutoff $\Lambda$. This makes inference challenging. In particular, we could always find modes where the $^1S_0$ state was bound at an energy of approximately 1 MeV.
    
    \item We computed the model evidence for each mode in the Laplace approximation. At each cutoff we initialized a final round of MCMC sampling around the mode(s) with the largest evidence (see \cref{tab:modes_high_ev}). Modes harboring a bound $^1S_0$ state were always found to be insignificant. However, pairs of modes in the $\tP$ channel for $\Lambda=700$ MeV and in the $^3P_0$ channel for $\Lambda=4000$ MeV had similar evidences and could not be discerned. We expect that the poor constraints on the LECs might be improved with a better error model and with the inclusion of more types of scattering observables in the calibration data.
    
    \item Our inference shows that $\bar{c}\approx 2.8$ across all cutoffs investigated. Since the main posterior modes have similar evidences (see \cref{tab:modes_high_ev}) this implies that the description of the low-energy scattering data is essentially equivalent for the different momentum cutoffs in the range 400 to 4000~MeV.

    \item Conditioning the LEC inference on scattering data leads to excessive repulsion in the $^3P_0$ and $^3P_2$ partial waves compared to the Nijmegen analysis~\cite{Stoks:1993tb} and to the phase-shift optimization performed by~\citet{Yang:2020pgi}. Despite the poor accuracy of $P$-wave phase shifts (see \cref{fig:ppd_phase}), we find that scattering observables are reproduced surprisingly well (see \cref{fig:ppd_450,fig:ppd_SGTandPB,fig:ppd_bindingD}) for a LO potential. We conclude that it might be problematic to use phase shifts for calibration of LO potentials in MWPC. These potentials, acting in the first few partial waves only, need to compensate for excluded higher-order contributions to reproduce observables. Consequently, the inference becomes crucially dependent on the specification of the EFT error model, of which we have limited information. This highlights the need for further studies at higher orders.
    
    \item We conclude that the ppds for various bound-state and scattering observables in the $np$ sector exhibit no significant cutoff dependence. 

\end{itemize}

It remains to be seen what happens as sub-leading orders are included perturbatively and when the model for the EFT truncation error is refined. The inferred LO potentials, with broad distributions for the $P$-wave LECs, will likely lead to significant predictive uncertainties in nuclei. Therefore, the ppds for ground-state energies and radii should also be quantified to assess the quality and physical relevance of existing RG-invariant $\chi$EFT formulations.

\begin{acknowledgments}
O.T.~thanks C.-J.~Yang for helpful discussions regarding the potential and conventions, as well as providing more detailed data from Ref.~\cite{Yang:2020pgi}. This work was supported by the European Research Council (ERC) under the European Unions Horizon 2020 research and innovation program (Grant Agreement No.~758027), the Swedish Research Council (Grants No.~2017-04234, No.~2020-05127 and No.~2021-04507). The computations were enabled by resources provided by the Swedish National Infrastructure for Computing (SNIC) partially funded by the Swedish Research Council through Grant Agreement No.~2018-05973.

\end{acknowledgments}
\bibliography{main.bib}         

\begin{thebibliography}{80}%
\makeatletter
\providecommand \@ifxundefined [1]{%
 \@ifx{#1\undefined}
}%
\providecommand \@ifnum [1]{%
 \ifnum #1\expandafter \@firstoftwo
 \else \expandafter \@secondoftwo
 \fi
}%
\providecommand \@ifx [1]{%
 \ifx #1\expandafter \@firstoftwo
 \else \expandafter \@secondoftwo
 \fi
}%
\providecommand \natexlab [1]{#1}%
\providecommand \enquote  [1]{``#1''}%
\providecommand \bibnamefont  [1]{#1}%
\providecommand \bibfnamefont [1]{#1}%
\providecommand \citenamefont [1]{#1}%
\providecommand \href@noop [0]{\@secondoftwo}%
\providecommand \href [0]{\begingroup \@sanitize@url \@href}%
\providecommand \@href[1]{\@@startlink{#1}\@@href}%
\providecommand \@@href[1]{\endgroup#1\@@endlink}%
\providecommand \@sanitize@url [0]{\catcode `\\12\catcode `\$12\catcode
  `\&12\catcode `\#12\catcode `\^12\catcode `\_12\catcode `\%12\relax}%
\providecommand \@@startlink[1]{}%
\providecommand \@@endlink[0]{}%
\providecommand \url  [0]{\begingroup\@sanitize@url \@url }%
\providecommand \@url [1]{\endgroup\@href {#1}{\urlprefix }}%
\providecommand \urlprefix  [0]{URL }%
\providecommand \Eprint [0]{\href }%
\providecommand \doibase [0]{https://doi.org/}%
\providecommand \selectlanguage [0]{\@gobble}%
\providecommand \bibinfo  [0]{\@secondoftwo}%
\providecommand \bibfield  [0]{\@secondoftwo}%
\providecommand \translation [1]{[#1]}%
\providecommand \BibitemOpen [0]{}%
\providecommand \bibitemStop [0]{}%
\providecommand \bibitemNoStop [0]{.\EOS\space}%
\providecommand \EOS [0]{\spacefactor3000\relax}%
\providecommand \BibitemShut  [1]{\csname bibitem#1\endcsname}%
\let\auto@bib@innerbib\@empty
\bibitem [{\citenamefont {Weinberg}(1979)}]{Weinberg:1978kz}%
  \BibitemOpen
  \bibfield  {author} {\bibinfo {author} {\bibfnamefont {S.}~\bibnamefont
  {Weinberg}},\ }\bibfield  {title} {\bibinfo {title} {{Phenomenological
  Lagrangians}},\ }\href {https://doi.org/10.1016/0378-4371(79)90223-1}
  {\bibfield  {journal} {\bibinfo  {journal} {Physica A}\ }\textbf {\bibinfo
  {volume} {96}},\ \bibinfo {pages} {327} (\bibinfo {year} {1979})}\BibitemShut
  {NoStop}%
\bibitem [{\citenamefont {Weinberg}(1990)}]{Weinberg:1990rz}%
  \BibitemOpen
  \bibfield  {author} {\bibinfo {author} {\bibfnamefont {S.}~\bibnamefont
  {Weinberg}},\ }\bibfield  {title} {\bibinfo {title} {{Nuclear forces from
  chiral Lagrangians}},\ }\href {https://doi.org/10.1016/0370-2693(90)90938-3}
  {\bibfield  {journal} {\bibinfo  {journal} {Phys. Lett. B}\ }\textbf
  {\bibinfo {volume} {251}},\ \bibinfo {pages} {288} (\bibinfo {year}
  {1990})}\BibitemShut {NoStop}%
\bibitem [{\citenamefont {Weinberg}(1991)}]{Weinberg:1991um}%
  \BibitemOpen
  \bibfield  {author} {\bibinfo {author} {\bibfnamefont {S.}~\bibnamefont
  {Weinberg}},\ }\bibfield  {title} {\bibinfo {title} {{Effective chiral
  Lagrangians for nucleon - pion interactions and nuclear forces}},\ }\href
  {https://doi.org/10.1016/0550-3213(91)90231-L} {\bibfield  {journal}
  {\bibinfo  {journal} {Nucl. Phys. B}\ }\textbf {\bibinfo {volume} {363}},\
  \bibinfo {pages} {3} (\bibinfo {year} {1991})}\BibitemShut {NoStop}%
\bibitem [{\citenamefont {van Kolck}(2020{\natexlab{a}})}]{vanKolck:2020llt}%
  \BibitemOpen
  \bibfield  {author} {\bibinfo {author} {\bibfnamefont {U.}~\bibnamefont {van
  Kolck}},\ }\bibfield  {title} {\bibinfo {title} {{The Problem of
  Renormalization of Chiral Nuclear Forces}},\ }\href
  {https://doi.org/10.3389/fphy.2020.00079} {\bibfield  {journal} {\bibinfo
  {journal} {Front. Phys.}\ }\textbf {\bibinfo {volume} {8}},\ \bibinfo {pages}
  {79} (\bibinfo {year} {2020}{\natexlab{a}})},\ \Eprint
  {https://arxiv.org/abs/2003.06721} {arXiv:2003.06721 [nucl-th]} \BibitemShut
  {NoStop}%
\bibitem [{\citenamefont {Yang}\ \emph
  {et~al.}(2021{\natexlab{a}})\citenamefont {Yang}, \citenamefont {Ekstr\"om},
  \citenamefont {Forss\'en},\ and\ \citenamefont {Hagen}}]{Yang:2020pgi}%
  \BibitemOpen
  \bibfield  {author} {\bibinfo {author} {\bibfnamefont {C.~J.}\ \bibnamefont
  {Yang}}, \bibinfo {author} {\bibfnamefont {A.}~\bibnamefont {Ekstr\"om}},
  \bibinfo {author} {\bibfnamefont {C.}~\bibnamefont {Forss\'en}},\ and\
  \bibinfo {author} {\bibfnamefont {G.}~\bibnamefont {Hagen}},\ }\bibfield
  {title} {\bibinfo {title} {{Power counting in chiral effective field theory
  and nuclear binding}},\ }\href {https://doi.org/10.1103/PhysRevC.103.054304}
  {\bibfield  {journal} {\bibinfo  {journal} {Phys. Rev. C}\ }\textbf {\bibinfo
  {volume} {103}},\ \bibinfo {pages} {054304} (\bibinfo {year}
  {2021}{\natexlab{a}})},\ \Eprint {https://arxiv.org/abs/2011.11584}
  {arXiv:2011.11584 [nucl-th]} \BibitemShut {NoStop}%
\bibitem [{\citenamefont {Carlsson}\ \emph {et~al.}(2016)\citenamefont
  {Carlsson}, \citenamefont {Ekstr\"om}, \citenamefont {Forss\'en},
  \citenamefont {Str\"omberg}, \citenamefont {Jansen}, \citenamefont {Lilja},
  \citenamefont {Lindby}, \citenamefont {Mattsson},\ and\ \citenamefont
  {Wendt}}]{Carlsson:2015vda}%
  \BibitemOpen
  \bibfield  {author} {\bibinfo {author} {\bibfnamefont {B.~D.}\ \bibnamefont
  {Carlsson}}, \bibinfo {author} {\bibfnamefont {A.}~\bibnamefont {Ekstr\"om}},
  \bibinfo {author} {\bibfnamefont {C.}~\bibnamefont {Forss\'en}}, \bibinfo
  {author} {\bibfnamefont {D.~F.}\ \bibnamefont {Str\"omberg}}, \bibinfo
  {author} {\bibfnamefont {G.~R.}\ \bibnamefont {Jansen}}, \bibinfo {author}
  {\bibfnamefont {O.}~\bibnamefont {Lilja}}, \bibinfo {author} {\bibfnamefont
  {M.}~\bibnamefont {Lindby}}, \bibinfo {author} {\bibfnamefont {B.~A.}\
  \bibnamefont {Mattsson}},\ and\ \bibinfo {author} {\bibfnamefont {K.~A.}\
  \bibnamefont {Wendt}},\ }\bibfield  {title} {\bibinfo {title} {{Uncertainty
  analysis and order-by-order optimization of chiral nuclear interactions}},\
  }\href {https://doi.org/10.1103/PhysRevX.6.011019} {\bibfield  {journal}
  {\bibinfo  {journal} {Phys. Rev. X}\ }\textbf {\bibinfo {volume} {6}},\
  \bibinfo {pages} {011019} (\bibinfo {year} {2016})},\ \Eprint
  {https://arxiv.org/abs/1506.02466} {arXiv:1506.02466 [nucl-th]} \BibitemShut
  {NoStop}%
\bibitem [{\citenamefont {Bedaque}\ and\ \citenamefont {van
  Kolck}(2002)}]{Bedaque:2002mn}%
  \BibitemOpen
  \bibfield  {author} {\bibinfo {author} {\bibfnamefont {P.~F.}\ \bibnamefont
  {Bedaque}}\ and\ \bibinfo {author} {\bibfnamefont {U.}~\bibnamefont {van
  Kolck}},\ }\bibfield  {title} {\bibinfo {title} {{Effective field theory for
  few nucleon systems}},\ }\href
  {https://doi.org/10.1146/annurev.nucl.52.050102.090637} {\bibfield  {journal}
  {\bibinfo  {journal} {Ann. Rev. Nucl. Part. Sci.}\ }\textbf {\bibinfo
  {volume} {52}},\ \bibinfo {pages} {339} (\bibinfo {year} {2002})},\ \Eprint
  {https://arxiv.org/abs/nucl-th/0203055} {arXiv:nucl-th/0203055} \BibitemShut
  {NoStop}%
\bibitem [{\citenamefont {Epelbaum}\ \emph {et~al.}(2009)\citenamefont
  {Epelbaum}, \citenamefont {Hammer},\ and\ \citenamefont
  {Meissner}}]{Epelbaum:2008ga}%
  \BibitemOpen
  \bibfield  {author} {\bibinfo {author} {\bibfnamefont {E.}~\bibnamefont
  {Epelbaum}}, \bibinfo {author} {\bibfnamefont {H.-W.}\ \bibnamefont
  {Hammer}},\ and\ \bibinfo {author} {\bibfnamefont {U.-G.}\ \bibnamefont
  {Meissner}},\ }\bibfield  {title} {\bibinfo {title} {{Modern Theory of
  Nuclear Forces}},\ }\href {https://doi.org/10.1103/RevModPhys.81.1773}
  {\bibfield  {journal} {\bibinfo  {journal} {Rev. Mod. Phys.}\ }\textbf
  {\bibinfo {volume} {81}},\ \bibinfo {pages} {1773} (\bibinfo {year}
  {2009})},\ \Eprint {https://arxiv.org/abs/0811.1338} {arXiv:0811.1338
  [nucl-th]} \BibitemShut {NoStop}%
\bibitem [{\citenamefont {Machleidt}\ and\ \citenamefont
  {Entem}(2011)}]{Machleidt:2011zz}%
  \BibitemOpen
  \bibfield  {author} {\bibinfo {author} {\bibfnamefont {R.}~\bibnamefont
  {Machleidt}}\ and\ \bibinfo {author} {\bibfnamefont {D.~R.}\ \bibnamefont
  {Entem}},\ }\bibfield  {title} {\bibinfo {title} {{Chiral effective field
  theory and nuclear forces}},\ }\href
  {https://doi.org/10.1016/j.physrep.2011.02.001} {\bibfield  {journal}
  {\bibinfo  {journal} {Phys. Rep.}\ }\textbf {\bibinfo {volume} {503}},\
  \bibinfo {pages} {1} (\bibinfo {year} {2011})},\ \Eprint
  {https://arxiv.org/abs/1105.2919} {arXiv:1105.2919 [nucl-th]} \BibitemShut
  {NoStop}%
\bibitem [{\citenamefont {Hammer}\ \emph {et~al.}(2020)\citenamefont {Hammer},
  \citenamefont {K\"onig},\ and\ \citenamefont {van Kolck}}]{Hammer:2019poc}%
  \BibitemOpen
  \bibfield  {author} {\bibinfo {author} {\bibfnamefont {H.~W.}\ \bibnamefont
  {Hammer}}, \bibinfo {author} {\bibfnamefont {S.}~\bibnamefont {K\"onig}},\
  and\ \bibinfo {author} {\bibfnamefont {U.}~\bibnamefont {van Kolck}},\
  }\bibfield  {title} {\bibinfo {title} {{Nuclear effective field theory:
  status and perspectives}},\ }\href
  {https://doi.org/10.1103/RevModPhys.92.025004} {\bibfield  {journal}
  {\bibinfo  {journal} {Rev. Mod. Phys.}\ }\textbf {\bibinfo {volume} {92}},\
  \bibinfo {pages} {025004} (\bibinfo {year} {2020})},\ \Eprint
  {https://arxiv.org/abs/1906.12122} {arXiv:1906.12122 [nucl-th]} \BibitemShut
  {NoStop}%
\bibitem [{\citenamefont {Stetcu}\ \emph {et~al.}(2007)\citenamefont {Stetcu},
  \citenamefont {Barrett},\ and\ \citenamefont {van Kolck}}]{Stetcu:2006ey}%
  \BibitemOpen
  \bibfield  {author} {\bibinfo {author} {\bibfnamefont {I.}~\bibnamefont
  {Stetcu}}, \bibinfo {author} {\bibfnamefont {B.~R.}\ \bibnamefont
  {Barrett}},\ and\ \bibinfo {author} {\bibfnamefont {U.}~\bibnamefont {van
  Kolck}},\ }\bibfield  {title} {\bibinfo {title} {{No-core shell model in an
  effective-field-theory framework}},\ }\href
  {https://doi.org/10.1016/j.physletb.2007.07.065} {\bibfield  {journal}
  {\bibinfo  {journal} {Phys. Lett. B}\ }\textbf {\bibinfo {volume} {653}},\
  \bibinfo {pages} {358} (\bibinfo {year} {2007})},\ \Eprint
  {https://arxiv.org/abs/nucl-th/0609023} {arXiv:nucl-th/0609023} \BibitemShut
  {NoStop}%
\bibitem [{\citenamefont {Contessi}\ \emph {et~al.}(2017)\citenamefont
  {Contessi}, \citenamefont {Lovato}, \citenamefont {Pederiva}, \citenamefont
  {Roggero}, \citenamefont {Kirscher},\ and\ \citenamefont {van
  Kolck}}]{Contessi:2017rww}%
  \BibitemOpen
  \bibfield  {author} {\bibinfo {author} {\bibfnamefont {L.}~\bibnamefont
  {Contessi}}, \bibinfo {author} {\bibfnamefont {A.}~\bibnamefont {Lovato}},
  \bibinfo {author} {\bibfnamefont {F.}~\bibnamefont {Pederiva}}, \bibinfo
  {author} {\bibfnamefont {A.}~\bibnamefont {Roggero}}, \bibinfo {author}
  {\bibfnamefont {J.}~\bibnamefont {Kirscher}},\ and\ \bibinfo {author}
  {\bibfnamefont {U.}~\bibnamefont {van Kolck}},\ }\bibfield  {title} {\bibinfo
  {title} {{Ground-state properties of $^{4}$He and $^{16}$O extrapolated from
  lattice QCD with pionless EFT}},\ }\href
  {https://doi.org/10.1016/j.physletb.2017.07.048} {\bibfield  {journal}
  {\bibinfo  {journal} {Phys. Lett. B}\ }\textbf {\bibinfo {volume} {772}},\
  \bibinfo {pages} {839} (\bibinfo {year} {2017})},\ \Eprint
  {https://arxiv.org/abs/1701.06516} {arXiv:1701.06516 [nucl-th]} \BibitemShut
  {NoStop}%
\bibitem [{\citenamefont {Bansal}\ \emph {et~al.}(2018)\citenamefont {Bansal},
  \citenamefont {Binder}, \citenamefont {Ekstr\"om}, \citenamefont {Hagen},
  \citenamefont {Jansen},\ and\ \citenamefont {Papenbrock}}]{Bansal:2017pwn}%
  \BibitemOpen
  \bibfield  {author} {\bibinfo {author} {\bibfnamefont {A.}~\bibnamefont
  {Bansal}}, \bibinfo {author} {\bibfnamefont {S.}~\bibnamefont {Binder}},
  \bibinfo {author} {\bibfnamefont {A.}~\bibnamefont {Ekstr\"om}}, \bibinfo
  {author} {\bibfnamefont {G.}~\bibnamefont {Hagen}}, \bibinfo {author}
  {\bibfnamefont {G.~R.}\ \bibnamefont {Jansen}},\ and\ \bibinfo {author}
  {\bibfnamefont {T.}~\bibnamefont {Papenbrock}},\ }\bibfield  {title}
  {\bibinfo {title} {{Pion-less effective field theory for atomic nuclei and
  lattice nuclei}},\ }\href {https://doi.org/10.1103/PhysRevC.98.054301}
  {\bibfield  {journal} {\bibinfo  {journal} {Phys. Rev. C}\ }\textbf {\bibinfo
  {volume} {98}},\ \bibinfo {pages} {054301} (\bibinfo {year} {2018})},\
  \Eprint {https://arxiv.org/abs/1712.10246} {arXiv:1712.10246 [nucl-th]}
  \BibitemShut {NoStop}%
\bibitem [{\citenamefont {Sch\"afer}\ \emph {et~al.}(2021)\citenamefont
  {Sch\"afer}, \citenamefont {Contessi}, \citenamefont {Kirscher},\ and\
  \citenamefont {Mare\v{s}}}]{Schafer:2020ivj}%
  \BibitemOpen
  \bibfield  {author} {\bibinfo {author} {\bibfnamefont {M.}~\bibnamefont
  {Sch\"afer}}, \bibinfo {author} {\bibfnamefont {L.}~\bibnamefont {Contessi}},
  \bibinfo {author} {\bibfnamefont {J.}~\bibnamefont {Kirscher}},\ and\
  \bibinfo {author} {\bibfnamefont {J.}~\bibnamefont {Mare\v{s}}},\ }\bibfield
  {title} {\bibinfo {title} {{Multi-fermion systems with contact theories}},\
  }\href {https://doi.org/10.1016/j.physletb.2021.136194} {\bibfield  {journal}
  {\bibinfo  {journal} {Phys. Lett. B}\ }\textbf {\bibinfo {volume} {816}},\
  \bibinfo {pages} {136194} (\bibinfo {year} {2021})},\ \Eprint
  {https://arxiv.org/abs/2003.09862} {arXiv:2003.09862 [nucl-th]} \BibitemShut
  {NoStop}%
\bibitem [{\citenamefont {Mishra}\ \emph {et~al.}(2022)\citenamefont {Mishra},
  \citenamefont {Ekstr\"om}, \citenamefont {Hagen}, \citenamefont
  {Papenbrock},\ and\ \citenamefont {Platter}}]{Mishra:2021luw}%
  \BibitemOpen
  \bibfield  {author} {\bibinfo {author} {\bibfnamefont {C.}~\bibnamefont
  {Mishra}}, \bibinfo {author} {\bibfnamefont {A.}~\bibnamefont {Ekstr\"om}},
  \bibinfo {author} {\bibfnamefont {G.}~\bibnamefont {Hagen}}, \bibinfo
  {author} {\bibfnamefont {T.}~\bibnamefont {Papenbrock}},\ and\ \bibinfo
  {author} {\bibfnamefont {L.}~\bibnamefont {Platter}},\ }\bibfield  {title}
  {\bibinfo {title} {{Two-pion exchange as a leading-order contribution in
  chiral effective field theory}},\ }\href
  {https://doi.org/10.1103/PhysRevC.106.024004} {\bibfield  {journal} {\bibinfo
   {journal} {Phys. Rev. C}\ }\textbf {\bibinfo {volume} {106}},\ \bibinfo
  {pages} {024004} (\bibinfo {year} {2022})},\ \Eprint
  {https://arxiv.org/abs/2111.15515} {arXiv:2111.15515 [nucl-th]} \BibitemShut
  {NoStop}%
\bibitem [{\citenamefont {Kievsky}\ \emph {et~al.}(2017)\citenamefont
  {Kievsky}, \citenamefont {Viviani}, \citenamefont {Gattobigio},\ and\
  \citenamefont {Girlanda}}]{Kievsky:2016kzb}%
  \BibitemOpen
  \bibfield  {author} {\bibinfo {author} {\bibfnamefont {A.}~\bibnamefont
  {Kievsky}}, \bibinfo {author} {\bibfnamefont {M.}~\bibnamefont {Viviani}},
  \bibinfo {author} {\bibfnamefont {M.}~\bibnamefont {Gattobigio}},\ and\
  \bibinfo {author} {\bibfnamefont {L.}~\bibnamefont {Girlanda}},\ }\bibfield
  {title} {\bibinfo {title} {{Implications of Efimov physics for the
  description of three and four nucleons in chiral effective field theory}},\
  }\href {https://doi.org/10.1103/PhysRevC.95.024001} {\bibfield  {journal}
  {\bibinfo  {journal} {Phys. Rev. C}\ }\textbf {\bibinfo {volume} {95}},\
  \bibinfo {pages} {024001} (\bibinfo {year} {2017})},\ \Eprint
  {https://arxiv.org/abs/1610.09858} {arXiv:1610.09858 [nucl-th]} \BibitemShut
  {NoStop}%
\bibitem [{\citenamefont {Bazak}\ \emph {et~al.}(2019)\citenamefont {Bazak},
  \citenamefont {Kirscher}, \citenamefont {K\"onig}, \citenamefont
  {Pav\'on~Valderrama}, \citenamefont {Barnea},\ and\ \citenamefont {van
  Kolck}}]{Bazak:2018qnu}%
  \BibitemOpen
  \bibfield  {author} {\bibinfo {author} {\bibfnamefont {B.}~\bibnamefont
  {Bazak}}, \bibinfo {author} {\bibfnamefont {J.}~\bibnamefont {Kirscher}},
  \bibinfo {author} {\bibfnamefont {S.}~\bibnamefont {K\"onig}}, \bibinfo
  {author} {\bibfnamefont {M.}~\bibnamefont {Pav\'on~Valderrama}}, \bibinfo
  {author} {\bibfnamefont {N.}~\bibnamefont {Barnea}},\ and\ \bibinfo {author}
  {\bibfnamefont {U.}~\bibnamefont {van Kolck}},\ }\bibfield  {title} {\bibinfo
  {title} {{Four-Body Scale in Universal Few-Boson Systems}},\ }\href
  {https://doi.org/10.1103/PhysRevLett.122.143001} {\bibfield  {journal}
  {\bibinfo  {journal} {Phys. Rev. Lett.}\ }\textbf {\bibinfo {volume} {122}},\
  \bibinfo {pages} {143001} (\bibinfo {year} {2019})},\ \Eprint
  {https://arxiv.org/abs/1812.00387} {arXiv:1812.00387 [cond-mat.quant-gas]}
  \BibitemShut {NoStop}%
\bibitem [{\citenamefont {Yang}(2020)}]{Yang:2019hkn}%
  \BibitemOpen
  \bibfield  {author} {\bibinfo {author} {\bibfnamefont {C.~J.}\ \bibnamefont
  {Yang}},\ }\bibfield  {title} {\bibinfo {title} {{Do we know how to count
  powers in pionless and pionful effective field theory?}},\ }\href
  {https://doi.org/10.1140/epja/s10050-020-00104-0} {\bibfield  {journal}
  {\bibinfo  {journal} {Eur. Phys. J. A}\ }\textbf {\bibinfo {volume} {56}},\
  \bibinfo {pages} {96} (\bibinfo {year} {2020})},\ \Eprint
  {https://arxiv.org/abs/1905.12510} {arXiv:1905.12510 [nucl-th]} \BibitemShut
  {NoStop}%
\bibitem [{\citenamefont {Yang}\ \emph
  {et~al.}(2021{\natexlab{b}})\citenamefont {Yang}, \citenamefont {Ekstr\"om},
  \citenamefont {Forss\'en}, \citenamefont {Hagen}, \citenamefont {Rupak},\
  and\ \citenamefont {van Kolck}}]{Yang:2021vxa}%
  \BibitemOpen
  \bibfield  {author} {\bibinfo {author} {\bibfnamefont {C.~J.}\ \bibnamefont
  {Yang}}, \bibinfo {author} {\bibfnamefont {A.}~\bibnamefont {Ekstr\"om}},
  \bibinfo {author} {\bibfnamefont {C.}~\bibnamefont {Forss\'en}}, \bibinfo
  {author} {\bibfnamefont {G.}~\bibnamefont {Hagen}}, \bibinfo {author}
  {\bibfnamefont {G.}~\bibnamefont {Rupak}},\ and\ \bibinfo {author}
  {\bibfnamefont {U.}~\bibnamefont {van Kolck}},\ }\href@noop {} {\bibinfo
  {title} {{The importance of few-nucleon forces in chiral effective field
  theory}}} (\bibinfo {year} {2021}{\natexlab{b}}),\ \Eprint
  {https://arxiv.org/abs/2109.13303} {arXiv:2109.13303 [nucl-th]} \BibitemShut
  {NoStop}%
\bibitem [{\citenamefont {Furnstahl}\ \emph {et~al.}(2015)\citenamefont
  {Furnstahl}, \citenamefont {Klco}, \citenamefont {Phillips},\ and\
  \citenamefont {Wesolowski}}]{Furnstahl:2015rha}%
  \BibitemOpen
  \bibfield  {author} {\bibinfo {author} {\bibfnamefont {R.~J.}\ \bibnamefont
  {Furnstahl}}, \bibinfo {author} {\bibfnamefont {N.}~\bibnamefont {Klco}},
  \bibinfo {author} {\bibfnamefont {D.~R.}\ \bibnamefont {Phillips}},\ and\
  \bibinfo {author} {\bibfnamefont {S.}~\bibnamefont {Wesolowski}},\ }\bibfield
   {title} {\bibinfo {title} {{Quantifying truncation errors in effective field
  theory}},\ }\href {https://doi.org/10.1103/PhysRevC.92.024005} {\bibfield
  {journal} {\bibinfo  {journal} {Phys. Rev. C}\ }\textbf {\bibinfo {volume}
  {92}},\ \bibinfo {pages} {024005} (\bibinfo {year} {2015})},\ \Eprint
  {https://arxiv.org/abs/1506.01343} {arXiv:1506.01343 [nucl-th]} \BibitemShut
  {NoStop}%
\bibitem [{\citenamefont {Melendez}\ \emph {et~al.}(2017)\citenamefont
  {Melendez}, \citenamefont {Wesolowski},\ and\ \citenamefont
  {Furnstahl}}]{Melendez:2017phj}%
  \BibitemOpen
  \bibfield  {author} {\bibinfo {author} {\bibfnamefont {J.~A.}\ \bibnamefont
  {Melendez}}, \bibinfo {author} {\bibfnamefont {S.}~\bibnamefont
  {Wesolowski}},\ and\ \bibinfo {author} {\bibfnamefont {R.~J.}\ \bibnamefont
  {Furnstahl}},\ }\bibfield  {title} {\bibinfo {title} {{Bayesian truncation
  errors in chiral effective field theory: nucleon-nucleon observables}},\
  }\href {https://doi.org/10.1103/PhysRevC.96.024003} {\bibfield  {journal}
  {\bibinfo  {journal} {Phys. Rev. C}\ }\textbf {\bibinfo {volume} {96}},\
  \bibinfo {pages} {024003} (\bibinfo {year} {2017})},\ \Eprint
  {https://arxiv.org/abs/1704.03308} {arXiv:1704.03308 [nucl-th]} \BibitemShut
  {NoStop}%
\bibitem [{\citenamefont {Melendez}\ \emph {et~al.}(2019)\citenamefont
  {Melendez}, \citenamefont {Furnstahl}, \citenamefont {Phillips},
  \citenamefont {Pratola},\ and\ \citenamefont
  {Wesolowski}}]{Melendez:2019izc}%
  \BibitemOpen
  \bibfield  {author} {\bibinfo {author} {\bibfnamefont {J.~A.}\ \bibnamefont
  {Melendez}}, \bibinfo {author} {\bibfnamefont {R.~J.}\ \bibnamefont
  {Furnstahl}}, \bibinfo {author} {\bibfnamefont {D.~R.}\ \bibnamefont
  {Phillips}}, \bibinfo {author} {\bibfnamefont {M.~T.}\ \bibnamefont
  {Pratola}},\ and\ \bibinfo {author} {\bibfnamefont {S.}~\bibnamefont
  {Wesolowski}},\ }\bibfield  {title} {\bibinfo {title} {{Quantifying
  Correlated Truncation Errors in Effective Field Theory}},\ }\href
  {https://doi.org/10.1103/PhysRevC.100.044001} {\bibfield  {journal} {\bibinfo
   {journal} {Phys. Rev. C}\ }\textbf {\bibinfo {volume} {100}},\ \bibinfo
  {pages} {044001} (\bibinfo {year} {2019})},\ \Eprint
  {https://arxiv.org/abs/1904.10581} {arXiv:1904.10581 [nucl-th]} \BibitemShut
  {NoStop}%
\bibitem [{\citenamefont {Wesolowski}\ \emph {et~al.}(2021)\citenamefont
  {Wesolowski}, \citenamefont {Svensson}, \citenamefont {Ekstr\"om},
  \citenamefont {Forss\'en}, \citenamefont {Furnstahl}, \citenamefont
  {Melendez},\ and\ \citenamefont {Phillips}}]{Wesolowski:2021cni}%
  \BibitemOpen
  \bibfield  {author} {\bibinfo {author} {\bibfnamefont {S.}~\bibnamefont
  {Wesolowski}}, \bibinfo {author} {\bibfnamefont {I.}~\bibnamefont
  {Svensson}}, \bibinfo {author} {\bibfnamefont {A.}~\bibnamefont {Ekstr\"om}},
  \bibinfo {author} {\bibfnamefont {C.}~\bibnamefont {Forss\'en}}, \bibinfo
  {author} {\bibfnamefont {R.~J.}\ \bibnamefont {Furnstahl}}, \bibinfo {author}
  {\bibfnamefont {J.~A.}\ \bibnamefont {Melendez}},\ and\ \bibinfo {author}
  {\bibfnamefont {D.~R.}\ \bibnamefont {Phillips}},\ }\bibfield  {title}
  {\bibinfo {title} {{Rigorous constraints on three-nucleon forces in chiral
  effective field theory from fast and accurate calculations of few-body
  observables}},\ }\href {https://doi.org/10.1103/PhysRevC.104.064001}
  {\bibfield  {journal} {\bibinfo  {journal} {Phys. Rev. C}\ }\textbf {\bibinfo
  {volume} {104}},\ \bibinfo {pages} {064001} (\bibinfo {year} {2021})},\
  \Eprint {https://arxiv.org/abs/2104.04441} {arXiv:2104.04441 [nucl-th]}
  \BibitemShut {NoStop}%
\bibitem [{\citenamefont {Svensson}\ \emph {et~al.}(2022)\citenamefont
  {Svensson}, \citenamefont {Ekstr\"om},\ and\ \citenamefont
  {Forss\'en}}]{Svensson:2022}%
  \BibitemOpen
  \bibfield  {author} {\bibinfo {author} {\bibfnamefont {I.}~\bibnamefont
  {Svensson}}, \bibinfo {author} {\bibfnamefont {A.}~\bibnamefont
  {Ekstr\"om}},\ and\ \bibinfo {author} {\bibfnamefont {C.}~\bibnamefont
  {Forss\'en}},\ }\bibfield  {title} {\bibinfo {title} {Bayesian parameter
  estimation in chiral effective field theory using the hamiltonian monte carlo
  method},\ }\href {https://doi.org/10.1103/PhysRevC.105.014004} {\bibfield
  {journal} {\bibinfo  {journal} {Phys. Rev. C}\ }\textbf {\bibinfo {volume}
  {105}},\ \bibinfo {pages} {014004} (\bibinfo {year} {2022})}\BibitemShut
  {NoStop}%
\bibitem [{\citenamefont {Svensson}\ \emph {et~al.}(2023)\citenamefont
  {Svensson}, \citenamefont {Ekstr\"om},\ and\ \citenamefont
  {Forss\'en}}]{Svensson:2022kkj}%
  \BibitemOpen
  \bibfield  {author} {\bibinfo {author} {\bibfnamefont {I.}~\bibnamefont
  {Svensson}}, \bibinfo {author} {\bibfnamefont {A.}~\bibnamefont
  {Ekstr\"om}},\ and\ \bibinfo {author} {\bibfnamefont {C.}~\bibnamefont
  {Forss\'en}},\ }\bibfield  {title} {\bibinfo {title} {{Bayesian estimation of
  the low-energy constants up to fourth order in the nucleon-nucleon sector of
  chiral effective field theory}},\ }\href
  {https://doi.org/10.1103/PhysRevC.107.014001} {\bibfield  {journal} {\bibinfo
   {journal} {Phys. Rev. C}\ }\textbf {\bibinfo {volume} {107}},\ \bibinfo
  {pages} {014001} (\bibinfo {year} {2023})},\ \Eprint
  {https://arxiv.org/abs/2206.08250} {arXiv:2206.08250 [nucl-th]} \BibitemShut
  {NoStop}%
\bibitem [{\citenamefont {Drischler}\ \emph {et~al.}(2020)\citenamefont
  {Drischler}, \citenamefont {Furnstahl}, \citenamefont {Melendez},\ and\
  \citenamefont {Phillips}}]{Drischler:2020hwi}%
  \BibitemOpen
  \bibfield  {author} {\bibinfo {author} {\bibfnamefont {C.}~\bibnamefont
  {Drischler}}, \bibinfo {author} {\bibfnamefont {R.~J.}\ \bibnamefont
  {Furnstahl}}, \bibinfo {author} {\bibfnamefont {J.~A.}\ \bibnamefont
  {Melendez}},\ and\ \bibinfo {author} {\bibfnamefont {D.~R.}\ \bibnamefont
  {Phillips}},\ }\bibfield  {title} {\bibinfo {title} {{How Well Do We Know the
  Neutron-Matter Equation of State at the Densities Inside Neutron Stars? A
  Bayesian Approach with Correlated Uncertainties}},\ }\href
  {https://doi.org/10.1103/PhysRevLett.125.202702} {\bibfield  {journal}
  {\bibinfo  {journal} {Phys. Rev. Lett.}\ }\textbf {\bibinfo {volume} {125}},\
  \bibinfo {pages} {202702} (\bibinfo {year} {2020})},\ \Eprint
  {https://arxiv.org/abs/2004.07232} {arXiv:2004.07232 [nucl-th]} \BibitemShut
  {NoStop}%
\bibitem [{\citenamefont {Dj\"arv}\ \emph {et~al.}(2022)\citenamefont
  {Dj\"arv}, \citenamefont {Ekstr\"om}, \citenamefont {Forss\'en},\ and\
  \citenamefont {Johansson}}]{Djarv:2021hcj}%
  \BibitemOpen
  \bibfield  {author} {\bibinfo {author} {\bibfnamefont {T.}~\bibnamefont
  {Dj\"arv}}, \bibinfo {author} {\bibfnamefont {A.}~\bibnamefont {Ekstr\"om}},
  \bibinfo {author} {\bibfnamefont {C.}~\bibnamefont {Forss\'en}},\ and\
  \bibinfo {author} {\bibfnamefont {H.~T.}\ \bibnamefont {Johansson}},\
  }\bibfield  {title} {\bibinfo {title} {{Bayesian predictions for A=6 nuclei
  using eigenvector continuation emulators}},\ }\href
  {https://doi.org/10.1103/PhysRevC.105.014005} {\bibfield  {journal} {\bibinfo
   {journal} {Phys. Rev. C}\ }\textbf {\bibinfo {volume} {105}},\ \bibinfo
  {pages} {014005} (\bibinfo {year} {2022})},\ \Eprint
  {https://arxiv.org/abs/2108.13313} {arXiv:2108.13313 [nucl-th]} \BibitemShut
  {NoStop}%
\bibitem [{\citenamefont {Hu}\ \emph {et~al.}(2022)\citenamefont {Hu} \emph
  {et~al.}}]{Hu:2021trw}%
  \BibitemOpen
  \bibfield  {author} {\bibinfo {author} {\bibfnamefont {B.}~\bibnamefont {Hu}}
  \emph {et~al.},\ }\bibfield  {title} {\bibinfo {title} {{Ab initio
  predictions link the neutron skin of $^{208}$Pb to nuclear forces}},\ }\href
  {https://doi.org/10.1038/s41567-022-01715-8} {\bibfield  {journal} {\bibinfo
  {journal} {Nature Phys.}\ }\textbf {\bibinfo {volume} {18}},\ \bibinfo
  {pages} {1196} (\bibinfo {year} {2022})},\ \Eprint
  {https://arxiv.org/abs/2112.01125} {arXiv:2112.01125 [nucl-th]} \BibitemShut
  {NoStop}%
\bibitem [{\citenamefont {Manohar}\ and\ \citenamefont
  {Georgi}(1984)}]{Manohar:1983md}%
  \BibitemOpen
  \bibfield  {author} {\bibinfo {author} {\bibfnamefont {A.}~\bibnamefont
  {Manohar}}\ and\ \bibinfo {author} {\bibfnamefont {H.}~\bibnamefont
  {Georgi}},\ }\bibfield  {title} {\bibinfo {title} {{Chiral Quarks and the
  Nonrelativistic Quark Model}},\ }\href
  {https://doi.org/10.1016/0550-3213(84)90231-1} {\bibfield  {journal}
  {\bibinfo  {journal} {Nucl. Phys. B}\ }\textbf {\bibinfo {volume} {234}},\
  \bibinfo {pages} {189} (\bibinfo {year} {1984})}\BibitemShut {NoStop}%
\bibitem [{\citenamefont {Nogga}\ \emph {et~al.}(2005)\citenamefont {Nogga},
  \citenamefont {Timmermans},\ and\ \citenamefont {van Kolck}}]{Nogga:2005hy}%
  \BibitemOpen
  \bibfield  {author} {\bibinfo {author} {\bibfnamefont {A.}~\bibnamefont
  {Nogga}}, \bibinfo {author} {\bibfnamefont {R.~G.~E.}\ \bibnamefont
  {Timmermans}},\ and\ \bibinfo {author} {\bibfnamefont {U.}~\bibnamefont {van
  Kolck}},\ }\bibfield  {title} {\bibinfo {title} {{Renormalization of one-pion
  exchange and power counting}},\ }\href
  {https://doi.org/10.1103/PhysRevC.72.054006} {\bibfield  {journal} {\bibinfo
  {journal} {Phys. Rev. C}\ }\textbf {\bibinfo {volume} {72}},\ \bibinfo
  {pages} {054006} (\bibinfo {year} {2005})},\ \Eprint
  {https://arxiv.org/abs/nucl-th/0506005} {arXiv:nucl-th/0506005} \BibitemShut
  {NoStop}%
\bibitem [{\citenamefont {Long}\ and\ \citenamefont
  {Yang}(2012{\natexlab{a}})}]{Long:2012ve}%
  \BibitemOpen
  \bibfield  {author} {\bibinfo {author} {\bibfnamefont {B.}~\bibnamefont
  {Long}}\ and\ \bibinfo {author} {\bibfnamefont {C.~J.}\ \bibnamefont
  {Yang}},\ }\bibfield  {title} {\bibinfo {title} {{Short-range nuclear forces
  in singlet channels}},\ }\href {https://doi.org/10.1103/PhysRevC.86.024001}
  {\bibfield  {journal} {\bibinfo  {journal} {Phys. Rev. C}\ }\textbf {\bibinfo
  {volume} {86}},\ \bibinfo {pages} {024001} (\bibinfo {year}
  {2012}{\natexlab{a}})},\ \Eprint {https://arxiv.org/abs/1202.4053}
  {arXiv:1202.4053 [nucl-th]} \BibitemShut {NoStop}%
\bibitem [{\citenamefont {Long}(2013)}]{Long:2013cya}%
  \BibitemOpen
  \bibfield  {author} {\bibinfo {author} {\bibfnamefont {B.}~\bibnamefont
  {Long}},\ }\bibfield  {title} {\bibinfo {title} {{Improved convergence of
  chiral effective field theory for 1S0 of NN scattering}},\ }\href
  {https://doi.org/10.1103/PhysRevC.88.014002} {\bibfield  {journal} {\bibinfo
  {journal} {Phys. Rev. C}\ }\textbf {\bibinfo {volume} {88}},\ \bibinfo
  {pages} {014002} (\bibinfo {year} {2013})},\ \Eprint
  {https://arxiv.org/abs/1304.7382} {arXiv:1304.7382 [nucl-th]} \BibitemShut
  {NoStop}%
\bibitem [{\citenamefont {Long}\ and\ \citenamefont
  {Yang}(2012{\natexlab{b}})}]{PhysRevC.85.034002}%
  \BibitemOpen
  \bibfield  {author} {\bibinfo {author} {\bibfnamefont {B.}~\bibnamefont
  {Long}}\ and\ \bibinfo {author} {\bibfnamefont {C.-J.}\ \bibnamefont
  {Yang}},\ }\bibfield  {title} {\bibinfo {title} {Renormalizing chiral nuclear
  forces: Triplet channels},\ }\href
  {https://doi.org/10.1103/PhysRevC.85.034002} {\bibfield  {journal} {\bibinfo
  {journal} {Phys. Rev. C}\ }\textbf {\bibinfo {volume} {85}},\ \bibinfo
  {pages} {034002} (\bibinfo {year} {2012}{\natexlab{b}})}\BibitemShut
  {NoStop}%
\bibitem [{\citenamefont {Valderrama}\ and\ \citenamefont
  {Ruiz~Arriola}(2009)}]{Valderrama:2008kj}%
  \BibitemOpen
  \bibfield  {author} {\bibinfo {author} {\bibfnamefont {M.~P.}\ \bibnamefont
  {Valderrama}}\ and\ \bibinfo {author} {\bibfnamefont {E.}~\bibnamefont
  {Ruiz~Arriola}},\ }\bibfield  {title} {\bibinfo {title} {{Renormalization of
  chiral two-pion exchange NN interactions with Delta-excitations: Central
  Phases and the Deuteron}},\ }\href
  {https://doi.org/10.1103/PhysRevC.79.044001} {\bibfield  {journal} {\bibinfo
  {journal} {Phys. Rev. C}\ }\textbf {\bibinfo {volume} {79}},\ \bibinfo
  {pages} {044001} (\bibinfo {year} {2009})},\ \Eprint
  {https://arxiv.org/abs/0809.3186} {arXiv:0809.3186 [nucl-th]} \BibitemShut
  {NoStop}%
\bibitem [{\citenamefont {Birse}(2006)}]{Birse:2005um}%
  \BibitemOpen
  \bibfield  {author} {\bibinfo {author} {\bibfnamefont {M.~C.}\ \bibnamefont
  {Birse}},\ }\bibfield  {title} {\bibinfo {title} {{Power counting with
  one-pion exchange}},\ }\href {https://doi.org/10.1103/PhysRevC.74.014003}
  {\bibfield  {journal} {\bibinfo  {journal} {Phys. Rev. C}\ }\textbf {\bibinfo
  {volume} {74}},\ \bibinfo {pages} {014003} (\bibinfo {year} {2006})},\
  \Eprint {https://arxiv.org/abs/nucl-th/0507077} {arXiv:nucl-th/0507077}
  \BibitemShut {NoStop}%
\bibitem [{\citenamefont {Kaplan}\ \emph {et~al.}(1998)\citenamefont {Kaplan},
  \citenamefont {Savage},\ and\ \citenamefont {Wise}}]{Kaplan:1998tg}%
  \BibitemOpen
  \bibfield  {author} {\bibinfo {author} {\bibfnamefont {D.~B.}\ \bibnamefont
  {Kaplan}}, \bibinfo {author} {\bibfnamefont {M.~J.}\ \bibnamefont {Savage}},\
  and\ \bibinfo {author} {\bibfnamefont {M.~B.}\ \bibnamefont {Wise}},\
  }\bibfield  {title} {\bibinfo {title} {{A New expansion for nucleon-nucleon
  interactions}},\ }\href {https://doi.org/10.1016/S0370-2693(98)00210-X}
  {\bibfield  {journal} {\bibinfo  {journal} {Phys. Lett. B}\ }\textbf
  {\bibinfo {volume} {424}},\ \bibinfo {pages} {390} (\bibinfo {year}
  {1998})},\ \Eprint {https://arxiv.org/abs/nucl-th/9801034}
  {arXiv:nucl-th/9801034} \BibitemShut {NoStop}%
\bibitem [{\citenamefont {Long}\ and\ \citenamefont {van
  Kolck}(2008)}]{Long:2007vp}%
  \BibitemOpen
  \bibfield  {author} {\bibinfo {author} {\bibfnamefont {B.}~\bibnamefont
  {Long}}\ and\ \bibinfo {author} {\bibfnamefont {U.}~\bibnamefont {van
  Kolck}},\ }\bibfield  {title} {\bibinfo {title} {{Renormalization of Singular
  Potentials and Power Counting}},\ }\href
  {https://doi.org/10.1016/j.aop.2008.01.003} {\bibfield  {journal} {\bibinfo
  {journal} {Ann. Phys.}\ }\textbf {\bibinfo {volume} {323}},\ \bibinfo {pages}
  {1304} (\bibinfo {year} {2008})},\ \Eprint {https://arxiv.org/abs/0707.4325}
  {arXiv:0707.4325 [quant-ph]} \BibitemShut {NoStop}%
\bibitem [{\citenamefont {Epelbaum}\ and\ \citenamefont
  {Meissner}(2013)}]{Epelbaum:2006pt}%
  \BibitemOpen
  \bibfield  {author} {\bibinfo {author} {\bibfnamefont {E.}~\bibnamefont
  {Epelbaum}}\ and\ \bibinfo {author} {\bibfnamefont {U.~G.}\ \bibnamefont
  {Meissner}},\ }\bibfield  {title} {\bibinfo {title} {{On the Renormalization
  of the One-Pion Exchange Potential and the Consistency of Weinberg`s Power
  Counting}},\ }\href {https://doi.org/10.1007/s00601-012-0492-1} {\bibfield
  {journal} {\bibinfo  {journal} {Few Body Syst.}\ }\textbf {\bibinfo {volume}
  {54}},\ \bibinfo {pages} {2175} (\bibinfo {year} {2013})},\ \Eprint
  {https://arxiv.org/abs/nucl-th/0609037} {arXiv:nucl-th/0609037} \BibitemShut
  {NoStop}%
\bibitem [{\citenamefont {Epelbaum}\ and\ \citenamefont
  {Gegelia}(2009)}]{Epelbaum:2009sd}%
  \BibitemOpen
  \bibfield  {author} {\bibinfo {author} {\bibfnamefont {E.}~\bibnamefont
  {Epelbaum}}\ and\ \bibinfo {author} {\bibfnamefont {J.}~\bibnamefont
  {Gegelia}},\ }\bibfield  {title} {\bibinfo {title} {{Regularization,
  renormalization and 'peratization' in effective field theory for two
  nucleons}},\ }\href {https://doi.org/10.1140/epja/i2009-10833-3} {\bibfield
  {journal} {\bibinfo  {journal} {Eur. Phys. J. A}\ }\textbf {\bibinfo {volume}
  {41}},\ \bibinfo {pages} {341} (\bibinfo {year} {2009})},\ \Eprint
  {https://arxiv.org/abs/0906.3822} {arXiv:0906.3822 [nucl-th]} \BibitemShut
  {NoStop}%
\bibitem [{\citenamefont {Epelbaum}\ \emph {et~al.}(2018)\citenamefont
  {Epelbaum}, \citenamefont {Gasparyan}, \citenamefont {Gegelia},\ and\
  \citenamefont {Mei\ss{}ner}}]{Epelbaum:2018zli}%
  \BibitemOpen
  \bibfield  {author} {\bibinfo {author} {\bibfnamefont {E.}~\bibnamefont
  {Epelbaum}}, \bibinfo {author} {\bibfnamefont {A.~M.}\ \bibnamefont
  {Gasparyan}}, \bibinfo {author} {\bibfnamefont {J.}~\bibnamefont {Gegelia}},\
  and\ \bibinfo {author} {\bibfnamefont {U.-G.}\ \bibnamefont {Mei\ss{}ner}},\
  }\bibfield  {title} {\bibinfo {title} {{How (not) to renormalize integral
  equations with singular potentials in effective field theory}},\ }\href
  {https://doi.org/10.1140/epja/i2018-12632-1} {\bibfield  {journal} {\bibinfo
  {journal} {Eur. Phys. J. A}\ }\textbf {\bibinfo {volume} {54}},\ \bibinfo
  {pages} {186} (\bibinfo {year} {2018})},\ \Eprint
  {https://arxiv.org/abs/1810.02646} {arXiv:1810.02646 [nucl-th]} \BibitemShut
  {NoStop}%
\bibitem [{\citenamefont {Gasparyan}\ and\ \citenamefont
  {Epelbaum}(2022)}]{Gasparyan:2022isg}%
  \BibitemOpen
  \bibfield  {author} {\bibinfo {author} {\bibfnamefont {A.~M.}\ \bibnamefont
  {Gasparyan}}\ and\ \bibinfo {author} {\bibfnamefont {E.}~\bibnamefont
  {Epelbaum}},\ }\bibfield  {title} {\bibinfo {title} {{Is the RG-invariant EFT
  for few-nucleon systems cutoff independent?}},\ }\href@noop {} {\bibfield
  {journal} {\bibinfo  {journal} {arxiv}\ } (\bibinfo {year} {2022})},\ \Eprint
  {https://arxiv.org/abs/2210.16225} {arXiv:2210.16225 [nucl-th]} \BibitemShut
  {NoStop}%
\bibitem [{\citenamefont {Hergert}(2020)}]{Hergert:2020bxy}%
  \BibitemOpen
  \bibfield  {author} {\bibinfo {author} {\bibfnamefont {H.}~\bibnamefont
  {Hergert}},\ }\bibfield  {title} {\bibinfo {title} {{A Guided Tour of $ab$
  $initio$ Nuclear Many-Body Theory}},\ }\href
  {https://doi.org/10.3389/fphy.2020.00379} {\bibfield  {journal} {\bibinfo
  {journal} {Front. Phys.}\ }\textbf {\bibinfo {volume} {8}},\ \bibinfo {pages}
  {379} (\bibinfo {year} {2020})},\ \Eprint {https://arxiv.org/abs/2008.05061}
  {arXiv:2008.05061 [nucl-th]} \BibitemShut {NoStop}%
\bibitem [{\citenamefont {Hebeler}(2021)}]{Hebeler:2020ocj}%
  \BibitemOpen
  \bibfield  {author} {\bibinfo {author} {\bibfnamefont {K.}~\bibnamefont
  {Hebeler}},\ }\bibfield  {title} {\bibinfo {title} {{Three-nucleon forces:
  Implementation and applications to atomic nuclei and dense matter}},\ }\href
  {https://doi.org/10.1016/j.physrep.2020.08.009} {\bibfield  {journal}
  {\bibinfo  {journal} {Phys. Rep.}\ }\textbf {\bibinfo {volume} {890}},\
  \bibinfo {pages} {1} (\bibinfo {year} {2021})},\ \Eprint
  {https://arxiv.org/abs/2002.09548} {arXiv:2002.09548 [nucl-th]} \BibitemShut
  {NoStop}%
\bibitem [{\citenamefont {Griesshammer}(2020)}]{Griesshammer:2020fwr}%
  \BibitemOpen
  \bibfield  {author} {\bibinfo {author} {\bibfnamefont {H.~W.}\ \bibnamefont
  {Griesshammer}},\ }\bibfield  {title} {\bibinfo {title} {{A Consistency Test
  of EFT Power Countings from Residual Cutoff Dependence}},\ }\href
  {https://doi.org/10.1140/epja/s10050-020-00129-5} {\bibfield  {journal}
  {\bibinfo  {journal} {Eur. Phys. J. A}\ }\textbf {\bibinfo {volume} {56}},\
  \bibinfo {pages} {118} (\bibinfo {year} {2020})},\ \Eprint
  {https://arxiv.org/abs/2004.00411} {arXiv:2004.00411 [nucl-th]} \BibitemShut
  {NoStop}%
\bibitem [{\citenamefont {Epelbaum}\ \emph {et~al.}(2015)\citenamefont
  {Epelbaum}, \citenamefont {Krebs},\ and\ \citenamefont
  {Mei\ss{}ner}}]{Epelbaum:2014efa}%
  \BibitemOpen
  \bibfield  {author} {\bibinfo {author} {\bibfnamefont {E.}~\bibnamefont
  {Epelbaum}}, \bibinfo {author} {\bibfnamefont {H.}~\bibnamefont {Krebs}},\
  and\ \bibinfo {author} {\bibfnamefont {U.~G.}\ \bibnamefont {Mei\ss{}ner}},\
  }\bibfield  {title} {\bibinfo {title} {{Improved chiral nucleon-nucleon
  potential up to next-to-next-to-next-to-leading order}},\ }\href
  {https://doi.org/10.1140/epja/i2015-15053-8} {\bibfield  {journal} {\bibinfo
  {journal} {Eur. Phys. J. A}\ }\textbf {\bibinfo {volume} {51}},\ \bibinfo
  {pages} {53} (\bibinfo {year} {2015})},\ \Eprint
  {https://arxiv.org/abs/1412.0142} {arXiv:1412.0142 [nucl-th]} \BibitemShut
  {NoStop}%
\bibitem [{\citenamefont {van Kolck}(2020{\natexlab{b}})}]{vanKolck_2020}%
  \BibitemOpen
  \bibfield  {author} {\bibinfo {author} {\bibfnamefont {U.}~\bibnamefont {van
  Kolck}},\ }\bibfield  {title} {\bibinfo {title} {Naturalness in nuclear
  effective field theories},\ }\href
  {https://doi.org/10.1140/epja/s10050-020-00092-1} {\bibfield  {journal}
  {\bibinfo  {journal} {Eur. Phys. J. A}\ }\textbf {\bibinfo {volume} {56}},\
  \bibinfo {pages} {97} (\bibinfo {year} {2020}{\natexlab{b}})}\BibitemShut
  {NoStop}%
\bibitem [{\citenamefont {Barford}\ and\ \citenamefont
  {Birse}(2001)}]{Barford:2001sx}%
  \BibitemOpen
  \bibfield  {author} {\bibinfo {author} {\bibfnamefont {T.}~\bibnamefont
  {Barford}}\ and\ \bibinfo {author} {\bibfnamefont {M.~C.}\ \bibnamefont
  {Birse}},\ }\bibfield  {title} {\bibinfo {title} {{A Renormalization-group
  approach to two-body scattering with long range forces}},\ }\href
  {https://doi.org/10.1063/1.1436603} {\bibfield  {journal} {\bibinfo
  {journal} {AIP Conf. Proc.}\ }\textbf {\bibinfo {volume} {603}},\ \bibinfo
  {pages} {229} (\bibinfo {year} {2001})},\ \Eprint
  {https://arxiv.org/abs/nucl-th/0108024} {arXiv:nucl-th/0108024} \BibitemShut
  {NoStop}%
\bibitem [{\citenamefont {Barford}\ and\ \citenamefont
  {Birse}(2003)}]{Barford:2002je}%
  \BibitemOpen
  \bibfield  {author} {\bibinfo {author} {\bibfnamefont {T.}~\bibnamefont
  {Barford}}\ and\ \bibinfo {author} {\bibfnamefont {M.~C.}\ \bibnamefont
  {Birse}},\ }\bibfield  {title} {\bibinfo {title} {{A Renormalization group
  approach to two-body scattering in the presence of long range forces}},\
  }\href {https://doi.org/10.1103/PhysRevC.67.064006} {\bibfield  {journal}
  {\bibinfo  {journal} {Phys. Rev. C}\ }\textbf {\bibinfo {volume} {67}},\
  \bibinfo {pages} {064006} (\bibinfo {year} {2003})},\ \Eprint
  {https://arxiv.org/abs/hep-ph/0206146} {arXiv:hep-ph/0206146} \BibitemShut
  {NoStop}%
\bibitem [{\citenamefont {Valderrama}(2011)}]{Valderrama:2009ei}%
  \BibitemOpen
  \bibfield  {author} {\bibinfo {author} {\bibfnamefont {M.~P.}\ \bibnamefont
  {Valderrama}},\ }\bibfield  {title} {\bibinfo {title} {{Perturbative
  renormalizability of chiral two pion exchange in nucleon-nucleon
  scattering}},\ }\href {https://doi.org/10.1103/PhysRevC.83.024003} {\bibfield
   {journal} {\bibinfo  {journal} {Phys. Rev. C}\ }\textbf {\bibinfo {volume}
  {83}},\ \bibinfo {pages} {024003} (\bibinfo {year} {2011})},\ \Eprint
  {https://arxiv.org/abs/0912.0699} {arXiv:0912.0699 [nucl-th]} \BibitemShut
  {NoStop}%
\bibitem [{\citenamefont {Pavon~Valderrama}(2011)}]{PavonValderrama:2011fcz}%
  \BibitemOpen
  \bibfield  {author} {\bibinfo {author} {\bibfnamefont {M.}~\bibnamefont
  {Pavon~Valderrama}},\ }\bibfield  {title} {\bibinfo {title} {{Perturbative
  Renormalizability of Chiral Two Pion Exchange in Nucleon-Nucleon Scattering:
  P- and D-waves}},\ }\href {https://doi.org/10.1103/PhysRevC.84.064002}
  {\bibfield  {journal} {\bibinfo  {journal} {Phys. Rev. C}\ }\textbf {\bibinfo
  {volume} {84}},\ \bibinfo {pages} {064002} (\bibinfo {year} {2011})},\
  \Eprint {https://arxiv.org/abs/1108.0872} {arXiv:1108.0872 [nucl-th]}
  \BibitemShut {NoStop}%
\bibitem [{\citenamefont {Long}\ and\ \citenamefont
  {Yang}(2011)}]{PhysRevC.84.057001}%
  \BibitemOpen
  \bibfield  {author} {\bibinfo {author} {\bibfnamefont {B.}~\bibnamefont
  {Long}}\ and\ \bibinfo {author} {\bibfnamefont {C.-J.}\ \bibnamefont
  {Yang}},\ }\bibfield  {title} {\bibinfo {title} {Renormalizing chiral nuclear
  forces: A case study of ${}^{3}\phantom{\rule{-0.16em}{0ex}}{P}_{0}$},\
  }\href {https://doi.org/10.1103/PhysRevC.84.057001} {\bibfield  {journal}
  {\bibinfo  {journal} {Phys. Rev. C}\ }\textbf {\bibinfo {volume} {84}},\
  \bibinfo {pages} {057001} (\bibinfo {year} {2011})}\BibitemShut {NoStop}%
\bibitem [{\citenamefont {Workman}\ \emph {et~al.}()\citenamefont {Workman}
  \emph {et~al.}}]{PDG_2022}%
  \BibitemOpen
  \bibfield  {author} {\bibinfo {author} {\bibfnamefont {R.}~\bibnamefont
  {Workman}} \emph {et~al.} (\bibinfo {collaboration} {Particle Data Group}),\
  }\href@noop {} {\bibinfo {title} {Review of particle physics}},\ \bibinfo
  {note} {to be published (2022)}\BibitemShut {NoStop}%
\bibitem [{\citenamefont {Erkelenz}\ \emph {et~al.}(1971)\citenamefont
  {Erkelenz}, \citenamefont {Alzetta},\ and\ \citenamefont
  {Holinde}}]{Erkelenz:1971caz}%
  \BibitemOpen
  \bibfield  {author} {\bibinfo {author} {\bibfnamefont {K.}~\bibnamefont
  {Erkelenz}}, \bibinfo {author} {\bibfnamefont {R.}~\bibnamefont {Alzetta}},\
  and\ \bibinfo {author} {\bibfnamefont {K.}~\bibnamefont {Holinde}},\
  }\bibfield  {title} {\bibinfo {title} {{Momentum space calculations and
  helicity formalism in nuclear physics}},\ }\href
  {https://doi.org/10.1016/0375-9474(71)90279-X} {\bibfield  {journal}
  {\bibinfo  {journal} {Nucl. Phys. A}\ }\textbf {\bibinfo {volume} {176}},\
  \bibinfo {pages} {413} (\bibinfo {year} {1971})}\BibitemShut {NoStop}%
\bibitem [{\citenamefont {Brown}\ \emph {et~al.}(1969)\citenamefont {Brown},
  \citenamefont {Jackson},\ and\ \citenamefont {Kuo}}]{Brown:1969tfp}%
  \BibitemOpen
  \bibfield  {author} {\bibinfo {author} {\bibfnamefont {G.~E.}\ \bibnamefont
  {Brown}}, \bibinfo {author} {\bibfnamefont {A.~D.}\ \bibnamefont {Jackson}},\
  and\ \bibinfo {author} {\bibfnamefont {T.~T.~S.}\ \bibnamefont {Kuo}},\
  }\bibfield  {title} {\bibinfo {title} {{Nucleon-nucleon potential and minimal
  relativity}},\ }\href {https://doi.org/10.1016/0375-9474(69)90549-1}
  {\bibfield  {journal} {\bibinfo  {journal} {Nucl. Phys. A}\ }\textbf
  {\bibinfo {volume} {133}},\ \bibinfo {pages} {481} (\bibinfo {year}
  {1969})}\BibitemShut {NoStop}%
\bibitem [{\citenamefont {Haftel}\ and\ \citenamefont
  {Tabakin}(1970)}]{Haftel:1970zz}%
  \BibitemOpen
  \bibfield  {author} {\bibinfo {author} {\bibfnamefont {M.~I.}\ \bibnamefont
  {Haftel}}\ and\ \bibinfo {author} {\bibfnamefont {F.}~\bibnamefont
  {Tabakin}},\ }\bibfield  {title} {\bibinfo {title} {{Nuclear saturation and
  the smoothness of nucleon-nucleon potentials}},\ }\href
  {https://doi.org/10.1016/0375-9474(70)90047-3} {\bibfield  {journal}
  {\bibinfo  {journal} {Nucl. Phys. A}\ }\textbf {\bibinfo {volume} {158}},\
  \bibinfo {pages} {1} (\bibinfo {year} {1970})}\BibitemShut {NoStop}%
\bibitem [{\citenamefont {Bystricky}\ \emph {et~al.}(1978)\citenamefont
  {Bystricky}, \citenamefont {Lehar},\ and\ \citenamefont
  {Winternitz}}]{Bystricky:1976jr}%
  \BibitemOpen
  \bibfield  {author} {\bibinfo {author} {\bibfnamefont {J.}~\bibnamefont
  {Bystricky}}, \bibinfo {author} {\bibfnamefont {F.}~\bibnamefont {Lehar}},\
  and\ \bibinfo {author} {\bibfnamefont {P.}~\bibnamefont {Winternitz}},\
  }\bibfield  {title} {\bibinfo {title} {{Formalism of Nucleon-Nucleon Elastic
  Scattering Experiments}},\ }\href
  {https://doi.org/10.1051/jphys:019780039010100} {\bibfield  {journal}
  {\bibinfo  {journal} {J. Phys. (France)}\ }\textbf {\bibinfo {volume} {39}},\
  \bibinfo {pages} {1} (\bibinfo {year} {1978})}\BibitemShut {NoStop}%
\bibitem [{\citenamefont {Case}(1950)}]{Case:1950an}%
  \BibitemOpen
  \bibfield  {author} {\bibinfo {author} {\bibfnamefont {K.~M.}\ \bibnamefont
  {Case}},\ }\bibfield  {title} {\bibinfo {title} {{Singular potentials}},\
  }\href {https://doi.org/10.1103/PhysRev.80.797} {\bibfield  {journal}
  {\bibinfo  {journal} {Phys. Rev.}\ }\textbf {\bibinfo {volume} {80}},\
  \bibinfo {pages} {797} (\bibinfo {year} {1950})}\BibitemShut {NoStop}%
\bibitem [{\citenamefont {Frank}\ \emph {et~al.}(1971)\citenamefont {Frank},
  \citenamefont {Land},\ and\ \citenamefont {Spector}}]{Frank:1971xx}%
  \BibitemOpen
  \bibfield  {author} {\bibinfo {author} {\bibfnamefont {W.}~\bibnamefont
  {Frank}}, \bibinfo {author} {\bibfnamefont {D.~J.}\ \bibnamefont {Land}},\
  and\ \bibinfo {author} {\bibfnamefont {R.~M.}\ \bibnamefont {Spector}},\
  }\bibfield  {title} {\bibinfo {title} {{Singular potentials}},\ }\href
  {https://doi.org/10.1103/RevModPhys.43.36} {\bibfield  {journal} {\bibinfo
  {journal} {Rev. Mod. Phys.}\ }\textbf {\bibinfo {volume} {43}},\ \bibinfo
  {pages} {36} (\bibinfo {year} {1971})}\BibitemShut {NoStop}%
\bibitem [{\citenamefont {Beane}\ \emph {et~al.}(2001)\citenamefont {Beane},
  \citenamefont {Bedaque}, \citenamefont {Childress}, \citenamefont
  {Kryjevski}, \citenamefont {McGuire},\ and\ \citenamefont {van
  Kolck}}]{PhysRevA.64.042103}%
  \BibitemOpen
  \bibfield  {author} {\bibinfo {author} {\bibfnamefont {S.~R.}\ \bibnamefont
  {Beane}}, \bibinfo {author} {\bibfnamefont {P.~F.}\ \bibnamefont {Bedaque}},
  \bibinfo {author} {\bibfnamefont {L.}~\bibnamefont {Childress}}, \bibinfo
  {author} {\bibfnamefont {A.}~\bibnamefont {Kryjevski}}, \bibinfo {author}
  {\bibfnamefont {J.}~\bibnamefont {McGuire}},\ and\ \bibinfo {author}
  {\bibfnamefont {U.}~\bibnamefont {van Kolck}},\ }\bibfield  {title} {\bibinfo
  {title} {Singular potentials and limit cycles},\ }\href
  {https://doi.org/10.1103/PhysRevA.64.042103} {\bibfield  {journal} {\bibinfo
  {journal} {Phys. Rev. A}\ }\textbf {\bibinfo {volume} {64}},\ \bibinfo
  {pages} {042103} (\bibinfo {year} {2001})}\BibitemShut {NoStop}%
\bibitem [{\citenamefont {Hammer}\ and\ \citenamefont
  {Swingle}(2006)}]{HAMMER2006306}%
  \BibitemOpen
  \bibfield  {author} {\bibinfo {author} {\bibfnamefont {H.-W.}\ \bibnamefont
  {Hammer}}\ and\ \bibinfo {author} {\bibfnamefont {B.~G.}\ \bibnamefont
  {Swingle}},\ }\bibfield  {title} {\bibinfo {title} {On the limit cycle for
  the 1/r2 potential in momentum space},\ }\href
  {https://doi.org/https://doi.org/10.1016/j.aop.2005.04.017} {\bibfield
  {journal} {\bibinfo  {journal} {Ann. Phys.}\ }\textbf {\bibinfo {volume}
  {321}},\ \bibinfo {pages} {306} (\bibinfo {year} {2006})}\BibitemShut
  {NoStop}%
\bibitem [{\citenamefont {Wu}\ and\ \citenamefont
  {Long}(2019)}]{PhysRevC.99.024003}%
  \BibitemOpen
  \bibfield  {author} {\bibinfo {author} {\bibfnamefont {S.}~\bibnamefont
  {Wu}}\ and\ \bibinfo {author} {\bibfnamefont {B.}~\bibnamefont {Long}},\
  }\bibfield  {title} {\bibinfo {title} {Perturbative $nn$ scattering in chiral
  effective field theory},\ }\href {https://doi.org/10.1103/PhysRevC.99.024003}
  {\bibfield  {journal} {\bibinfo  {journal} {Phys. Rev. C}\ }\textbf {\bibinfo
  {volume} {99}},\ \bibinfo {pages} {024003} (\bibinfo {year}
  {2019})}\BibitemShut {NoStop}%
\bibitem [{\citenamefont {Stoks}\ \emph {et~al.}(1993)\citenamefont {Stoks},
  \citenamefont {Klomp}, \citenamefont {Rentmeester},\ and\ \citenamefont
  {de~Swart}}]{Stoks:1993tb}%
  \BibitemOpen
  \bibfield  {author} {\bibinfo {author} {\bibfnamefont {V.~G.~J.}\
  \bibnamefont {Stoks}}, \bibinfo {author} {\bibfnamefont {R.~A.~M.}\
  \bibnamefont {Klomp}}, \bibinfo {author} {\bibfnamefont {M.~C.~M.}\
  \bibnamefont {Rentmeester}},\ and\ \bibinfo {author} {\bibfnamefont {J.~J.}\
  \bibnamefont {de~Swart}},\ }\bibfield  {title} {\bibinfo {title} {{Partial
  wave analaysis of all nucleon-nucleon scattering data below 350-MeV}},\
  }\href {https://doi.org/10.1103/PhysRevC.48.792} {\bibfield  {journal}
  {\bibinfo  {journal} {Phys. Rev. C}\ }\textbf {\bibinfo {volume} {48}},\
  \bibinfo {pages} {792} (\bibinfo {year} {1993})}\BibitemShut {NoStop}%
\bibitem [{\citenamefont {Taylor}(1972)}]{Taylor72}%
  \BibitemOpen
  \bibfield  {author} {\bibinfo {author} {\bibfnamefont {J.~R.}\ \bibnamefont
  {Taylor}},\ }\href@noop {} {\emph {\bibinfo {title} {{S}cattering {T}heory:
  {T}he quantum {T}heory on {N}onrelativistic {C}ollisions}}}\ (\bibinfo
  {publisher} {Wiley, New York},\ \bibinfo {year} {1972})\BibitemShut {NoStop}%
\bibitem [{\citenamefont {Wesolowski}\ \emph {et~al.}(2019)\citenamefont
  {Wesolowski}, \citenamefont {Furnstahl}, \citenamefont {Melendez},\ and\
  \citenamefont {Phillips}}]{Wesolowski_2019}%
  \BibitemOpen
  \bibfield  {author} {\bibinfo {author} {\bibfnamefont {S.}~\bibnamefont
  {Wesolowski}}, \bibinfo {author} {\bibfnamefont {R.~J.}\ \bibnamefont
  {Furnstahl}}, \bibinfo {author} {\bibfnamefont {J.~A.}\ \bibnamefont
  {Melendez}},\ and\ \bibinfo {author} {\bibfnamefont {D.~R.}\ \bibnamefont
  {Phillips}},\ }\bibfield  {title} {\bibinfo {title} {Exploring bayesian
  parameter estimation for chiral effective field theory using nucleon-nucleon
  phase shifts},\ }\href {https://doi.org/10.1088/1361-6471/aaf5fc} {\bibfield
  {journal} {\bibinfo  {journal} {J. Phys. G}\ }\textbf {\bibinfo {volume}
  {46}},\ \bibinfo {pages} {045102} (\bibinfo {year} {2019})}\BibitemShut
  {NoStop}%
\bibitem [{\citenamefont {Navarro~P\'erez}\ \emph {et~al.}(2013)\citenamefont
  {Navarro~P\'erez}, \citenamefont {Amaro},\ and\ \citenamefont
  {Ruiz~Arriola}}]{Granada_1}%
  \BibitemOpen
  \bibfield  {author} {\bibinfo {author} {\bibfnamefont {R.}~\bibnamefont
  {Navarro~P\'erez}}, \bibinfo {author} {\bibfnamefont {J.~E.}\ \bibnamefont
  {Amaro}},\ and\ \bibinfo {author} {\bibfnamefont {E.}~\bibnamefont
  {Ruiz~Arriola}},\ }\bibfield  {title} {\bibinfo {title} {Partial-wave
  analysis of nucleon-nucleon scattering below the pion-production threshold},\
  }\href@noop {} {\bibfield  {journal} {\bibinfo  {journal} {Phys. Rev. C}\
  }\textbf {\bibinfo {volume} {88}},\ \bibinfo {pages} {024002} (\bibinfo
  {year} {2013})}\BibitemShut {NoStop}%
\bibitem [{\citenamefont {P\'erez}\ \emph {et~al.}(2013)\citenamefont
  {P\'erez}, \citenamefont {Amaro},\ and\ \citenamefont {Arriola}}]{Granada_2}%
  \BibitemOpen
  \bibfield  {author} {\bibinfo {author} {\bibfnamefont {R.~N.}\ \bibnamefont
  {P\'erez}}, \bibinfo {author} {\bibfnamefont {J.~E.}\ \bibnamefont {Amaro}},\
  and\ \bibinfo {author} {\bibfnamefont {E.~R.}\ \bibnamefont {Arriola}},\
  }\bibfield  {title} {\bibinfo {title} {Coarse-grained potential analysis of
  neutron-proton and proton-proton scattering below the pion production
  threshold},\ }\href@noop {} {\bibfield  {journal} {\bibinfo  {journal} {Phys.
  Rev. C}\ }\textbf {\bibinfo {volume} {88}},\ \bibinfo {pages} {064002}
  (\bibinfo {year} {2013})}\BibitemShut {NoStop}%
\bibitem [{\citenamefont {Vernon}\ \emph {et~al.}(2010)\citenamefont {Vernon},
  \citenamefont {Goldstein},\ and\ \citenamefont {Bower}}]{Vernon:2010}%
  \BibitemOpen
  \bibfield  {author} {\bibinfo {author} {\bibfnamefont {I.}~\bibnamefont
  {Vernon}}, \bibinfo {author} {\bibfnamefont {M.}~\bibnamefont {Goldstein}},\
  and\ \bibinfo {author} {\bibfnamefont {R.}~\bibnamefont {Bower}},\ }\bibfield
   {title} {\bibinfo {title} {Galaxy formation: a bayesian uncertainty
  analysis},\ }\href {https://doi.org/https://doi.org/10.1214/10-BA524}
  {\bibfield  {journal} {\bibinfo  {journal} {Bayesian Anal.}\ }\textbf
  {\bibinfo {volume} {5(4)}},\ \bibinfo {pages} {619} (\bibinfo {year}
  {2010})}\BibitemShut {NoStop}%
\bibitem [{\citenamefont {Vernon}\ \emph {et~al.}(2014)\citenamefont {Vernon},
  \citenamefont {Goldstein},\ and\ \citenamefont {Bower}}]{Vernon:2014}%
  \BibitemOpen
  \bibfield  {author} {\bibinfo {author} {\bibfnamefont {I.}~\bibnamefont
  {Vernon}}, \bibinfo {author} {\bibfnamefont {M.}~\bibnamefont {Goldstein}},\
  and\ \bibinfo {author} {\bibfnamefont {R.}~\bibnamefont {Bower}},\ }\bibfield
   {title} {\bibinfo {title} {Galaxy formation: Bayesian history matching for
  the observable universe},\ }\href
  {https://doi.org/https://doi.org/10.1214/12-STS412} {\bibfield  {journal}
  {\bibinfo  {journal} {Statist. Sci.}\ }\textbf {\bibinfo {volume} {29}},\
  \bibinfo {pages} {81} (\bibinfo {year} {2014})}\BibitemShut {NoStop}%
\bibitem [{\citenamefont {Vernon}\ \emph {et~al.}(2018)\citenamefont {Vernon},
  \citenamefont {Liu}, \citenamefont {Goldstein}, \citenamefont {Rowe},
  \citenamefont {Topping},\ and\ \citenamefont {Lindsey}}]{Vernon:2018}%
  \BibitemOpen
  \bibfield  {author} {\bibinfo {author} {\bibfnamefont {I.}~\bibnamefont
  {Vernon}}, \bibinfo {author} {\bibfnamefont {J.}~\bibnamefont {Liu}},
  \bibinfo {author} {\bibfnamefont {M.}~\bibnamefont {Goldstein}}, \bibinfo
  {author} {\bibfnamefont {J.}~\bibnamefont {Rowe}}, \bibinfo {author}
  {\bibfnamefont {J.}~\bibnamefont {Topping}},\ and\ \bibinfo {author}
  {\bibfnamefont {K.}~\bibnamefont {Lindsey}},\ }\bibfield  {title} {\bibinfo
  {title} {Bayesian uncertainty analysis for complex systems biology models:
  emulation, global parameter searches and evaluation of gene functions},\
  }\href {https://doi.org/https://doi.org/10.1186/s12918-017-0484-3} {\bibfield
   {journal} {\bibinfo  {journal} {BMC Syst. Biol.}\ }\textbf {\bibinfo
  {volume} {12}},\ \bibinfo {pages} {1} (\bibinfo {year} {2018})}\BibitemShut
  {NoStop}%
\bibitem [{\citenamefont {Foreman-Mackey}\ \emph {et~al.}(2013)\citenamefont
  {Foreman-Mackey}, \citenamefont {Hogg}, \citenamefont {Lang},\ and\
  \citenamefont {Goodman}}]{emcee}%
  \BibitemOpen
  \bibfield  {author} {\bibinfo {author} {\bibfnamefont {D.}~\bibnamefont
  {Foreman-Mackey}}, \bibinfo {author} {\bibfnamefont {D.~W.}\ \bibnamefont
  {Hogg}}, \bibinfo {author} {\bibfnamefont {D.}~\bibnamefont {Lang}},\ and\
  \bibinfo {author} {\bibfnamefont {J.}~\bibnamefont {Goodman}},\ }\bibfield
  {title} {\bibinfo {title} {emcee: The {MCMC} hammer},\ }\href@noop {}
  {\bibfield  {journal} {\bibinfo  {journal} {Publ. Astron. Soc. Pac.}\
  }\textbf {\bibinfo {volume} {125}},\ \bibinfo {pages} {306} (\bibinfo {year}
  {2013})}\BibitemShut {NoStop}%
\bibitem [{\citenamefont {Kondo}\ \emph {et~al.}(2023)\citenamefont {Kondo}
  \emph {et~al.}}]{Kondo:2023lty}%
  \BibitemOpen
  \bibfield  {author} {\bibinfo {author} {\bibfnamefont {Y.}~\bibnamefont
  {Kondo}} \emph {et~al.},\ }\bibfield  {title} {\bibinfo {title} {{First
  observation of $^{28}$O}},\ }\href
  {https://doi.org/10.1038/s41586-023-06352-6} {\bibfield  {journal} {\bibinfo
  {journal} {Nature}\ }\textbf {\bibinfo {volume} {620}},\ \bibinfo {pages}
  {965} (\bibinfo {year} {2023})}\BibitemShut {NoStop}%
\bibitem [{\citenamefont {Joseph}(2016)}]{JosephRoshan}%
  \BibitemOpen
  \bibfield  {author} {\bibinfo {author} {\bibfnamefont {V.}~\bibnamefont
  {Joseph}},\ }\bibfield  {title} {\bibinfo {title} {Space-filling designs for
  computer experiments: A review},\ }\href
  {https://doi.org/https://doi.org/10.1080/08982112.2015.1100447} {\bibfield
  {journal} {\bibinfo  {journal} {Qual. Eng.}\ }\textbf {\bibinfo {volume}
  {28}},\ \bibinfo {pages} {28} (\bibinfo {year} {2016})}\BibitemShut {NoStop}%
\bibitem [{\citenamefont {Pukelsheim}(1994)}]{pukelsheim}%
  \BibitemOpen
  \bibfield  {author} {\bibinfo {author} {\bibfnamefont {F.}~\bibnamefont
  {Pukelsheim}},\ }\bibfield  {title} {\bibinfo {title} {The three sigma
  rule},\ }\href
  {https://doi.org/https://doi.org/10.1080/00031305.1994.10476030} {\bibfield
  {journal} {\bibinfo  {journal} {Am. Stat.}\ }\textbf {\bibinfo {volume}
  {48}},\ \bibinfo {pages} {88–} (\bibinfo {year} {1994})}\BibitemShut
  {NoStop}%
\bibitem [{\citenamefont {Gregory}(2005)}]{gregory_2005}%
  \BibitemOpen
  \bibfield  {author} {\bibinfo {author} {\bibfnamefont {P.}~\bibnamefont
  {Gregory}},\ }\href {https://doi.org/10.1017/CBO9780511791277} {\emph
  {\bibinfo {title} {Bayesian Logical Data Analysis for the Physical Sciences:
  A Comparative Approach with Mathematica® Support}}}\ (\bibinfo  {publisher}
  {Cambridge University Press},\ \bibinfo {year} {2005})\BibitemShut {NoStop}%
\bibitem [{\citenamefont {Glöckle}(1983)}]{Glockle}%
  \BibitemOpen
  \bibfield  {author} {\bibinfo {author} {\bibfnamefont {W.}~\bibnamefont
  {Glöckle}},\ }\href@noop {} {\emph {\bibinfo {title} {{The Quantum
  Mechanical Few-body Problem}}}}\ (\bibinfo  {publisher} {Springer-Verlag},\
  \bibinfo {address} {Berlin Heidelberg},\ \bibinfo {year} {1983})\BibitemShut
  {NoStop}%
\bibitem [{\citenamefont {Landau}(1996)}]{Landau:1990qp}%
  \BibitemOpen
  \bibfield  {author} {\bibinfo {author} {\bibfnamefont {R.~H.}\ \bibnamefont
  {Landau}},\ }\href@noop {} {\emph {\bibinfo {title} {Quantum mechanics II: a
  second course in quantum theory}}},\ \bibinfo {edition} {2nd}\ ed.\ (\bibinfo
   {publisher} {Wiley},\ \bibinfo {address} {New York},\ \bibinfo {year}
  {1996})\BibitemShut {NoStop}%
\bibitem [{\citenamefont {Hoppe}\ \emph {et~al.}(2017)\citenamefont {Hoppe},
  \citenamefont {Drischler}, \citenamefont {Furnstahl}, \citenamefont
  {Hebeler},\ and\ \citenamefont {Schwenk}}]{Hoppe:2017lok}%
  \BibitemOpen
  \bibfield  {author} {\bibinfo {author} {\bibfnamefont {J.}~\bibnamefont
  {Hoppe}}, \bibinfo {author} {\bibfnamefont {C.}~\bibnamefont {Drischler}},
  \bibinfo {author} {\bibfnamefont {R.~J.}\ \bibnamefont {Furnstahl}}, \bibinfo
  {author} {\bibfnamefont {K.}~\bibnamefont {Hebeler}},\ and\ \bibinfo {author}
  {\bibfnamefont {A.}~\bibnamefont {Schwenk}},\ }\bibfield  {title} {\bibinfo
  {title} {{Weinberg eigenvalues for chiral nucleon-nucleon interactions}},\
  }\href {https://doi.org/10.1103/PhysRevC.96.054002} {\bibfield  {journal}
  {\bibinfo  {journal} {Phys. Rev. C}\ }\textbf {\bibinfo {volume} {96}},\
  \bibinfo {pages} {054002} (\bibinfo {year} {2017})},\ \Eprint
  {https://arxiv.org/abs/1707.06438} {arXiv:1707.06438 [nucl-th]} \BibitemShut
  {NoStop}%
\bibitem [{\citenamefont {Machleidt}(2001)}]{Machleidt:2000ge}%
  \BibitemOpen
  \bibfield  {author} {\bibinfo {author} {\bibfnamefont {R.}~\bibnamefont
  {Machleidt}},\ }\bibfield  {title} {\bibinfo {title} {{The High precision,
  charge dependent Bonn nucleon-nucleon potential (CD-Bonn)}},\ }\href
  {https://doi.org/10.1103/PhysRevC.63.024001} {\bibfield  {journal} {\bibinfo
  {journal} {Phys. Rev. C}\ }\textbf {\bibinfo {volume} {63}},\ \bibinfo
  {pages} {024001} (\bibinfo {year} {2001})},\ \Eprint
  {https://arxiv.org/abs/nucl-th/0006014} {arXiv:nucl-th/0006014} \BibitemShut
  {NoStop}%
\bibitem [{\citenamefont {Blatt}\ and\ \citenamefont
  {Biedenharn}(1952)}]{Blatt:1952zz}%
  \BibitemOpen
  \bibfield  {author} {\bibinfo {author} {\bibfnamefont {J.~M.}\ \bibnamefont
  {Blatt}}\ and\ \bibinfo {author} {\bibfnamefont {L.~C.}\ \bibnamefont
  {Biedenharn}},\ }\bibfield  {title} {\bibinfo {title} {{The Angular
  Distribution of Scattering and Reaction Cross Sections}},\ }\href
  {https://doi.org/10.1103/RevModPhys.24.258} {\bibfield  {journal} {\bibinfo
  {journal} {Rev. Mod. Phys.}\ }\textbf {\bibinfo {volume} {24}},\ \bibinfo
  {pages} {258} (\bibinfo {year} {1952})}\BibitemShut {NoStop}%
\bibitem [{\citenamefont {Newton}(1982)}]{Newton:1982qc}%
  \BibitemOpen
  \bibfield  {author} {\bibinfo {author} {\bibfnamefont {R.~G.}\ \bibnamefont
  {Newton}},\ }\href@noop {} {\emph {\bibinfo {title} {{Scattering theory of
  waves and particles}}}}\ (\bibinfo  {publisher} {Springer-Verlag New York,
  Inc.},\ \bibinfo {address} {175 Fifth Avenue, New York, New York, 10010,
  U.S.A.},\ \bibinfo {year} {1982})\BibitemShut {NoStop}%
\end{thebibliography}%

\appendix
\section{Neutron-proton scattering\label{app:np_scattering}}
\begin{figure}[b]
    \centering
    \usetikzlibrary{calc,patterns,angles,quotes,shapes,snakes}

\begin{tikzpicture}
    \coordinate (Origo)   at (0,0);
    \coordinate (a) at (-4,0);
    \coordinate (b) at (-1,0);
    \coordinate (c) at (3,0);
    \coordinate (d) at (4,0);
    \coordinate (e) at (0.7,0.7);
    \coordinate (f) at (3,3);
    \coordinate (g) at (4*0.7,4*0.7);
    
    \draw [ultra thick,-latex] (c)
        -- (d) node [above left] {$\bm{\hat{z}}$};
        
    \draw [thick,-latex] (a)
        -- (b) node [above left] {$\bm{p}, (s,m_s)$};
    
     \draw [thick,-latex] (e)
        -- (g) node [above left] {$\bm{p}', (s,m'_s)$};
    
    \draw [style=help lines,dashed] (a)
        -- (d) node [above left] {};
        
    \node[draw,star,star points=7,star point ratio=0.8,fill=gray!20,minimum size=0.8cm] at (Origo) {};
    
    \pic [draw, ->, "$\theta_{\mathrm{c.m.}}$", angle eccentricity=1.2,angle radius=2cm] {angle = c--Origo--f};
\end{tikzpicture}
    \caption{The geometry of $NN$ scattering in relative coordinates. An incoming nucleon with relative momentum $\bm{p}$ and spin $(s,m_s)$ scatters an angle $\thetacm$ to a relative momentum $\bm{p}'$ and spin $(s,m'_s)$. The azimuthal angle $\phi$ is conventionally set to zero.}
    \label{fig:scattering_kinematics}
\end{figure}
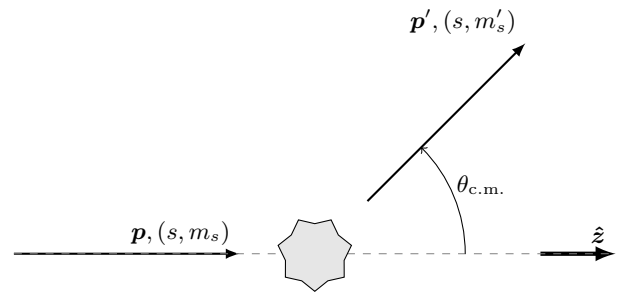

The scattering geometry in relative-momentum coordinates is illustrated in \cref{fig:scattering_kinematics}.
For a stationary proton and an incoming neutron with kinetic energy $T_\mathrm{lab}$ in the laboratory  frame of reference, the modulus of the relative momentum, $p$, is given by
\begin{equation}
    p^2 = \frac{m_p^2 T_\mathrm{lab} (2m_n + T_\mathrm{lab})}{(m_n+m_p)^2 + 2m_p T_\mathrm{lab}}.
\end{equation}

We partial-wave decompose the potential $V$ in~\cref{eq:LO_potential} using the helicity formalism introduced in Ref. \cite{Erkelenz:1971caz},
and employ partial wave states $\ket{lm_l,p}$ with the normalization
\begin{equation}
    \bra{\bm{p}'}\ket{lm_l,p} = \frac{\delta(p'-p)}{p^2} Y^{l}_{m_l}(\hat{p}'),
    \label{eq:p_klm}
\end{equation}
where $Y^l_{m_l}(\hat{p})$ is the spherical harmonics, including the Condon-Shortley phase, normalized as 
\begin{eqnarray}
    \int d\hat{p} \ Y^{*l'}_{\ m'_l}(\hat{p})Y^l_{m_l}(\hat{p}) = \delta_{l'l}\delta_{m'_lm_l}. 
\end{eqnarray}

We are interested in elastic scattering and express the $T$-matrix as a real-valued $R$-matrix defined as
\begin{equation}
       R = T + i\pi R \delta(E-H_0) T.
       \label{eq:R_T}
\end{equation}
where $H_0$ is the free Hamiltonian and $E= p^2/m_N$ \cite{Glockle}. Consequently,
we solve the Lippmann-Schwinger equation given by
\begin{align}
     &R^{sj}_{l'l}(p', p) = V^{sj}_{l'l}(p',p) \ + \nonumber \\ &+ \mathcal{P} \sum_{l''}\int_0^\infty dk \ k^2 \ V^{sj}_{l'l''}(p',k) \frac{m_N}{p^2-k^2} R^{sj}_{l''l}(k,p),
    \label{eq:LS_R_pw}          %
\end{align}
where $\mathcal{P}$ denotes the principal value. The momentum-space integral is discretized as described in Refs.~\cite{Haftel:1970zz,Landau:1990qp} using a Gauss-Legendre grid with $N_p=60$ points in the interval $[0,\Lambda + 300]$ MeV, i.e. on the  interval where the potential, and hence the part of the integral containing the potential is non-zero. The advantage of using a grid that does not extend to infinity is that the part of the integral that implements the principal value, and is outside the support of the potential, can be treated analytically, see, e.g., Ref. \cite{Hoppe:2017lok}. 

Following Refs.~\cite{Glockle,Machleidt:2000ge} we obtain scattering phase shifts in the Blatt and Bidenharn convention from on-shell $R$-matrix elements in uncoupled channels via
\begin{equation}
    \tan \delta^{sj}_{l}(p) = -\frac{\pi}{2} p \ m_N R^{sj}_{ll}(p,p),
\end{equation}
and for coupled channels via
\begin{align}
    &\tan \tilde{\delta}^{sj}_{\mp}(p) = -\frac{\pi}{4} p \ m_N\left[ R^{sj}_{--} + R^{sj}_{++} \pm \frac{R^{sj}_{--}-R^{sj}_{++}}{\cos(2\tilde{\epsilon})}\right], \nonumber \\
    &\tan\left[2\tilde{\epsilon}^{sj}(p)\right] = \frac{2R_{+-}}{R_{--} - R_{++}} ,
\end{align}
where  $R^{sj}_{\pm\pm} \equiv R^{sj}_{l'=j\pm 1, l=j\pm 1}(p,p)$. The more commonly employed Stapp- (bar-) phase shifts, which we use in this work, are given by the solution to the following equations~\cite{Glockle}
\begin{align}
    \tilde{\delta}^{sj}_{+} + \tilde{\delta}^{sj}_{-} &= \bar{\delta}^{sj}_{+} + \bar{\delta}^{sj}_{-} \nonumber \\
    \sin\left(\bar{\delta}^{sj}_{-} - \bar{\delta}^{sj}_{+}\right) &= \frac{\tan\left(2\bar{\epsilon}^{sj}\right)}{\tan\left(2\tilde{\epsilon}^{sj}\right)} \\
    \sin\left(\tilde{\delta}^{sj}_{-}- \tilde{\delta}^{sj}_{+}\right) &= \frac{\sin\left(2\bar{\epsilon}^{sj}\right)}{\sin\left(2\tilde{\epsilon}^{sj}\right)}. \nonumber 
\end{align}
Note that numerical instabilities can occur when the mixing angle ($\bar{\epsilon}^{sj}$) in the Stapp-convention changes sign.

The spin-scattering matrix, $M$, is given by~\cite{Blatt:1952zz,Glockle,Newton:1982qc}
\begin{align}
    &M^{s}_{m'_s m_s}(p,\thetacm,\phi) = \frac{\sqrt{4\pi}}{2ip} \sum_{j,l,l'} i^{l-l'} (2j+1)\sqrt{2l+1}  \nonumber\\ &Y^{l'}_{m_s -m'_s}(\thetacm,\phi)
     \begin{pmatrix} l' & s & j \nonumber \\ m_s-m'_s & m'_s & -m_s \end{pmatrix}  \begin{pmatrix} l & s & j \\ 0 & m_s& -m_s\end{pmatrix} \\
    & \left(S^{sj}_{l'l}(p,p)-\delta_{l'l}\right).
    \label{eq:M_matrix}
\end{align}
The angles $\thetacm\in[0,\pi]$ and $\phi\in[0,2\pi]$ are the polar and azimuthal scattering angles, respectively. We set $\phi=0$ by cylindrical symmetry. Numerical instabilities in \cref{eq:M_matrix} can occur at $\thetacm = 90 ^\circ$, but can be avoided by adding a small nugget.

Equipped with $M$, spin-scattering observables can be computed from traces in spin space. The spin averaged differential cross section is obtained as
\begin{eqnarray}
\frac{d\sigma}{d\Omega} = \frac{1}{4}\Tr{M^\dagger M}.
\end{eqnarray}
Other observables can be calculated in a similar fashion, or by decomposing the $M$-matrix into, e.g., Saclay-amplitudes as described in Ref. \cite{Bystricky:1976jr}.

\section{\label{app:tables_bound_states} Appearance of bound states}
\cref{tab:bound_channels} lists the threshold LEC values for which $np$ bound states appear at different cutoff values, $\Lambda$, for the potential discussed in \cref{eq:Full_LO_potential}.

\begin{table*}[b]
\caption{Threshold LEC values for which a bound state appears in the partial waves relevant for the contact potential at LO. For the $\tS$ channel and $\Lambda \geq 2000$ MeV it is the \emph{second} bound state that appears at the tabulated value, and the first appears for $\tilde{C}_{^3S_1}>0.4$. For the $^3P_0$ channel and $\Lambda = 4000$ MeV it is the \emph{second} bound state that appears at the tabulated value, and the first appears for $C_{^3P_0}>1$.}
\begin{ruledtabular}
\begin{tabular}{ccccc}
$\Lambda$ [MeV] &  $\tilde{C}_{^1S_0}$ [$10^4$ GeV$^{-2}$] &  $\tilde{C}_{^3S_1}$ [$10^4$ GeV$^{-2}$] & $C_{^3P_0}$ [$10^4$ GeV$^{-4}$] &  $C_{^3P_2}$  [$10^4$ GeV$^{-4}$]  \\
\colrule
 400 & $-0.1279$ & $-0.0825$ & $-1.6735$ & $-2.3178$ \\
 450 & $-0.1207$ & $-0.0636$ & $-1.1076$ & $-1.6375$ \\
 500 & $-0.1151$ & $-0.0449$ & $-0.7578$ & $-1.2016$ \\
 550 & $-0.1106$ & $-0.0254$ & $-0.5316$ & $-0.9095$ \\
 600 & $-0.1067$ & $-0.0037$ & $-0.3802$ & $-0.7063$ \\
 700 & $-0.1010$ & $0.0515 $ & $-0.2015$ & $-0.4528$ \\
2000 & $-0.0794$ & $-0.0983$ & $0.0268 $ & $-0.0263$ \\
3000 & $-0.0755$ & $0.0010 $ & $0.0449 $ & $-0.0096$ \\
4000 & $-0.0735$ & $0.3229 $ & $-0.0310$ & $-0.0048$ \\
\end{tabular}
\end{ruledtabular}
\label{tab:bound_channels}
\end{table*}

\section{\label{app:post}Marginal posterior distributions}
\begin{figure*}
\subfloat[$\Lambda=400$~MeV.]{%
      \includegraphics[width=\columnwidth]{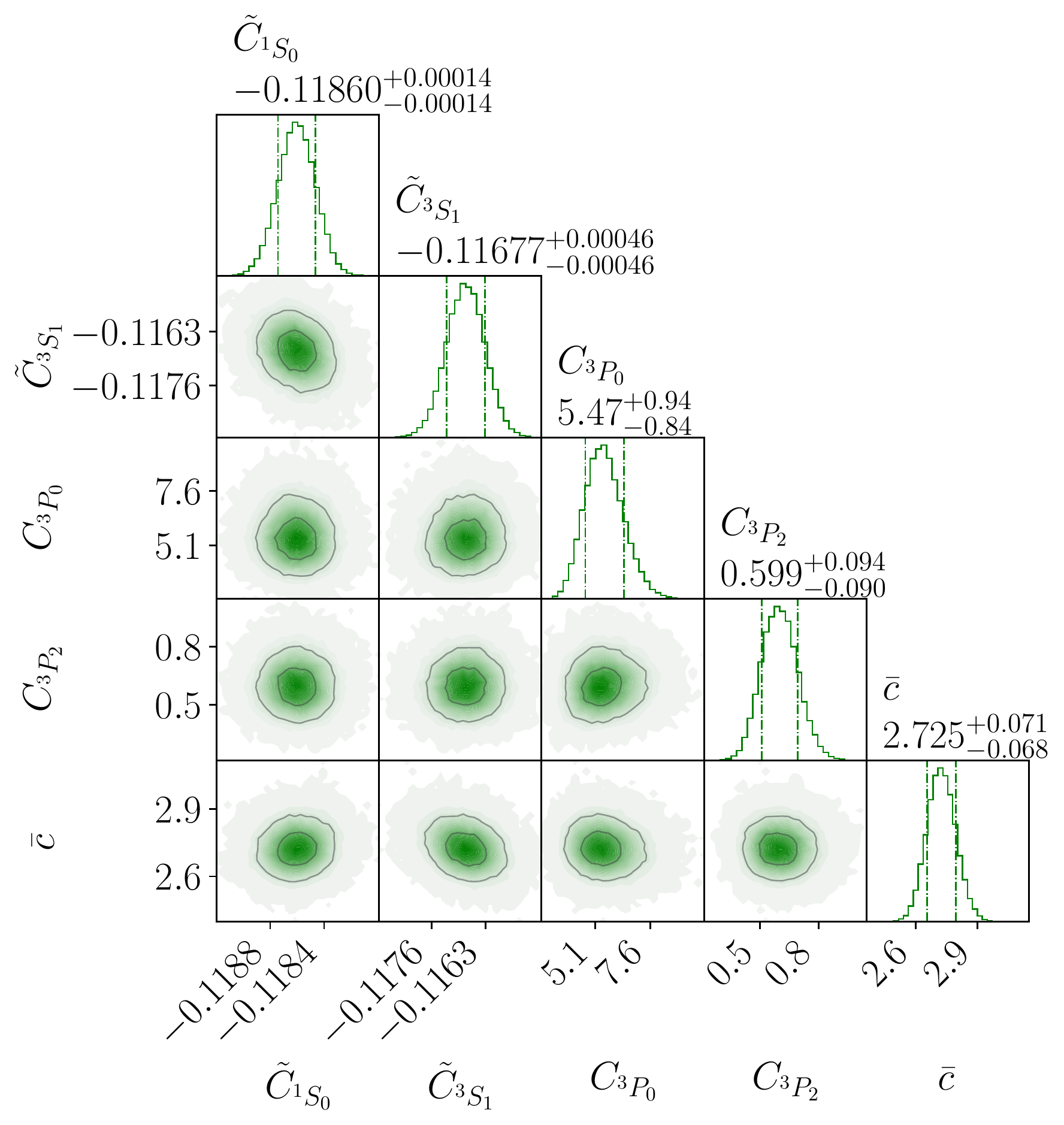}%
    }
\hfill
\subfloat[$\Lambda=500$~MeV.]{%
      \includegraphics[width=\columnwidth]{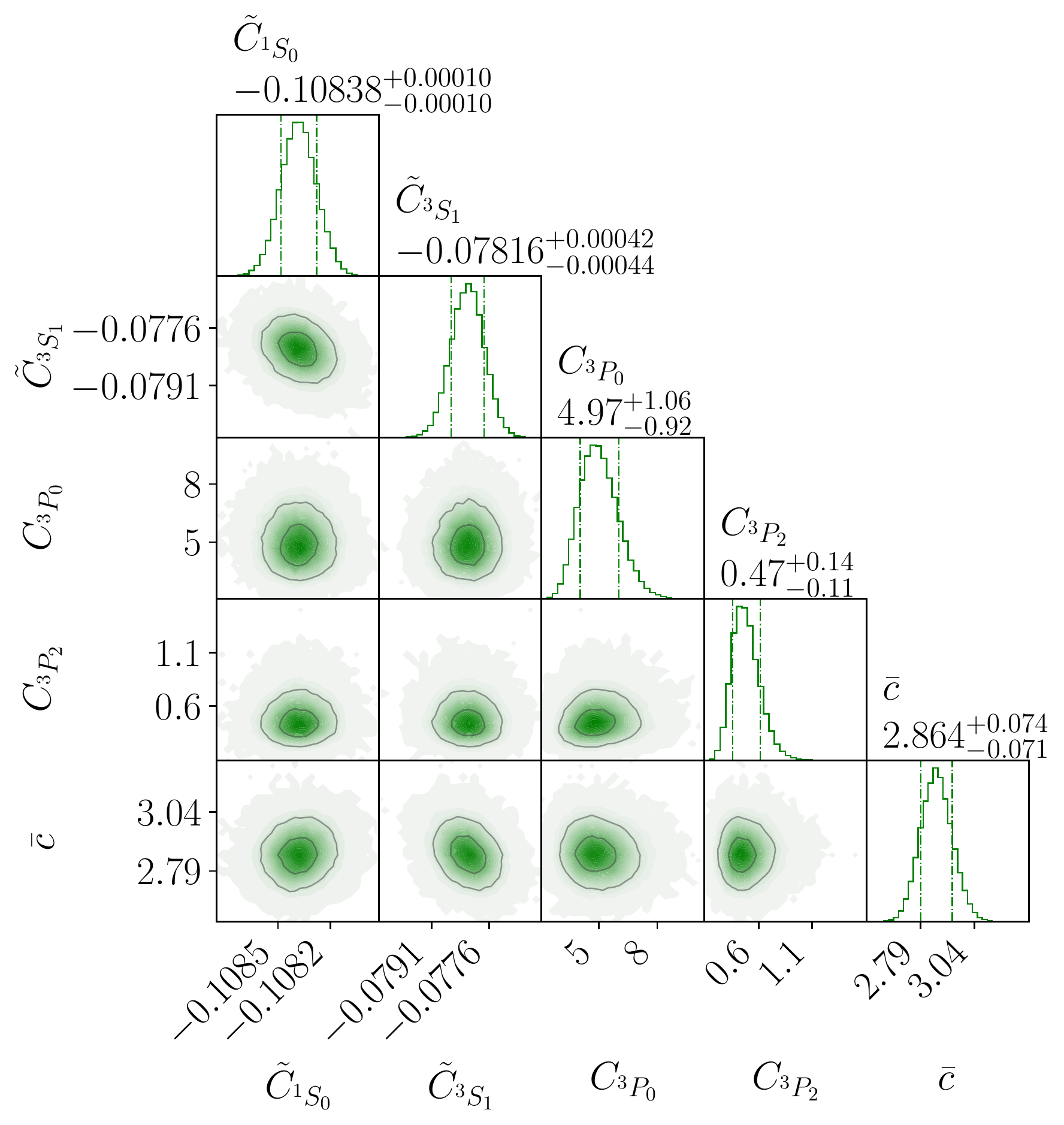}%
    }
\vfill 
\subfloat[$\Lambda=550$~MeV.]{%
      \includegraphics[width=\columnwidth]{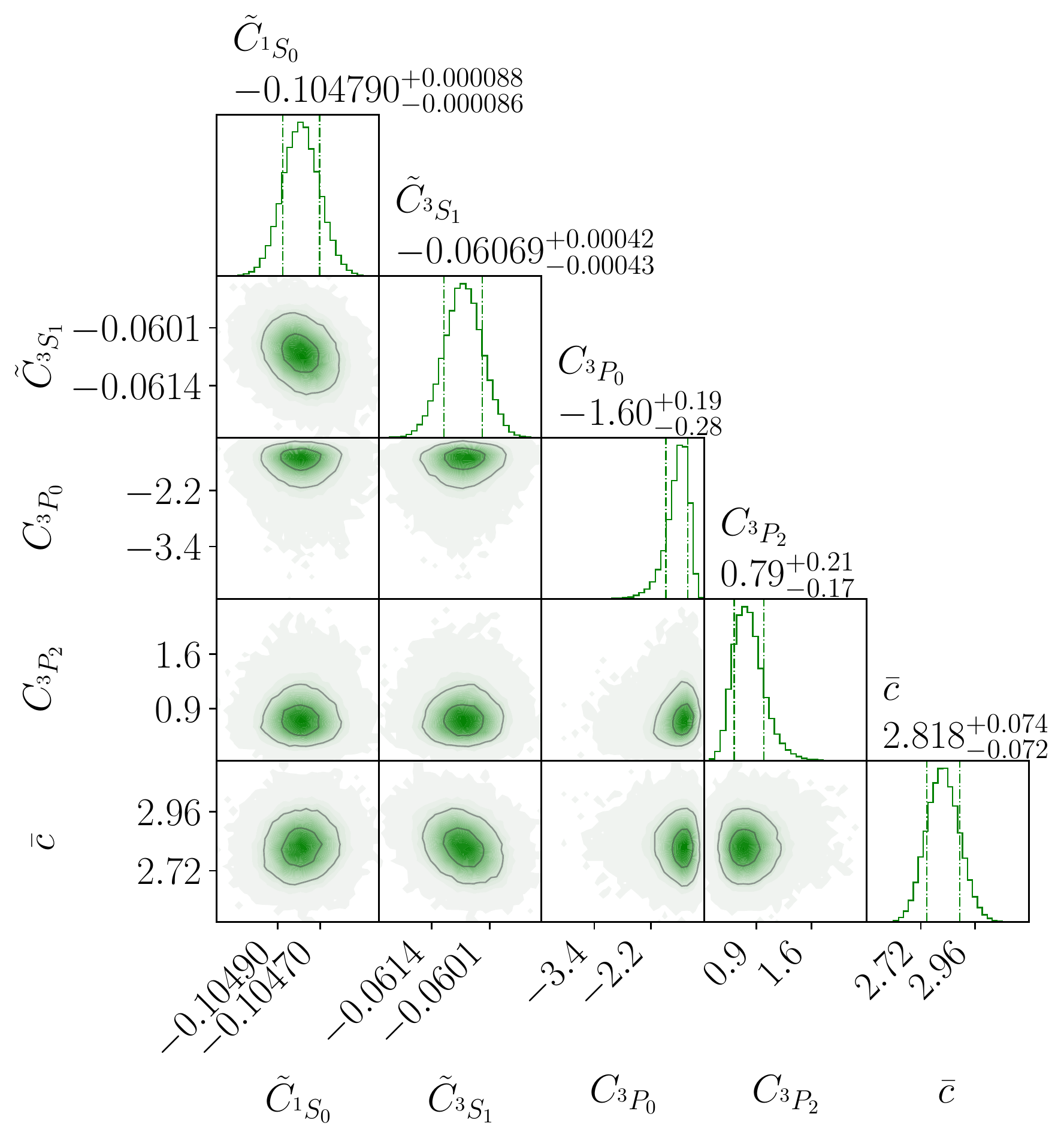}%
    }
\hfill
\subfloat[$\Lambda=600$~MeV.]{%
      \includegraphics[width=\columnwidth]{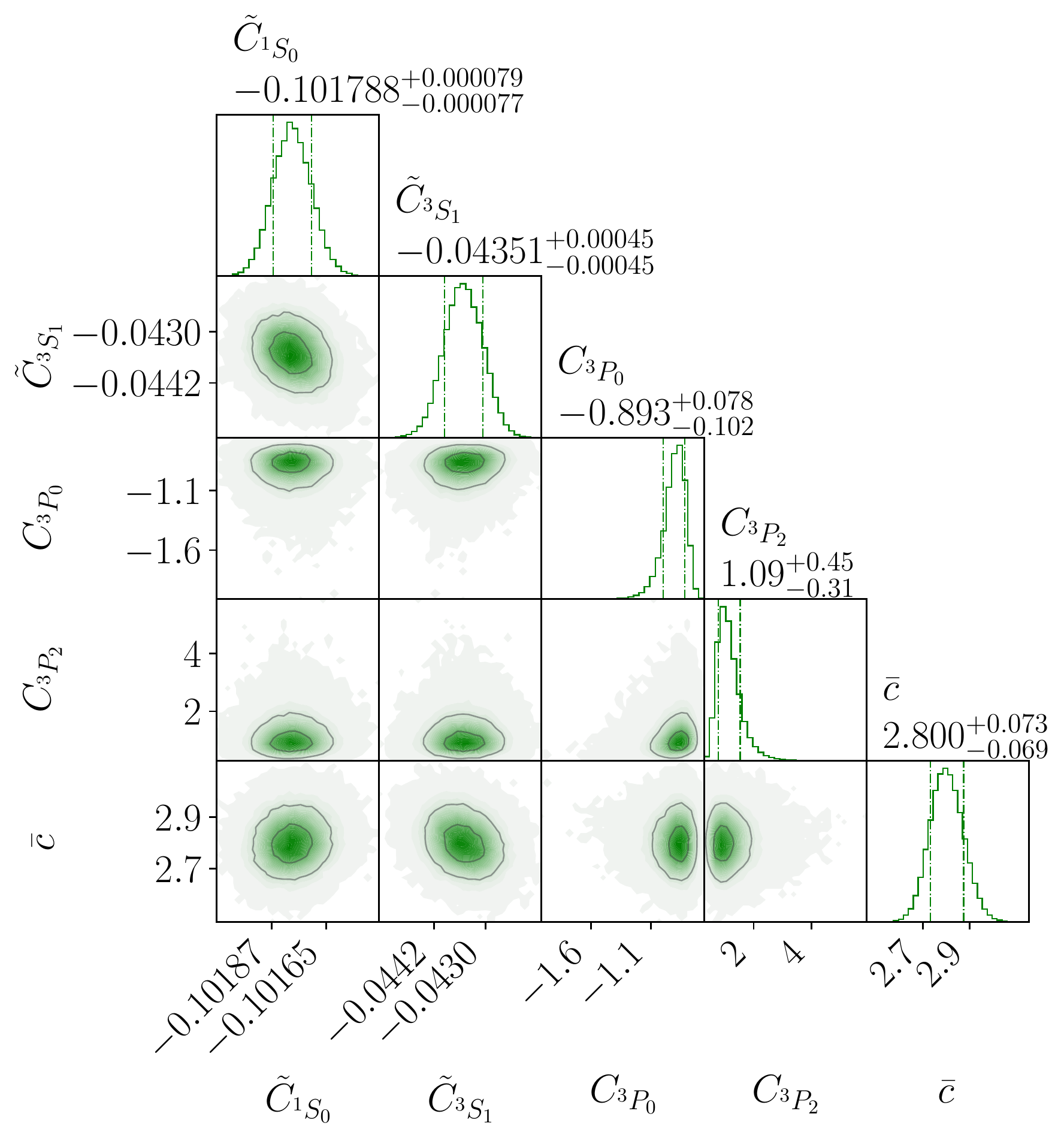}%
    }
\caption{\label{fig:pdf_set_1}Posterior pdfs for the parameters $\bm{\theta} = \left(\bm{\alpha},\bar{c}\right)$ at different cutoffs. The units of the LECs are $10^4$ GeV$^{-2}$ and $10^4$ GeV$^{-4}$ for the $S$- and $P$-waves respectively. The median and the 68\% equal-tailed credible interval are indicated for the univariate marginal pdfs..}
\end{figure*}

\begin{figure*}
\subfloat[$\Lambda=2000$~MeV.]{%
      \includegraphics[width=\columnwidth]{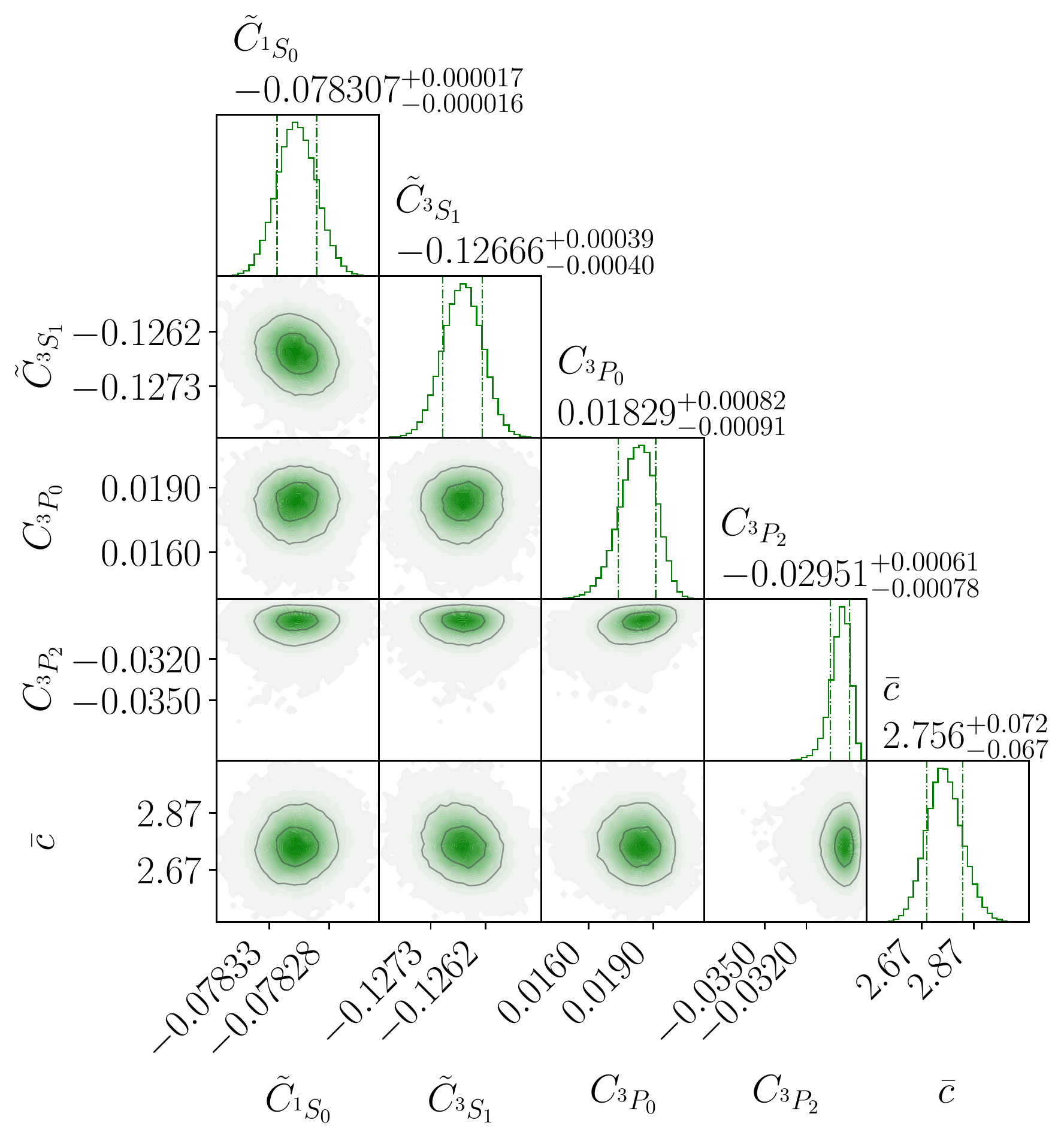}%
    }
\hfill
\subfloat[$\Lambda=3000$~MeV.]{%
      \includegraphics[width=\columnwidth]{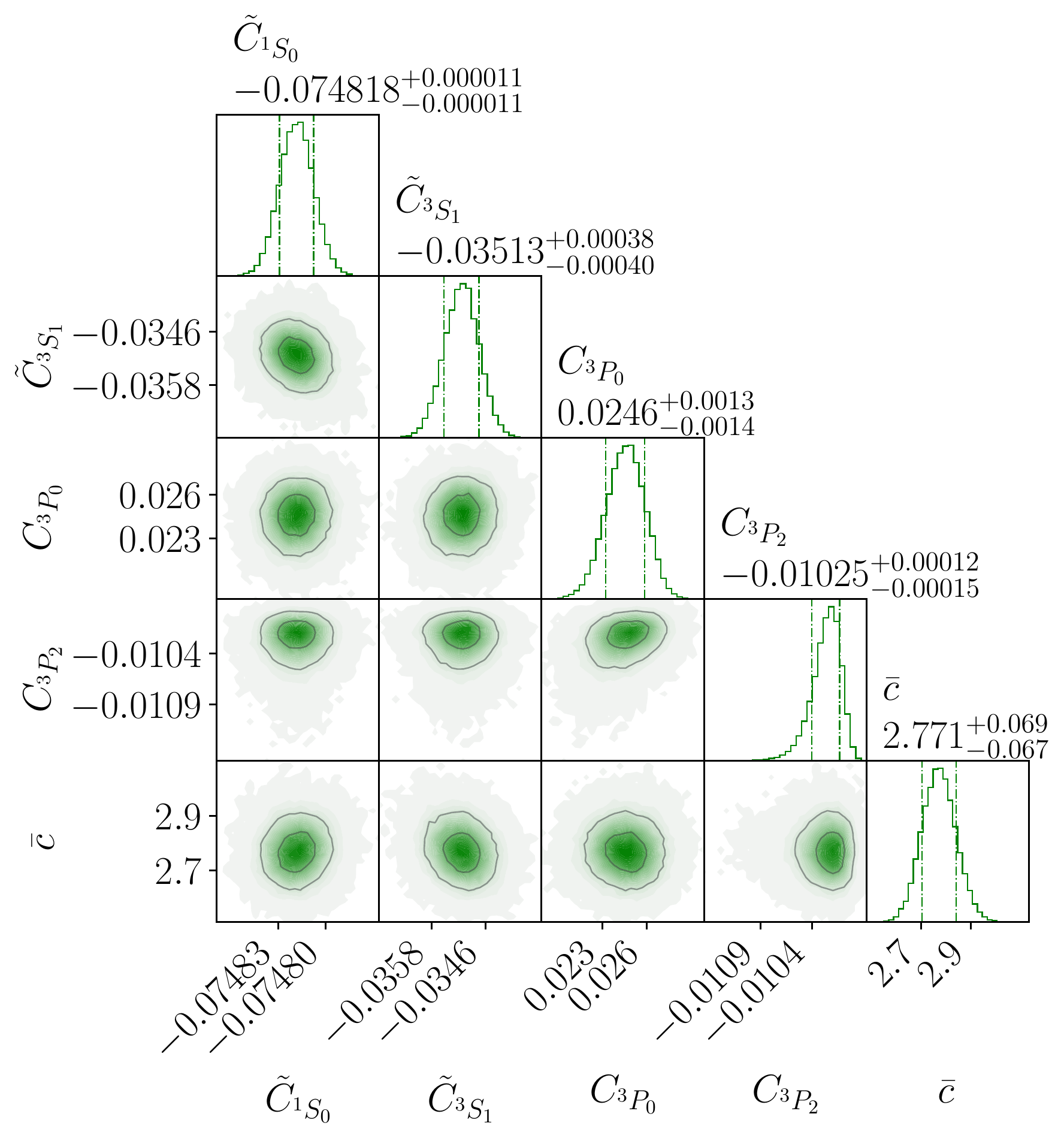}%
    }
\vfill
\subfloat[$\Lambda=4000$~MeV.]{%
      \includegraphics[width=\columnwidth]{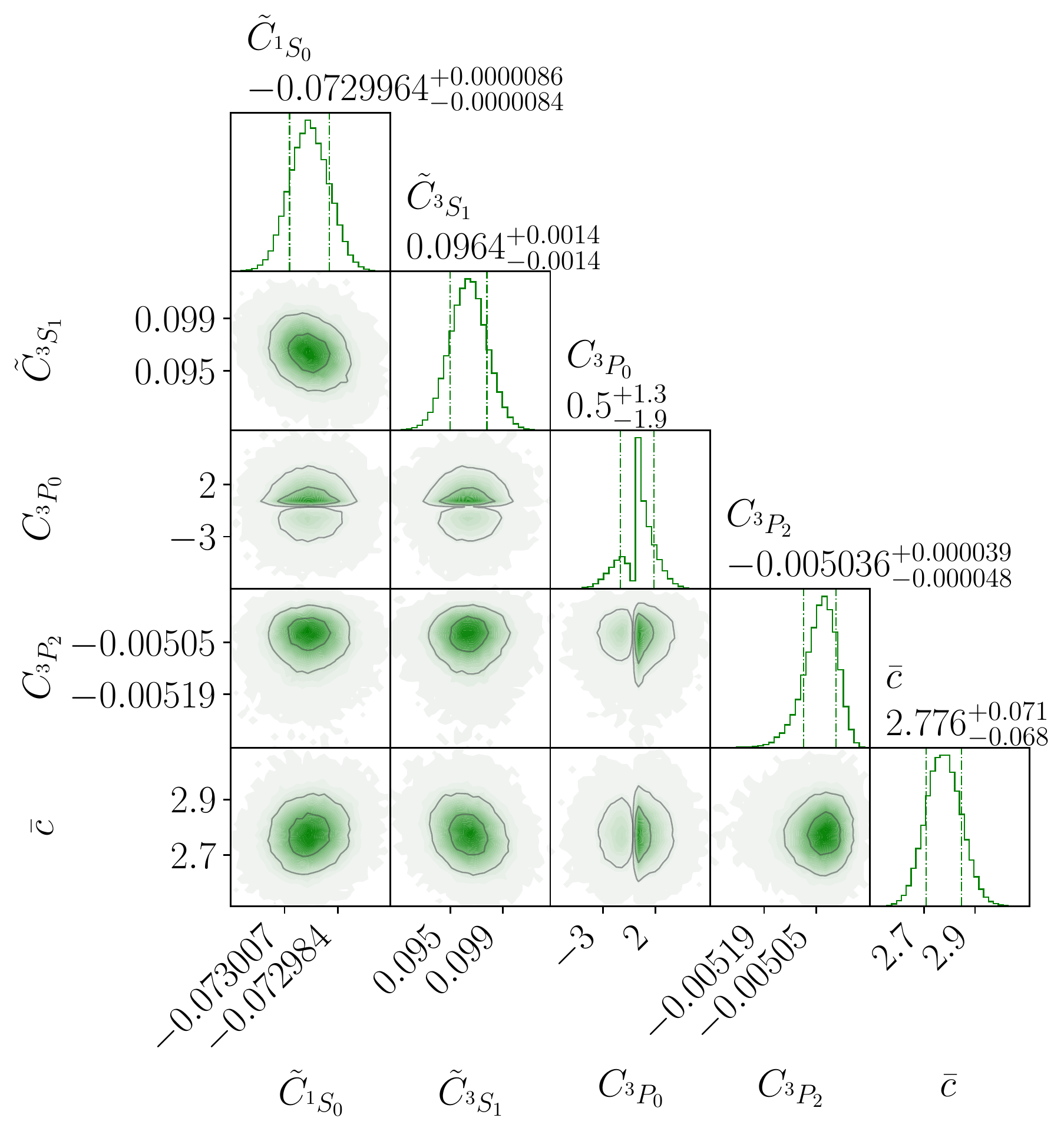}%
    }
\caption{\label{fig:pdf_set_2}Posterior pdfs for the parameters $\bm{\theta} = \left(\bm{\alpha},\bar{c}\right)$ at different cutoffs. The units of the LECs are $10^4$ GeV$^{-2}$ and $10^4$ GeV$^{-4}$ for the $S$- and $P$-waves respectively. The median and the 68\% equal-tailed credible interval are indicated for the univariate marginal pdfs.}
\end{figure*}

\cref{tab:modes_all} lists the MAP LEC values for the modes found during the initial MCMC. Figures \ref{fig:pdf_set_1} and \ref{fig:pdf_set_2} show the marginal posterior pdfs for the parameters $\bm{\theta}$ at cutoff values $\Lambda= 400,500,550,600,2000,3000,4000$~MeV.
\begin{table}[b]
\caption{Modes found in the initial MCMC sampling of the posterior pdfs. For each cutoff there are two starting regions ($s$), indicated in the second column, one where $^1S_0$ contains a bound state (b) and one where it does not (u).}
\begin{ruledtabular}
\begin{tabular}{cccc}
$\Lambda$ [MeV]& $s$ & $\ln(\mathcal{Z}_K)$ & MAP ($\bm{\theta}^*)$\\
\colrule
 400 & u & -3280 & (-0.1186,-0.1168, 5.3536, 0.5890, 2.7193) \\
 400 & b & -3887 & (-0.1594,-0.1214, 3.1047, 0.4464, 4.9501) \\
\colrule
 450 & u & -3312 & (-0.1129,-0.0964, 5.0422, 0.5193, 2.8031) \\  
 450 & b & -3879 & (-0.1448,-0.1000, 3.6050, 0.4650, 4.9108) \\
 450 & b & -3893 & (-0.1445,-0.0995,-6.2068, 0.5853, 4.9336) \\
\colrule
 500 & u & -3325 & (-0.1084,-0.0781, 4.9732, 0.4489, 2.8520) \\
 500 & b & -3861 & (-0.1340,-0.0807,-3.9206, 0.6015, 4.8125) \\
 500 & b & -3862 & (-0.1341,-0.0809, 4.0452, 0.5022, 4.8321) \\
 \colrule
 550 & u & -3305 & (-0.1048,-0.0607,-1.5204, 0.7656, 2.8144) \\
 550 & u & -3328 & (-0.1048,-0.0607, 4.4330, 0.3724, 2.8492) \\ 
 550 & b & -3832 & (-0.1259,-0.0629,-2.3130, 0.7294, 4.6720) \\
 550 & b & -3843 & (-0.1260,-0.0633, 4.1715, 0.5688, 4.7062) \\
 \colrule
 600 & u & -3296 & (-0.1018,-0.0436,-0.8677, 1.0069, 2.7920) \\
 600 & u & -3326 & (-0.1017,-0.0438, 4.2580, 0.3304, 2.8839) \\
 600 & b & -3822 & (-0.1196,-0.0453,-1.2360,-4.8906, 4.5958) \\
 600 & b & -3841 & (-0.1197,-0.0455, 4.2284,-4.7557, 4.6951) \\
 600 & b & -3828 & (-0.1196,-0.0455,-1.2646, 0.8954, 4.5454) \\
 600 & b & -3824 & (-0.1197,-0.0456, 4.2902, 0.6672, 4.6155) \\
 \colrule 
 700 & u & -3279 & (-0.0971,-0.0063,-0.3798, 1.9723, 2.7659) \\
 700 & u & -3282 & (-0.0971,-0.0064,-0.3771,-2.9677, 2.7711) \\
 700 & u & -3323 & (-0.0970,-0.0066, 3.7883, 0.2950, 2.8984) \\
 700 & b & -3766 & (-0.1105,-0.0078,-0.4676,-2.5175, 4.3600) \\
 700 & b & -3795 & (-0.1105,-0.0086, 3.8219, 1.1784, 4.5288) \\
 700 & b & -3764 & (-0.1105,-0.0080,-0.4887, 1.9523, 4.3593) \\
 \colrule
2000 & u & -3278 & (-0.0783,-0.1267, 0.0184,-0.0293, 2.7490) \\
2000 & b & -3688 & (-0.0812,-0.1266, 0.0165,-0.0285, 3.9715) \\
 \colrule
3000 & u & -3286 & (-0.0748,-0.0352, 0.0248,-0.0102, 2.7581) \\
3000 & b & -3688 & (-0.0767,-0.0350, 0.0222,-0.0100, 3.9804) \\
\colrule 
4000 & u & -3279 & (-0.0730, 0.0967,-1.0424,-0.0050, 2.7585) \\
4000 & u & -3282 & (-0.0730, 0.0964, 0.2322,-0.0050, 2.7627) \\
4000 & b & -3683 & (-0.0744, 0.0966, 0.0819,-0.0050, 3.9644) \\
\end{tabular}
\end{ruledtabular}
\label{tab:modes_all}
\end{table}

\section{\label{app:phase_shifts}Phase shift benchmark}
\cref{tab:phase_Lambda_450_map} lists selected phase shifts for the LO EFT potential defined in Sec.~\ref{sec:theory} with $\Lambda=450$ MeV and  the MAP LEC values $\left(\tilde{C}_{^1S_0},\tilde{C}_{^3S_1},C_{^3P_0}, C_{^3P_2}\right)$ = (-0.1128, -0.0964,  5.1151,  0.5118) in the units defined below \cref{eq:lecs_def}.
\begin{table*}[b]
\caption{Phase shifts and mixing angles in degrees for relevant partial waves.}
\begin{ruledtabular}
\begin{tabular}{ccccccccc}
$T_\mathrm{lab}$ [MeV]&  $\delta_{^1S_0}$ & $\delta_{^3S_1}$ & $\delta_{^3D_1}$ & 
$\delta_{\epsilon_1}$ & $\delta_{^3P_0}$ & $\delta_{^3P_2}$& $\delta_{^3F_2}$& $\delta_{\epsilon_2}$\\
\colrule
1     & $48.11269 $ & $144.56170$ & $-0.00641 $ & $0.17917  $ & $0.13249  $ & $-0.01180 $ & $0.00001  $ & $-0.00154 $ \\
50    & $55.43553 $ & $67.70410 $ & $-7.98257 $ & $4.45167  $ & $-2.94614 $ & $-3.27873 $ & $0.31327  $ & $-1.63134 $ \\
100   & $47.35263 $ & $51.13999 $ & $-16.12281$ & $6.19131  $ & $-18.21728$ & $-9.11571 $ & $0.70523  $ & $-2.73674 $ \\
200   & $32.27681 $ & $31.20600 $ & $-26.46882$ & $8.71350  $ & $-51.24600$ & $-21.64237$ & $1.00719  $ & $-3.46681 $ \\
300   & $19.60329 $ & $17.43370 $ & $-26.71548$ & $9.20362  $ & $-82.25737$ & $-29.45361$ & $0.80583  $ & $-3.12625 $
\end{tabular}
\end{ruledtabular}
\label{tab:phase_Lambda_450_map}
\end{table*}

\end{document}